\documentclass[twocolumn,showpacs,amsfonts,amssymb,floats,aps]{revtex4-1}
\usepackage{amsmath}
\usepackage{graphicx}
\def\expect#1{\left\langle#1 \right\rangle}
\def\e{\varepsilon}
\def\mueff{\mu^2_{\rm eff}}
\def\ie{i.e.\ }
\def\etal{{\em et al.}\ }
\def\Gn{\Gamma_{\rm 0}}
\def\Gi{\Gamma_{\rm i}}
\def\Gf{\Gamma_{\rm f}}
\def\Gc{\Gamma_{\rm c}}
\def\Ui{U_{\rm i}}
\def\Uf{U_{\rm f}}
\def\Uc{U_{\rm c}}
\def\tco{t_{\rm co}}
\def\TK{T_{\rm K}}

\begin{document}
\title{Real-time dynamics induced by quenches across the quantum critical points \\
in gapless Fermi systems with a magnetic impurity
}
\author{Christian Kleine}
\author{Julian Mu{\ss}hoff}
\author{Frithjof B. Anders}
\address{Lehrstuhl f\"ur Theoretische Physik II, Technische Universit\"at Dortmund,
Otto-Hahn-Stra{\ss}e 4, 44221 Dortmund, Germany}
\begin{abstract}
The energy-dependent
scattering of fermions from a localized orbital at an
energy-dependent rate $\Gamma(\e)\propto |\e|^r$ gives rise to quantum
critical points (QCPs) in the
pseudogap single-impurity Anderson model 
separating a local moment phase with an
unscreened spin moment from a strong-coupling phase which slightly
deviates from the screened phase of standard Kondo problem.  
Using the time-dependent numerical renormalization group (TD-NRG) approach
we show that local dynamic properties always equilibrate towards a
steady-state value even for quenches across the QCP but with
systematic deviations from the thermal equilibrium depending on the
distance to the critical coupling. Local non-equilibrium properties
are presented for interaction quenches and hybridization quenches.  We
augment our numerical data by an analytical calculation that becomes
exact at short times and find excellent agreement between the
numerics and the analytical theory.  For interaction quenches within
the screened phase we find a universal function for the
time-dependent local double occupancy.  We trace back the discrepancy
between our results and the data obtained by a time-dependent Gutzwiller variational
approach to restrictions of the wave-function ansatz in the Gutzwiller
theory: while the NRG ground states properly account for the
formation of an extended spin moment which decouples from the system
in the unscreened phase, the Gutzwiller ansatz only allows the
formation of the spin moment on the local impurity orbital.
\end{abstract}

\maketitle

\section{Introduction}

The investigation of the real-time dynamics in quantum-impurity
systems (QISs) is essential for our understanding of dissipation and
decoherence in qubits and electronic transport through nanodevices.
Such systems consist of a small subsystem comprising a finite
number of degrees of freedom, interacting with an infinitely large
environment of noninteracting particles.

Elzerman \etal \cite{Kouwenhoven2004} have reported the usage
of gate-voltage pulses for a single-shot readout of the spin
configuration of a single-electron transistor in a finite magnetic
field.  Such a system can be modeled by an Anderson impurity model
\cite{Anderson61} coupled to a noninteracting metallic host.  While
the normal single-impurity Anderson model (SIAM) has a well-established
and rather simple phase diagram
\cite{KrishWilWilson80a,*KrishWilWilson80b,BullaCostiPruschke2008},
it can be viewed as a special case ($r=0$) of a more general class of models
\cite{BullaPruschkeHewson1997,GonzalezBuxtonIngersent1998,GlossopLogan2003,Vojta2006}
whose coupling function $\Gamma(\e)$ to the local impurity contains a
pseudogap; \ie $\Gamma(\e) \propto |\e|^r$.  Withoff and Fradkin were
the first to point out the existence of a critical coupling
\cite{WithoffFradkin1990} using a perturbative renormalization group
argument.  This system exhibits a wide variety of different phases for
$r>0$. These phases are characterized by different fixed points whose
properties and occurrence depend on the bath exponent $r$ as well as
on particle-hole symmetry or the absence of it.  For $0<r<1/2$, there
exists
\cite{Chen1995,Ingersent1996,BullaPruschkeHewson1997,GonzalezBuxtonIngersent1998}
a critical coupling strength $\Gc$ governing the transition between a
local moment (LM) phase for a weak coupling and a strong-coupling (SC)
phase for a large coupling to the metallic host.

In this paper, we analyze the real-time dynamics of different quenches
within the LM or the SC phase but also quenches across the quantum
critical point (QCP) from one to the other phase for the particle-hole
symmetric pseudogap Anderson impurity model (pg-SIAM).  We employ a
recent extension of Wilson's numerical renormalization group (NRG)
\cite{Wilson75,BullaCostiPruschke2008} to the nonequilibrium quench
dynamics, the time-dependent NRG (TD-NRG)
\cite{AndersSchiller2005,AndersSchiller2006,NghiemCosti2014}.

While the quench dynamics in the SIAM ($r=0$) has been investigated
\cite{AndersSchiller2005} using the TD-NRG, the non-equilibrium
dynamics in the pg-SIAM within and across the QCP has only recently
been addressed by a time-dependent Gutzwiller ansatz
\cite{Schiro2012}.  Using this extension of the well-established
variational Gutzwiller technique \cite{Gutzwiller1965,Schickling2012}
to nonequilibrium \cite{SeiboldLorenzana2001,Schiro2012,LanataStrand2012} it has
been demonstrated that the pseudogap coupling function can yield
nontrivial dynamics as a consequence of the diverse low-energy fixed
points of the model \cite{Schiro2012}.

The pg-SIAM has been extensively investigated in the context of Kondo
impurities in unconventional superconductors
\cite{Chen1995,Ingersent1996,BullaPruschkeHewson1997,GonzalezBuxtonIngersent1998,Vojta2006,GlossopLogan2003}
or in the context of defects in graphene sheets
\cite{FritzVojta2013,Lo-graphene-2014}.  Some of the low-energy
properties of the fixed points have been worked out in detail
\cite{FritzVojta2004,VojtaFritz2004,Schneider2011} and it has been
shown that the universality class of the fixed point changes with the
coupling function exponent $r$.

One intriguing property of this model is the absence of the Kondo
screening at low coupling strength
\cite{WithoffFradkin1990,Chen1995,Ingersent1996,GonzalezBuxtonIngersent1998}
while for a large coupling a SC fixed point is found but with an only
partially screened moment for particle-hole symmetric models.  The
question arises how these orthogonal ground states 
of the different
phases influence the real-time dynamics of a system driven out of
equilibrium by a quantum quench.

Since an effective spin degree of freedom decouples from the impurity
in the LM phase, this fixed point (FP) property is expected to have a
strong influence on the steady-state formation and the thermalization
when quenched into the LM phase. We will show, however, that an
oversimplified picture does not hold and requires some modification.

In all our quenches we find a well-defined steady state at long times.
The distance from the thermal equilibrium for the same final Hamiltonian
serves as a measure for the degree of thermalization.
While for quenches within the SC phase, the system thermalizes within the
numerical accuracy of the TD-NRG \cite{AndersSchiller2005,AndersSchiller2006,NghiemCosti2014},
the deviations from the thermal equilibrium remain negligible even for quenches across the QCP 
very close to the QCP. The deviation appears to be a continuous function
of the distance to the critical coupling.

\subsection{Physical picture}

The surprising finding of a well-defined steady state even for 
quenches across the QCP into the LM phase can be understood in terms of
the known fixed point properties of the model.
It has been already pointed out in Ref.\ \cite{GonzalezBuxtonIngersent1998}
that the local thermodynamic properties such as double occupancy as well as
the fractional local moment on the impurity are continuous across the QCP.

In the strong-coupling (SC) phase, the Kondo temperature $\TK$
governing the excitations around the SC FP vanishes at the QCP and
increases with increasing coupling to the pseudogap metallic host.
The crossover scale $T^*$ taking the role of $\TK$ in the LM phase
characterizes the excitations around the LM FP,
decreases with increasing coupling and also vanishes at the QCP.

The associated length scales $\xi^* = v_{\rm F}/T^*$ ($\xi_{\rm K} =  v_{\rm F}/\TK$)
-- $v_{\rm F}$  being the average Fermi velocity of the host material --
can be interpreted as estimate for the spatial extension of the decoupled
local moment (or the Kondo singlet in the SC phase.)
Only for a very large local Coulomb repulsion $U$,
the crossover scale $T^*$  becomes  large  indicating that the local moment is mainly formed
closely to or on the impurity.

Bearing in mind these known equilibrium properties
\cite{GonzalezBuxtonIngersent1998} of the model it becomes apparent
that an ansatz for the ground state restricted to a local moment
formation on the impurity site only, as used in the Gutzwiller approach
\cite{Gutzwiller1965,SeiboldLorenzana2001,LanataStrand2012,Schiro2012},
significantly underestimates the critical coupling $\Gc$.
Then, the local moment formation can only occur in such an approach at
much lower coupling $\Gamma$ or much larger local Coulomb repulsion $U$
compared to the solution provided by the NRG.

This has also a profound consequence for the observed real-time dynamics.
Due to the extended nature of the decoupled local moment,
the local observable still can explore a larger phase space of
itinerant states and, therefore, shows signs of thermalization close to the QCP.

Reducing the size of the local moment by increasing $U$ away from the
critical $\Uc$ will still yield a steady state whose asymptotic properties
start to deviate significantly from the thermal expectation values:
Due to the increase of the nondecaying fraction of the expectation
value \cite{Mazur1969,Suzuki1971,UhrigHackmann2014} the difference
between the steady-state and the thermal expectation value increases
in the LM phase.

This difference, however, will strongly depend on the matrix element
of the operator with the decoupling degree of freedom (DOF).
While for the local double occupancy as a measure for the local correlations
we find an increasing deviation between the long-time steady-state value
and the thermal expectation value, the energy flow into the hybridization energy
seems to be unaffected when crossing over to the localized regime.
It is a rather surprising finding that hybridization energy 
shows thermalization even in the LM phase, although it has been conjectured
that the TD-NRG might have problems in describing properly
the energy flow \cite{Rosch2012}.

\subsection{Plan of the paper}

The main objective of this paper is to discuss the real-time dynamics of
the pseudogap SIAM with respect to interaction and hybridization quenches
within a given phase and across the quantum critical point.

To be more specific, we will introduce the model in Sec.\ \ref{sec:pg-siam}
and briefly the TD-NRG in Sec.\ \ref{sec:td-nrg}.
We continue with a short overview over the rich phase diagram in
Sec.\ \ref{sec:overview-phases} -- a much more comprehensive review can be found
in Refs.\ \cite{GonzalezBuxtonIngersent1998,Vojta2006} -- in order to define
the types of quenches that will be investigated in Sec.\ \ref {sec:results},
the main part of the paper.

For completeness and defining the parameter space, we present the known NRG phase diagram for
the symmetric pg-SIAM and also discuss the differences between the NRG and the equilibrium Gutzwiller
results in Sec.\ \ref{sec:equilibriumproperties}.
We start with analyzing our data for interaction quenches, \ie the sudden switching on the local Coulomb repulsion.
For small $U$ and finite hybridization, the system remains in the SC phase.
In Sec.\ \ref{sec:interactionquencheswithin}, we show that universality can be found for this type of quench
with a $U$-independent time scale.
Interaction quenches across the QCP are investigated in Sec.\ \ref{sec:interactionquenchesover}.
We address the difference between equilibration and thermalization.

Section \ref{sec:hybridisationquenches} is devoted to the two types of hybridization quenches.
We augment our TD-NRG results with a perturbative analysis --  details can be found in the Appendix --
and show an excellent agreement between the numerics and the analytics in Sec.\ \ref{sec:analytic-result}.
We also discuss the energy flow from the reservoir to the impurity after the quench in Sec.\ \ref{sec:energy-flow}.
We end the paper with a short conclusion.

\section{Theory}
\label{sec:theory}

\subsection{The pseudo-gap single impurity Anderson model (pg-SIAM)}
\label{sec:pg-siam}

We consider a magnetic impurity comprising a spin-degenerated level that is coupled to a single conduction band.
The Hamiltonian consists of three parts:
$H_{\rm c}$ accounts for the conduction band of noninteracting electrons
\begin{align}
  H_{\rm c} = \sum_{\vec{k}\sigma} \e_{\vec{k}\sigma} c_{\vec{k}\sigma}^\dagger c_{\vec{k}\sigma}^{\phantom{\dagger}}
\end{align}
where $c_{\vec{k}\sigma}^{\phantom{\dagger}}$ annihilates an electron 
with energy $\e_{\vec{k}\sigma}$ and spin $\sigma$, $H_{\rm imp}$ models the magnetic impurity 
\begin{align}
  H_{\rm imp} = \sum_{\sigma} \e_{d\sigma} d_{\sigma}^\dagger d_{\sigma}^{\phantom{\dagger}}
              + U n_{\mathrm{d}\uparrow} n_{\mathrm{d}\downarrow} ,
\end{align}
where $d^\dagger_\sigma$ creates an electron in the localized impurity orbital with spin $\sigma$,
the energy $\e_{d\sigma}=\e_d -H\sigma$, $H$ is the local magnetic field
measured in unit of energy, and $U>0$ denotes the Coulomb repulsion between two localized electrons with opposite spin.
Hereby $n_{\mathrm{d}\sigma}=d_\sigma^\dagger d_\sigma^{\phantom{\dagger}}$ is the occupation operator of the level with spin $\sigma$.
These two subsystems are coupled by the hybridization term 
\begin{align}
  H_{\rm hyb} = \sum_{\vec{k} \sigma} V_{\vec{k}} \left( c_{\vec{k}\sigma}^\dagger d_{\sigma}^{\phantom{\dagger}}
            + d_\sigma^\dagger c_{\vec{k}\sigma}^{\phantom{\dagger}} \right)
\end{align}
so that the full dynamics is given by $H= H_{\rm c} +H_{\rm imp} + H_{\rm hyb}$.

It has been realized \cite{BullaPruschkeHewson1997,GonzalezBuxtonIngersent1998,GlossopLogan2003} that the dynamics
of the magnetic impurity is fully determined by the coupling function
\begin{align}
  \Gamma_\sigma(\epsilon ) = \pi \sum_{\vec{k}} V_{\vec{k}}^2 \delta \left( \epsilon - \epsilon_{\vec{k}\sigma} \right)
\end{align}
which we will take as spin-independent in the following.
In real materials such as $d$ superconductors \cite{Vojta2006} or graphene sheets \cite{Kanao2012,VojtaFritzBulla2010,Lo-graphene-2014}
$\Gamma(\epsilon)$ is a complicated function of energy. It turned out, however, that only the
low-energy part of the spectrum close to the chemical potential is relevant for the structure of the low-energy fixed points.
Ignoring the high-energy details, $\Gamma(\epsilon)$ is replaced by the particle-hole symmetric power-law form
\begin{align}
  \Gamma \left( \epsilon \right) = \left(r + 1 \right) \Gamma_{\rm 0} \left| \frac{\epsilon}{D} \right|^r 
                                   \Theta \left( D - \left| \epsilon \right| \right)
\end{align}
where the cutoff $D$ defines the effective band width
\cite{WithoffFradkin1990,Chen1995,Ingersent1996,BullaPruschkeHewson1997,GonzalezBuxtonIngersent1998,GlossopLogan2003,Vojta2006}.
The normalization factor $(r+1)$ ensures that the integral over the coupling function,
\begin{align}
  \pi V_{\rm 0}^2 = \int d\e \Gamma(\e) =2 \Gamma_{\rm 0} D
  \quad ,
\end{align}
remains independent of the bath exponent $r \ge 0$.
The parameter $\Gn$ serves as the energy scale of the problem that turns into
the standard charge fluctuation scale for a constant density of states ($r=0$).
While $r=0$ and $r=1$ are the prototypical experimental realizations,
we take $r$ as an arbitrary parameter of the model.

The particle-hole asymmetry is governed by the deviation of $\Delta \epsilon = 2\epsilon_{\rm d} + U$ from zero energy.
Unless otherwise stated, we focus only on the particle-hole symmetric case in this paper.

\subsection{The time-dependent numerical renormalization group (TD-NRG)}
\label{sec:td-nrg}

In this work, we exploit the NRG approach
\cite{Wilson75,BullaCostiPruschke2008} for obtaining all
thermodynamic properties of the pg-SIAM.  The key idea of Wilson was
to map the logarithmically discretized coupling function $\Gamma(\e)$
onto an effective semi-infinite tight-binding chain where the impurity
is coupled only to the first chain link.  The energy hierarchy is
controlled by the discretization parameter $\Lambda>1$, and the
problem is solved by iterative diagonalization.

Due to the exponential growth of the Hilbert space, high-energy states
are discarded after each iteration.  Since the initial basis set is
known, the set of all discarded states form a complete basis set
\cite{AndersSchiller2005,AndersSchiller2006} and simultaneously serve
as an approximate eigenbasis of the Hamiltonian governing the time
evolution of the problem.

Then the time-dependent expectation value $\expect{O(t)}$ of a
general local operator $\hat{O}$ can be cast into the form
\begin{align}
  \expect{O(t)} = \sum_{m}^{N} \sum_{r,s}^{\rm trun} \; {\bf e}^{i t (E_{r}^m - E_{s}^m)}
                   O_{r,s}^m \rho^{\rm red}_{s,r}(m)
  \quad ,
  \label{eqn:time-evolution-intro}
\end{align}
where $E_{r}^m$ and $E_{s}^m$ are the dimensionful NRG eigenenergies of the Hamiltonian $H_{\rm f} = H(t>0)$
at iteration $m \le N$, $O_{r,s}^m$ is the matrix representation of $\hat{O}$ at that iteration,
and $\rho^{\rm red}_{s,r}(m)$ is the reduced density matrix defined as
\begin{align}
  \rho^{\rm red}_{s,r}(m) = \sum_{e} \langle s,e;m|\hat{\rho}_{\rm 0} |r,e;m \rangle
  \label{eqn:reduced-dm-def}
\end{align}
where $\hat{\rho}_{\rm 0}$ is the initial density operator of the problem prior to the quench.
The restricted sum over $r$ and $s$ in Eq.\ \eqref{eqn:time-evolution-intro} requires that at least one
of these states is discarded at iteration $m$.

Implementing the TD-NRG requires two NRG runs: one for the initial Hamiltonian $H_{\rm i} = H(t<0)$ to construct
the initial density operator $\hat{\rho}_0$ of the system and one for $H_{\rm f}$ to obtain the approximate eigenbasis
governing the time evolution in Eq.\ \eqref{eqn:time-evolution-intro}.
For more details on the TD-NRG see Refs.\ \cite{AndersSchiller2005,AndersSchiller2006}.
Recently, the TD-NRG has been extended to use the full density matrix including pulsed Hamiltonians
\cite{NghiemCosti2014} and periodic switching \cite{EidelsteinGuettgeSchillerAnders2012}.

In the original implementation of the TD-NRG
-- see Eq.\ (42) in  Ref.\ \cite{AndersSchiller2006} --
each phase factor in Eq.\ \eqref{eqn:time-evolution-intro}
\begin{align}
  e^{i t (E_{r}^m - E_{s}^m)} \to e^{i t (E_{r}^m - E_{s}^m) -\Gamma_m t}
  \label{eq:gamma_m}
\end{align}
was Lorentz-broadened with an energy-resolution-dependent damping factor 
$\Gamma_m = \alpha \omega_m$ proportional to the energy scale
$\omega_m = D \Lambda^{-(m-1)/2} ( 1 + 1/\Lambda ) / 2$
at iteration $m$ for all $E_{r}^m - E_{s}^m \not= 0$, and $\alpha = O(1)$
see Ref.\ \cite{AndersSchiller2005,AndersSchiller2006} for details.
Such a broadening smoothens the discretization-related oscillations in the same spirit as
the broadening of the NRG Lehmann representation of equilibrium spectral functions
\cite{CostiHewsonZlatic94,PetersPruschkeAnders2006,*WeichselbaumDelft2007,BullaCostiPruschke2008}.
If, however, localized states decoupled from the continuum contribute to the expectation value as
discussed in the local moment phase of the model \cite{GonzalezBuxtonIngersent1998,Schiro2012},
such a broadening could wrongly damp out oscillatory contributions at long times.
In order to avoid any prejudice, we usually set $\alpha = 0$
and use the $z$ averaging to minimize discretization-related oscillations.
Therefore, usually all our data contain some finite-size-related noise at very long times.
In order to illustrate the effect of the original TD-NRG broadening, we have
used $\alpha = 0.4$ in Fig.\ \ref{fig:hqLMtoLM_hyb}(c) and $\alpha = 1$ in Fig.\ \ref{fig:hqSz}
instead of $z$ averaging.

\subsection{Overview of the phases and the quench types}
\label{sec:overview-phases}

The equilibrium phase diagram of the pg-SIAM is very rich and has been
carefully explored by Gonzalez-Buxton and Ingersent
\cite{GonzalezBuxtonIngersent1998}.  Details can also be found in
Refs.\ \cite{Vojta2006} and \cite{BullaCostiPruschke2008}.

For a particle-hole (ph) symmetric model, an unstable intermediate
coupling fixed point governs the transition between the local moment
(LM) fixed point (FP) and the strong-coupling (SC) FP for $0 < r < 1/2$,
where the critical coupling ratio $\Uc/\Gn$ depends on the band
width $D$ and the exponent $r$.
The intermediate FP is also unstable with respect to ph-symmetry breaking
since potential scatting is a marginally relevant perturbation at this FP.

In contrast to the standard SC FP for $r = 0$, the SC FP of the
ph-symmetric pg-SIAM is characterized by a residual impurity entropy
$S_{\rm imp} = 2 r k_B \ln(2)$ and a residual unscreened effective
local moment \cite{GonzalezBuxtonIngersent1998} 
of $\mueff(0) = \lim_{T \to 0} \mueff(T) = r/8$.

For $1/2 < r$ the low-energy density of states is too small to screen
the local spin, and only the LM FP remains stable for all coupling
strengths at ph symmetry.  Breaking ph symmetry, an asymmetric SC (ASC)
fixed point is found for all $r$. Its thermodynamic properties are
closely related to the standard $r = 0$ SC FP since $S_{\rm imp} = 0$
and $\mueff(0) = 0$.

In addition, a second intermediate coupling fixed point at a finite
value of the potential scattering is found for $r^*\approx 0.375 < r$,
where the perturbation $V - V_{\rm c}$ with respect to the critical
potential scattering $V_{\rm c}$ is a marginal irrelevant operator
\cite{GonzalezBuxtonIngersent1998}.

In this paper, however, we will focus on the non-equilibrium dynamics
of the particle-hole symmetric model close to and across the
quantum-phase transition. Therefore, we mainly focus on
the parameter regime $0 < r < 1/2$.  There are two different ways to
drive the ph-symmetric system across the quantum critical point (QCP)
for $0 < r < 1/2$.  For a fixed value of $U$ we can vary the
hybridization strength $\Gn$, which will be called
hybridization quench (HQ) in the following, or for a fixed
$\Gn$ we can change $U$ that defines the interaction quench
(IQ).  In Table \ref{tb:phases} we summarize our notation for the
different quench types.
The naming of the phases refers to the equilibrium low-temperature FP
of the final Hamiltonian $H_{\rm f}$ after the quench.

In equilibrium, the SC FP can be reached in two ways:
choosing the charge-fluctuation scale $\Gn > \Gc(U)$ for fixed $U$ or
by setting $U < \Uc(\Gn)$ for fixed $\Gn$.
The resulting phase diagram for $\Uc(r)$ and $\Gc(r)$ for fixed $\Gn$
and $U$ respectively is shown in Fig.\ \ref{fig:phasediagram}(b).
The details of how the phase diagram is obtained from equilibrium NRG
calculations is presented in Sec.\ \ref{sec:equilibriumproperties} below.

An approximative description of the ground state wave function of the
model for a finite hybridization has been proposed using a Gutzwiller
wave-function ansatz \cite{Schiro2012}.  For $r=0$, the quasiparticle
renormalization factor $Z \propto \TK^G \propto \exp(- \pi U / 16
\Gn)$ is a smooth function of the Coulomb interaction $U$ and has been
interpreted as effective Kondo temperature within the Gutzwiller
approach \cite{Gutzwiller1965}.
Note, however, that (i) the exponent differs by a factor of $2$ from
the standard Kondo temperature and, therefore, underestimates the
exponential decay \cite{Lanata2010} and (ii) an incorrect Kondo scale
is found for the wide-band limit as has been pointed out in
Ref.\ \cite{Lanata2010}.

For $r > 0$, a finite critical $\Uc$ and a QCP is found
\cite{Schiro2012} for all $r$.
The prediction for $\Uc$ by the Gutzwiller approach has been added
to Fig.\ \ref{fig:phasediagram}(b) as an analytical curve for the wide-band limit.
That restricts the possible validity of the Gutzwiller approach to $r < 1/2$
since for $1/2 < r$ the LM FP is stable for all $U > 0$.
The prediction of a SC phase even for $1/2<r$ by the Gutzwiller approach
is directly related to the restriction of the wave function to 
the formation of a decoupled spin moment on the impurity site while the true
ground state properly accounts for the formation of an extended spin moment
as we will discuss in the next section. 
The comparison with the correct NRG phase boundary reveals again
an underestimation of the renormalization effects and, hence,
an overestimation of the critical $\Uc$.
We have to bare in mind these limitations of the Gutzwiller
approach of the equilibrium when comparing recent results
\cite{Schiro2012} obtained with a time-dependent Gutzwiller approach
\cite{SeiboldLorenzana2001,LanataStrand2012} with our TD-NRG data.

\begin{table}[tb]
  \caption{%
    Summary of the quench types.
    The critical parameters $\Gc$ and $\Uc$ are $r$ dependent and presented in Fig.\ \ref{fig:phasediagram}.
    The phase type refers to the low-temperature FP of the final Hamiltonian.
    The numerical values of the critical parameters for different band widths $D$ and exponents $r$ are given
    in Table \ref{tb:phasesValues}.
  }
  \centering
  \vspace{2ex}
  \begin{tabular}{lcc}
    \hline \hline
    Phase & Hybridization Quench & Interaction Quench \\ \hline
    SC & $\Gn > \Gc(U)$ & $U < \Uc(\Gn)$ \\
    LM & $\Gn < \Gc(U)$ & $U > \Uc(\Gn)$ \\ \hline \hline
  \end{tabular}
  \label{tb:phases}
\end{table}

\section{Results}
\label{sec:results}

We begin with a short review of the established thermodynamic
properties in Sec.\ \ref{sec:equilibriumproperties} and discuss the
differences between the NRG findings
\cite{GonzalezBuxtonIngersent1998} and the predictions of the
equilibrium Gutzwiller ansatz.  Then we proceed with the results of
our TD-NRG calculations for interaction quenches (IQs) in Sec.\
\ref{sec:interactionquenches} and for hybridization quenches (HQs) in
Sec.\ \ref{sec:hybridisationquenches}.

\subsection{Equilibrium properties}
\label{sec:equilibriumproperties}

\begin{figure}[tb]
  \centering
  \includegraphics[width=\linewidth]{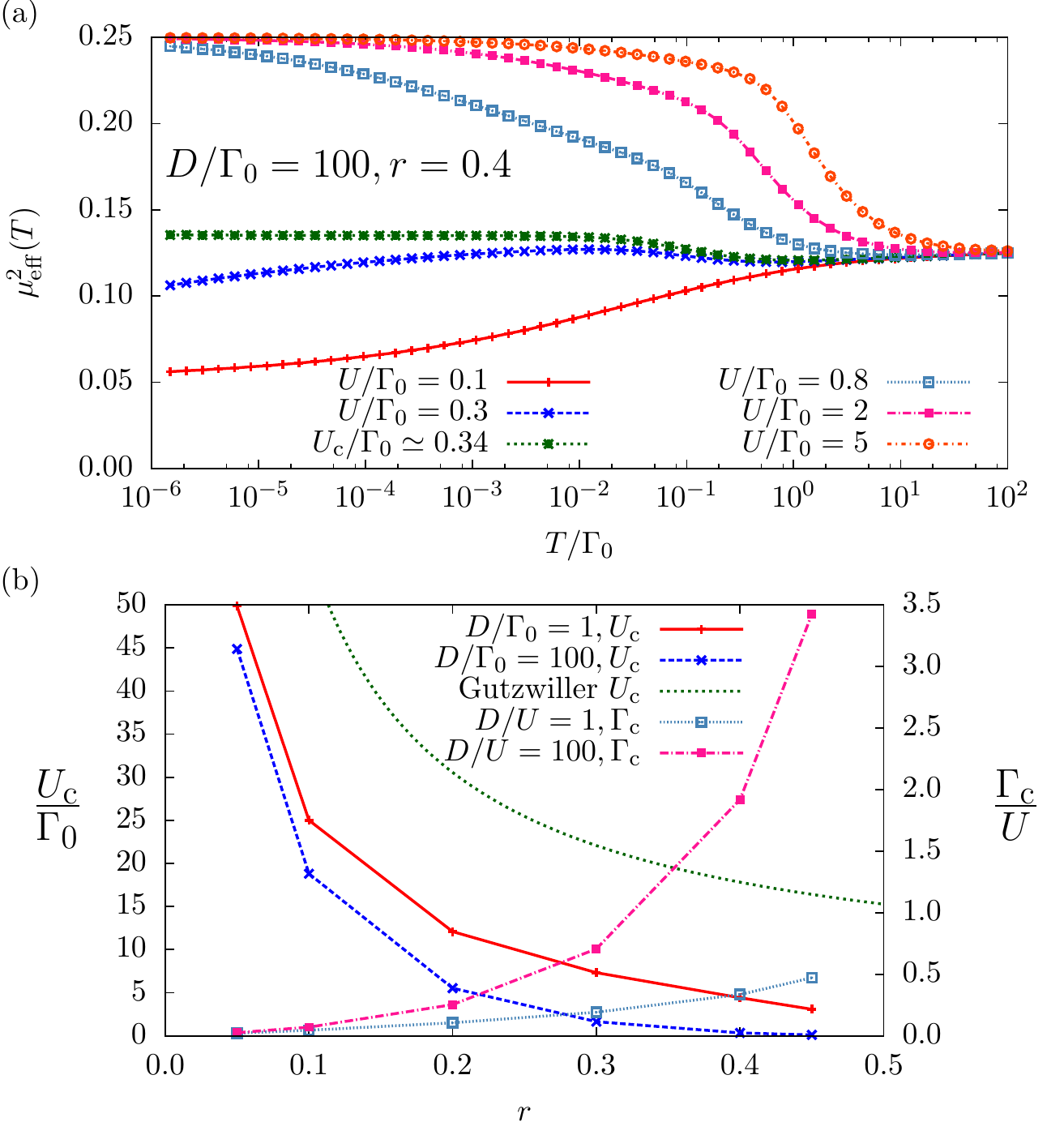}
  \caption{%
    (Color online)
    (a) Effective magnetic moment for different
    Coulomb interactions $U$ with $D/\Gn=100$ and $r=0.4$. The critical
    parameter for this setup is $\Uc/\Gn = 0.3392$. (b) Phase diagram
    for different bath exponents $r$ and bandwidths $D$.
    For comparison the analytical form of the Gutzwiller ansatz (cf. Ref.\
    \cite{Schiro2012}) is added.
  }
  \label{fig:phasediagram}
\end{figure}

In order to set the stage for the real-time dynamics, we extend the review
of the equilibrium properties to the temperature-dependent impurity observables
\cite{GonzalezBuxtonIngersent1998,Vojta2006,BullaCostiPruschke2008}.

One important quantity revealing the QCP is the effective local moment
\begin{align}
  \mueff(T)  = \Delta \left( \expect{S_{\rm z}^2 (T)} - \expect{S_{\rm z}(T)}^2 \right)
  \label{eq:chis}
\end{align}
whereby $\Delta \left( X \right)$ is measuring the observable $X$ in the
presence of the impurity and the Wilson chain and subtracts the effect
of the quantity $X$ of the pure Wilson chain without the impurity at
temperature $T$, and $S_{\rm z}$ denotes the $z$ component of the total
spin of the system.
By this definition \cite{Wilson75,GonzalezBuxtonIngersent1998} the
impurity contribution of the quantity $X$ is extracted.

Note, however, that $\mueff(T)$ does not measure the local impurity
spin observable but rather the difference in the total system
properties with and without the impurity.  Hence, the effective
impurity spin momentum $\mueff(T)$ is in general related to a degree
of freedom (DOF) comprising a linear combination of local and
conduction electron spin observables.  This will become important for
understanding the impurity expectation values in the different phases.

In the LM FP, the effective local moment of the impurity spin, \ie,
$\mueff(0) = \lim_{T \to 0} \mueff(T)$, is given by those of the free
spin $1/4$ while for the symmetric SC FP a residual moment $r/8$ has
been found \cite{GonzalezBuxtonIngersent1998} revealing the inability
of a power-law density of states to completely screen the Kondo spin.

In Fig.\ \ref{fig:phasediagram}(a), $\mueff(T)$ is shown as function
of temperature for different Coulomb interactions $U$ for the bath
exponent $r=0.4$ with the band width $D/\Gn=100$.  If not mentioned
otherwise we used the NRG parameter \cite{Wilson75} $\Lambda=2$ and
kept $N_{\rm S}=2000$ states after each iteration step.

For $U/\Uc < 1$ the effective local moment $\mueff(T)$ declines at low
temperatures towards the SC FP value $r/8$ whereas for $U/\Uc > 1$ it
inclines towards $1/4$ indicating a free local moment of the LM FP.
For $U/\Gn = 0.34 \approx \Uc$, $\mueff(T)$ approaches the value of
the unstable intermediate coupling FP.
We used the clear distinction between $\mueff(T)$ of
the symmetric SC FP and the LM FP to define the critical Coulomb
interaction $\Uc$ at which the value of the intermediate coupling FP
is obtained.

In Fig.\ \ref{fig:phasediagram}(b) we present the phase diagram of the
ph-symmetric pg-SIAM defined by the critical parameter $\Uc$ (or $\Gc$
respectively) for different band widths $D$.
The numerical values of the critical parameters are given in
Table \ref{tb:phasesValues}.
We observe a strong influence of the band width $D$ onto the
critical $\Uc$, and our $\Gc$ agrees excellently with Fig.\ 5
in Ref.\ \cite{GonzalezBuxtonIngersent1998}.

In order to qualitatively understand the phase diagram in
Fig.\ \ref{fig:phasediagram}(b), we start from $U=0$, where we always
find a SC FP for $r < 1/2$.
Since we systematically eliminate the high-energy degrees in a RG procedure,
a finite $U$ will matter only once the effective band width
$D \to D_{\rm eff}$ has reached the order $U$.
At those energies, the system starts detecting the differences between
the local doubly occupied state and the local moment states, and the
effective coupling to the remaining conduction band is given by $\Gamma(U)$.
Since $\Gamma(U)$ decreases with increasing $r$ and increasing $D$,
$\Uc$ also must decrease.

From the NRG calculation of the local moment as depicted in Fig.\
\ref{fig:phasediagram}(a) it is apparent that the approach to the LM
FP close to the QCP is governed by a small energy scale $T^*$
vanishing at $\Uc$ \footnote{The precise mathematical definition of
$T^*$ will be given in Sec.\ \ref{sec:interactionquenchesover}.}.
Consequently all energy scales contribute to the local moment
formation in the LM phase close to $\Uc$, similar to estimated
dimension of the Kondo cloud by $\xi_{\rm K} = v_{\rm F} / \TK$ in the SC
phase
\cite{Barzykin1996,Barzykin1998,Affleck2001,Affleck2005,Affleck2008,LechtenbergAnders2014}.
Therefore, we interpret $\xi^* \propto v_{\rm F} / T^*$ as an estimate
for the spatial extension of the local moment
decoupling from a free conduction band.

Additionally we have included  the analytical prediction for the
critical Coulomb interaction $\Uc/\Gn = 16 (r+1) / (\pi r)$ derived
by a Gutzwiller ansatz \cite{LanataStrand2012} for the pg-SIAM \cite{Schiro2012}
into Fig.\ \ref{fig:phasediagram}(b).
The Gutzwiller approach systematically overestimates $\Uc$ since its
wave-function ansatz is trying to strictly enforce a formation of the
free local moment on the impurity in the LM phase.
Such a picture is only valid for very large $U$ for that the system is
already deeply located in the LM phase.
The Gutzwiller ansatz has two shortcomings:
(i) it overestimates the critical $\Uc$ and, therefore, the SC regime 
since it cannot describe an extended local moment formation and
(ii) the local impurity decouples completely from the conduction band
in the LM phase which will have profound consequences for the real-time
dynamics within such an approach.

\begin{table}[tb]
  \caption{%
    Summary of the critical parameters $\Uc(\Gn)$ and $\Gc(U)$ as they are presented in Fig.\ \ref{fig:phasediagram}.
  }
  \centering
  \begin{tabular}{lcccc}
    \hline \hline
    r & $\dfrac{\Uc}{\Gn}$; $\dfrac{D}{\Gn}=1$ & $\dfrac{\Uc}{\Gn}$; $\dfrac{D}{\Gn}=100$ &
      $\dfrac{\Gc}{U}$; $\dfrac{D}{U}=1$ & $\dfrac{\Gc}{U}$; $\dfrac{D}{U}=100$  \\[1.5ex] \hline
    0.05           & 49.84\phantom{00} &           44.87\phantom{00} &           0.02154 &           0.0268\phantom{0} \\
    0.1\phantom{0} & 25.00\phantom{00} &           18.81\phantom{00} &           0.04586 &           0.07094           \\
    0.2\phantom{0} & 12.10\phantom{00} &           \phantom{0}5.534\phantom{0} & 0.10637 &           0.2535\phantom{0} \\
    0.3\phantom{0} & \phantom{0}7.331\phantom{0} & \phantom{0}1.6414 &           0.19246 &           0.7060\phantom{0} \\
    0.4\phantom{0} & \phantom{0}4.426\phantom{0} & \phantom{0}0.3392 &           0.3358\phantom{0} & 1.918\phantom{00} \\
    0.45           & \phantom{0}3.0826 &           \phantom{0}0.1075 &           0.4724\phantom{0} & 3.423\phantom{00} \\
    \hline \hline
  \end{tabular}
  \label{tb:phasesValues}
\end{table}

\begin{figure}[tb]
  \centering
  \includegraphics[width=\linewidth]{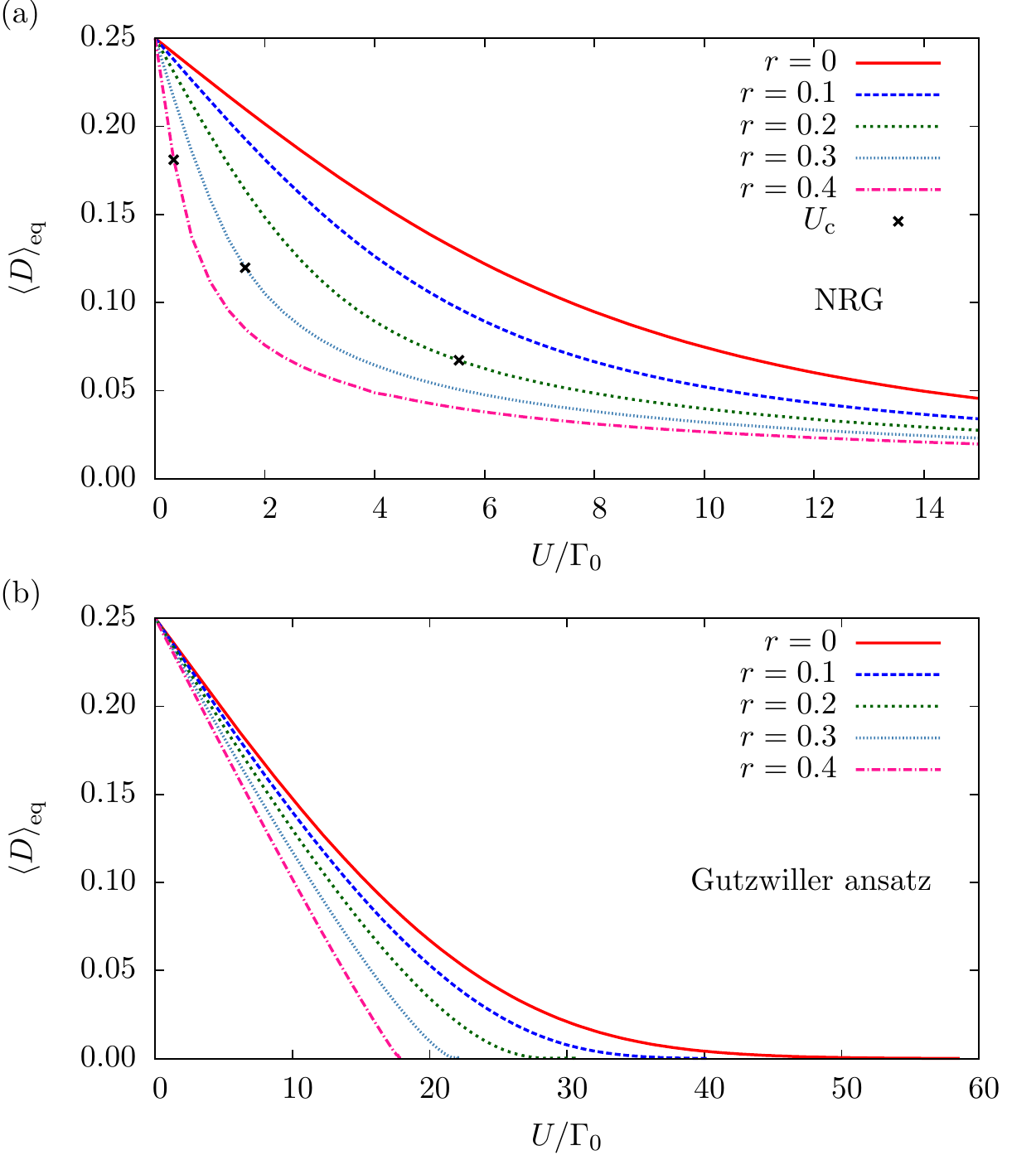}
  \caption{%
    (Color online)
    Equilibrium expectation value of the double occupancy $\expect{D}_{\rm eq}$ versus $U$ for different
    bath exponents $r$ and $D/\Gn = 100$ calculated  (a) using the equilibrium NRG and  (b)
    using the equilibrium Gutzwiller ansatz \cite{LanataStrand2012,Schiro2012}.
  }
 \label{fig:equi-double-occupancy}
\end{figure}

Since we investigate the real-time dynamics of the double occupancy
$\expect{D(t)}$ for different type of quenches, we also provide
results for the equilibrium double occupancy $\expect{D}_{\rm eq}$ vs
$U/\Gn$ at fixed $D/\Gn=100$ in Fig.\ \ref{fig:equi-double-occupancy}.
The NRG data depicted in Fig.\ \ref{fig:equi-double-occupancy}(a)
demonstrates that $\expect{D}_{\rm eq}$ is continuous across the QCP.
This continuity of local observables across the QCP was already
pointed out by Gonzalez-Buxton and Ingersent more the 15 years ago --
see Fig. 8 in Ref. \cite{GonzalezBuxtonIngersent1998}.
We also added Gutzwiller equilibrium data calculated for the same
hybridization function $\Gamma(\epsilon)$ as Fig.\ \ref{fig:equi-double-occupancy}(b)
to illustrate the difference to the NRG.
Within the Gutzwiller ansatz, $\expect{D(U_c)}_{\rm eq} = 0$ and
remains zero for $U > \Uc$.  Consequently, the physical properties of
the Gutzwiller wave function deviate significantly from the true ground
state as obtained by the NRG.  Apparently, the Gutzwiller wave
function ansatz cannot be applied in the LM phase close to the QCP
since it misses the spatial extension of the local moment that is
decoupling from the system.

\subsection{Interaction quenches}
\label{sec:interactionquenches}

In an interaction quench, we switch the Coulomb repulsion from its
initial value $\Ui$ at times $t < 0$ to the value $\Uf$ for $0 \le t$.
In order to maintain ph symmetry at all times, \ie $U(t) + 2
\epsilon_{\rm d}(t) = 0$, we enforce also a switching in the $d$-level
energy $-2 \epsilon_{\rm d}(t) = \Theta(-t) \Ui + \Theta(t) \Uf$.  The
hybridization strength $\Gamma(t) = \Gi = \Gf = \Gn$ is kept constant
and is used as a unit of energy.  Short, intermediate, and long times
will correspond to $t\Gn \ll 1$, $t\Gn \sim 1$ and $t\Gn \gg 1$.

We prepare the system initially in the uncorrelated state by setting
$\Ui = \epsilon_{\rm i} = 0$.  Since the impurity is coupled to the
conduction band, the system approaches the the SC FP for $T \to 0$.
Therefore, the initial double occupancy is given by the uncorrelated
value $\expect{D}_{\rm eq} = 1/4$.  In the LM FP, the double and the
empty states on the magnetic impurity remain unoccupied, and
$\expect{D}_{\rm eq} = 0$.

\subsubsection{Quenches within the SC phase}
\label{sec:interactionquencheswithin}

\begin{figure}[tb]
  \centering
  \includegraphics[width=\linewidth]{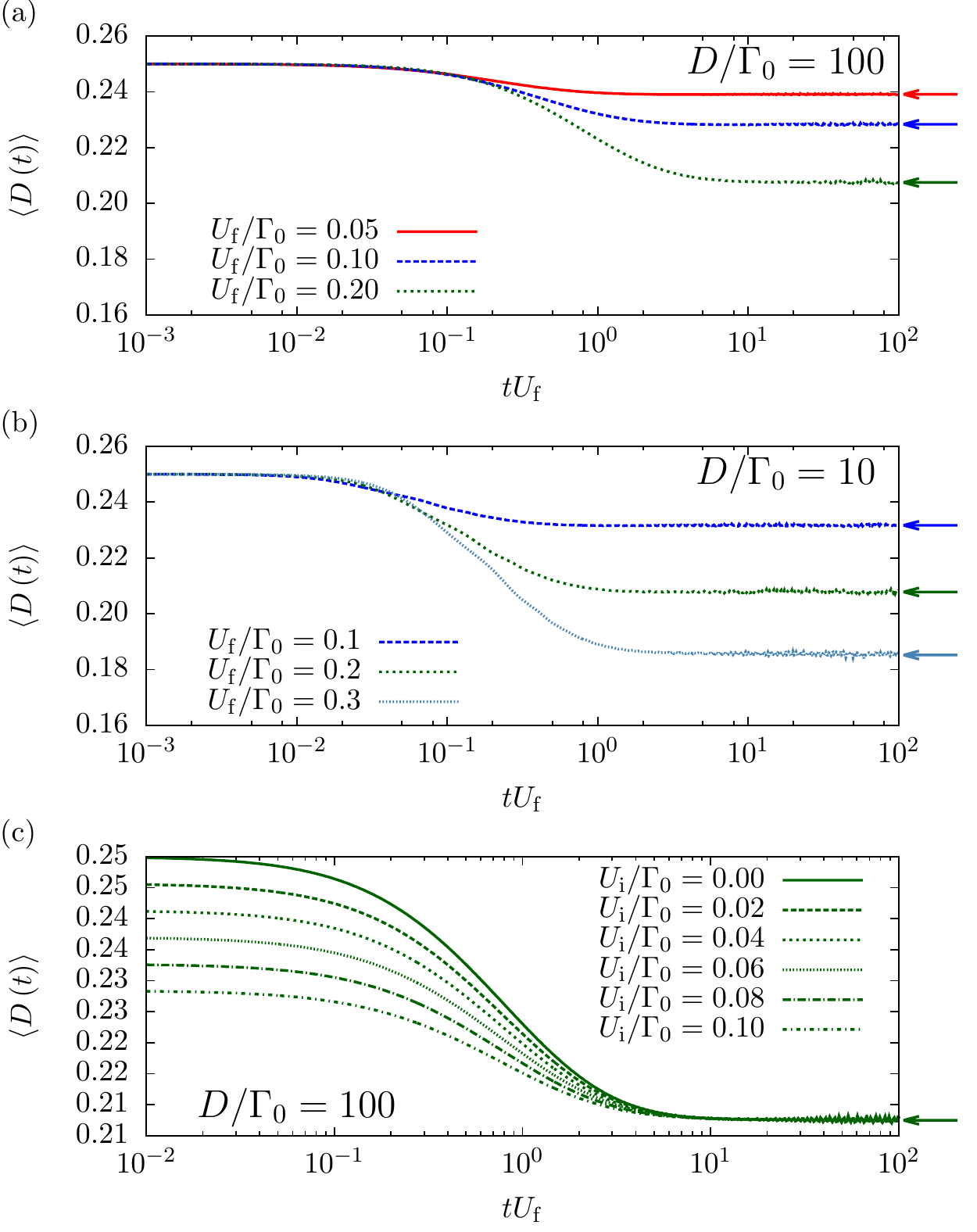}
  \caption{%
    (Color online)
    Time-dependent expectation value of the double occupancy $\expect{D(t)}$ for quenches within the SC phase.
    As a guide to the eye the equilibrium expectation values after the quench are marked by the arrows
    at the right side of the graph.
    (a) For the wide band $D/\Gn = 100$ with Coulomb repulsions $\Uf/\Gn = 0.05, 0.1, 0.2 < \Uc/\Gn \simeq 0.34$,
    (b) for $D/\Gn = 10$ with Coulomb repulsions $\Uf/\Gn = 0.1, 0.2, 0.3 < \Uc/\Gn \simeq 1.51$.
    For both bandwidths the system thermalizes at long times to its equilibrium value.
    (c) $\expect{D(t)}$ for different values of $\Ui$ and $\Uf/\Gn = 0.2$.
  }
  \label{fig:iqSCtoSC}
\end{figure}

For switching on the Coulomb repulsion at $t=0$, we can distinguish
two cases: (i) for $\Uf < \Uc$ the system remains in the SC phase
while (ii) for $\Uf > \Uc$ the equilibrium properties of the quenched
system belong to the LM phase.  Since we maintain ph symmetry, the
local occupancy always remains at half filling and is unaffected by
the quench.  Therefore, we focus on the dynamics of the local double
occupancy $\expect{D(t)}$.  For a clear energy separation of band
width $D$ and charge fluctuation scale $\Gn$, we have chosen $D/\Gn =
10,100 \gg 1$.

In Fig.\ \ref{fig:iqSCtoSC} we present the time-dependent local double
occupancy $\expect{D(t)}$ for quenches within the SC phase, $\Uf <
\Uc$, for two different band widths $D$ for $r=0.4$.  All curves start
at the non-interacting value $\expect{D(t=0)} = 1/4$ and reduced to
smaller values since the Coulomb interaction is suppressing the charge
fluctuations and the local double occupancy.

Quenches within the SC phase thermalize at long times to the
equilibrium value of the quenched system.
This thermalization is independent of the initial value $\Ui$
as shown in Fig.\ \ref{fig:iqSCtoSC}(c).
The equilibrium values obtained from an independent equilibrium
NRG calculation are indicated as arrows on the right side of the graph.
The decay is governed by the time scale set by the charge fluctuation
scale $\Gn$.
In contrary to results \cite{Schiro2012} obtained by a time-dependent
Gutzwiller ansatz -- see Fig.\ 2 in Ref.\ \cite{Schiro2012} --
we do not find any oscillations in the double occupancy.
However, the time-depended Gutzwiller approach predicts the thermalization
of $\expect{D(t)}$ for those type of quenches \cite{Schiro2012} which is a
non-trivial result.

\begin{figure}[tb]
  \centering
  \includegraphics[width=\linewidth]{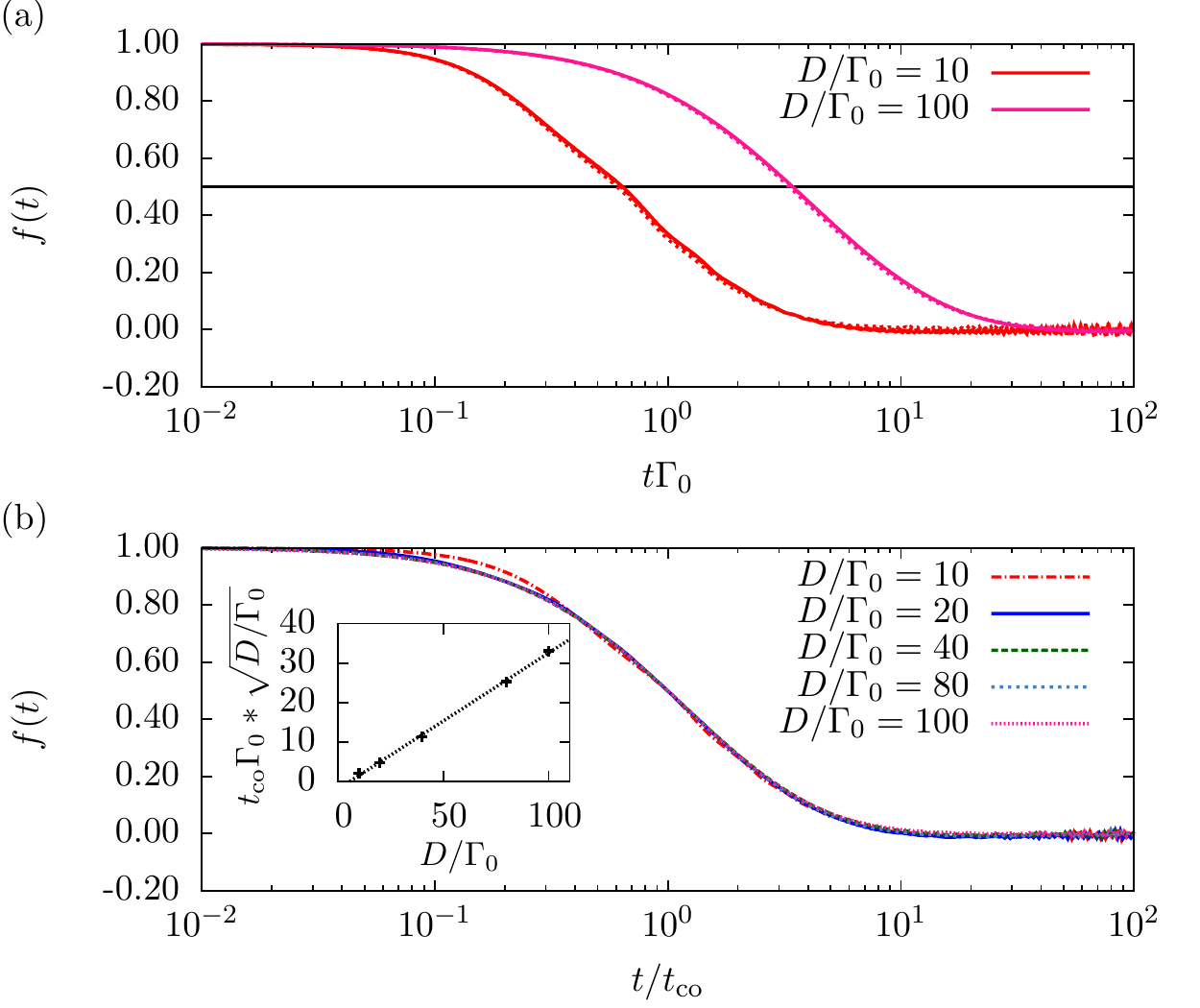}
  \caption{%
    (Color online)
    Scaling function $f(t)$  defined in  Eq.\ \ref{eq:f} versus $t$:
    (a) using the data of Fig.\ \ref{fig:iqSCtoSC} for different $\Uf$,
    (b) $f(t)$ versus $t/t_{\rm co}$  for different bandwidths.
    The inset in  (b) shows the time scale $\tco$ in dependence of the bandwidth $D$.
  }
  \label{fig:iqSCtoSC_tstar}
\end{figure}

It is interesting to note that we find universality for interaction quenches
within the SC phase.
In order to eliminate the influence of the different long-time expectation
values of $\expect{D(t \to \infty)}$ that includes the $\Uf$ dependency,
we define the function
\begin{align}
  f(t) = \frac{\expect{D(t)} - \expect{D(\infty)}}{\expect{D(0)} - \expect{D(\infty)}}
  \label{eq:f}
\end{align}
of the time-dependent double occupancy $\expect{D(t)}$.
The function $f$ starts at $f(0)=1$ and approaches $f=0$ at infinitely long
times independent of the parameters.

In Fig.\ \ref{fig:iqSCtoSC_tstar}(a) we demonstrate that all data
depicted in Fig.\ \ref{fig:iqSCtoSC} collapse onto one universal curve
that is only dependent on the ratio $D/\Gn$.  To eliminate the
$\Gn$ dependency, we define a crossover time $\tco$ as $f(\tco) = 0.5$.
Plotting $f(t)$ versus the dimensionless time scale $t/\tco$
maps all data for different ratios $D/\Gn$ onto one unique curve,
depicted in Fig.\ \ref{fig:iqSCtoSC_tstar}(b).  Only the curve for
$D/\Gn = 10$ deviates slightly from the others:
In this case the separation of energy scales is much less pronounced.
The inset of Fig.\ \ref{fig:iqSCtoSC_tstar}(b) illustrates the dependency
of $\tco$ on $D/\Gn$.
By fitting the numerical data we find that
\begin{align}
  \tco \propto \frac{1}{\Gn} \sqrt{\frac{D}{\Gn}}
  \quad .
  \label{eq:tcoD}
\end{align}
This establishes universality in the $f(t)$ dynamics:
the $\Uf$ dependency enters only via $\expect{D(\infty)}$ while the remaining real-time dynamics is only governed by the
time scale $\tco$ which depends on $D$ and $\Gn$.

\begin{figure}[tb]
  \centering
  \includegraphics[width=\linewidth]{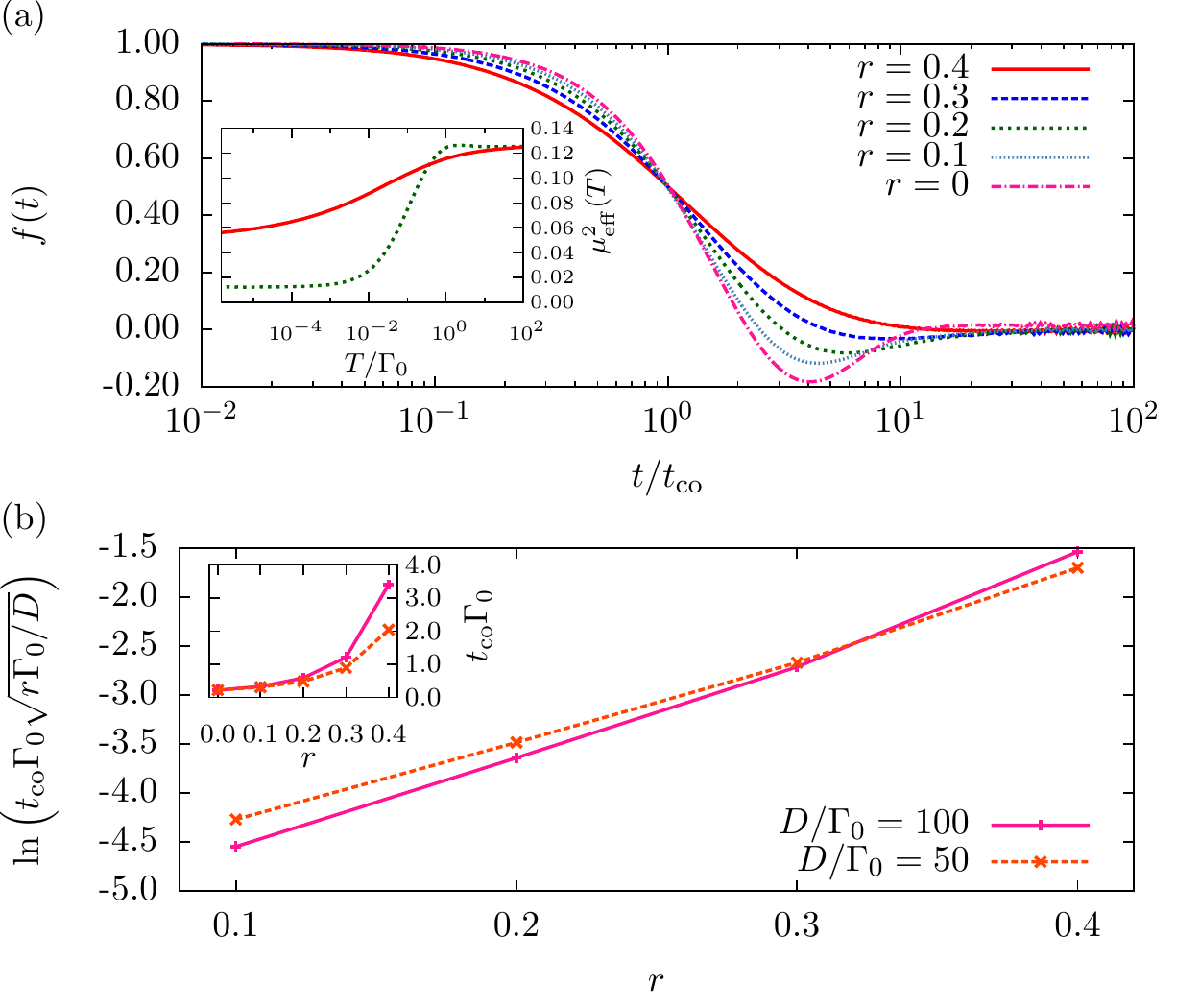}
  \caption{%
    (Color online)
    (a) The universal $f(t)$  versus $t/t_{\rm co}$  for different exponents $r = 0, 0.1, 0.2, 0.3, 0.4$.
    In the inset of (a) the temperature-dependent equilibrium effective local moment of the impurity spin
    $\mu_{\rm eff}^2(T)$ is depicted for $H_{\rm f}$, $D/\Gn = 100$, and $r = 0.2, 0.4$.
    (b) $\ln( \tco \Gn \sqrt{r \Gn / D} )$ vs $r$ while inset shows $\tco(r) \Gn$.
  }
  \label{fig:iqSCtoSC_rvar}
\end{figure}

So far, we only used a fixed bath exponent $r = 0.4$.  We extended our
investigation to interaction quenches within the SC phase to $r = 0,
0.1, 0.2, 0,3$.  Again, we find universality in $f_r(t/\tco)$ for
fixed $r$; the different universal functions for different $r$ are
depicted in Fig.\ \ref{fig:iqSCtoSC_rvar}(a).  With decreasing $r$ the
universal curve shows a dip at short times after which it increases
again to the long-time value. This correlates to a small peak
structure in the effective local moment $\mueff(T)$ for small $r$ as
depicted in the inset of Fig.\ \ref{fig:iqSCtoSC_rvar}(a).

In order to extract the $r$ dependence of the universality time scale,
we plot $\ln( \tco \Gn \sqrt{r \Gn/D} )$ vs $r$ in Fig.\
\ref{fig:iqSCtoSC_rvar}(b) where the inset of the panel shows the bare
time scale $\tco$.  Scaling $\tco$ with $\sqrt{r}$ for $r \ge 0.1$, we
found an exponential dependency of the time scale $\tco$ on $r$,
\begin{align}
  \tco \propto \frac{ e^{ m(D) r } }{\sqrt{r}} \sqrt{ \frac{D}{\Gn} }
  \label{eq:11}
\end{align}
with a bandwidth-dependent exponent $m(D)$ by fitting the numerically
obtained data to \eqref{eq:11}.
Note, however, that the phenomenological estimate \eqref{eq:11} for the time
scale does not interpolate to $r \to 0$ and, therefore, is only valid
for $0.1 \le r < 1/2$.

\begin{figure}[tb]
  \centering
  \includegraphics[width=\linewidth]{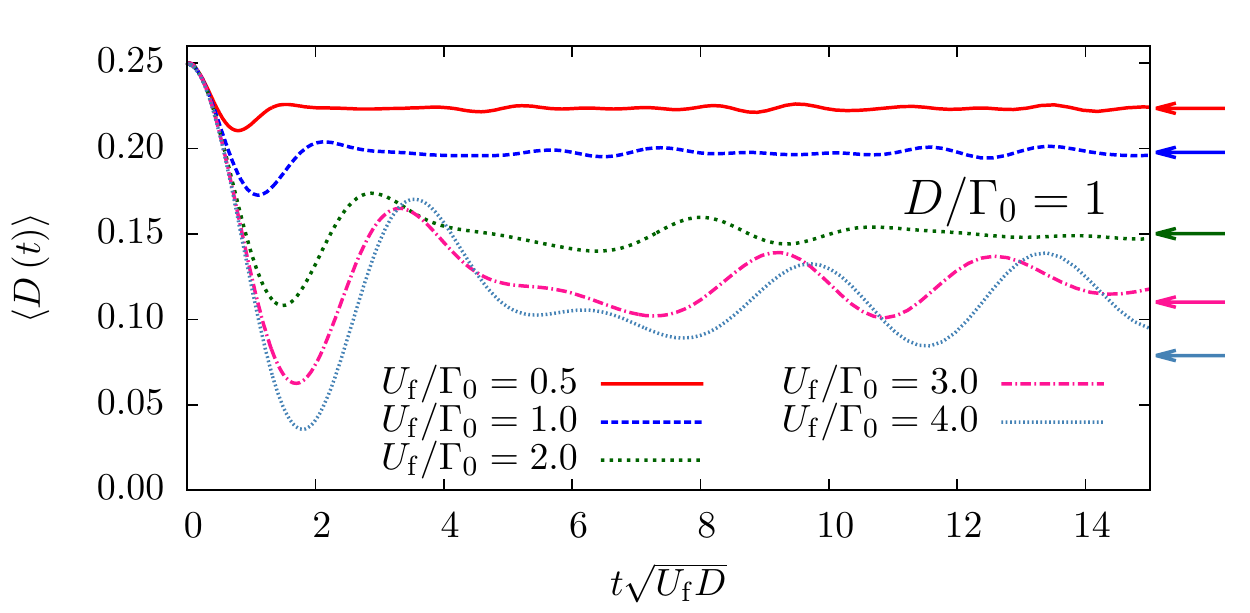}
  \caption{%
    (Color online)
    $\expect{D(t)}$ for quenches within the SC phase for hybridizations of the order
    of the band width $D/\Gn = 1$ and different Coulomb repulsions $\Uf/\Gn = 0.5, 1, 2, 3, 4 < \Uc/\Gn \simeq 4.426$
    at temperature $T/\Gn \sim 10^{-6}$.
    As a guide to the eye the equilibrium expectation values after the quench are marked by the arrows at
    the right side of the graph.
  }
  \label{fig:iqSCtoSCD1}
\end{figure}

When the charge-fluctuation scale $\Gn$ approaches the band width $D$,
the non equilibrium dynamics is dominated by local high-energy
oscillations and, therefore, is non-universal as already indicated
above.  We present the time evolution of the double occupancy for the
case $D/\Gn = 1$ in Fig.\ \ref{fig:iqSCtoSCD1}.  We have plotted
$\expect{D(t)}$ versus $t \sqrt{\Uf D}$ for $\Uf/\Gn = 0.5, 1, 2, 3,
4$ to reveal the quadratic decrease $\expect{D(t)} = 0.25 ( 1 - \alpha
( t / t_{\rm short} )^2 )$ where the short-time scale $1 / t_{\rm
short}^2 \propto \Uf$.  It steams for a linear contribution of $U
n_\uparrow n_\downarrow$ in a perturbative expansion of the
time-dependent density operator after the interaction quench.

For $\Uf/\Gn = 2$ the local single-particle excitation energies
$E_{\rm d}$ and $E_{\rm d}+U$ approach the band edges; for $\Uf/\Gn =
3, 4$ they exceed the band edges.  Then, these states are only weakly
coupled to the continuum with a finite energy gap preventing
thermalization.

The short-time dynamics of $\expect{D(t)}$ depicted in Fig.\
\ref{fig:iqSCtoSCD1} suggests a damped oscillation with a frequency
that is only weakly $\Uf$ dependent.  Asymptotically, the double
occupancy only thermalizes to the equilibrium value obtained directly
with $H_{\rm f}$ for smaller values of $\Uf$.  With increasing $\Uf$,
$\expect{D(\infty)}$ is decreasing and, therefore, the oscillation
amplitude must increase.

In order to gain some understanding on the origin of these short-time
oscillations we consider the limit of a vanishing band width at a
constant hybridization $\pi V_{\rm 0}^2 = \int d \e \Gamma(\e) =
const$.  In this case we are left with a purely local problem
described by the $r$-independent Hamiltonian $H_{\rm 0}$ of the first
Wilson shell \cite{KrishWilWilson80a} whose eigenenergies are
analytically known \cite{KrishWilWilson80a}.  In this limit, there
would be no damping, and we find a perfect oscillatory solution for
$\expect{D(t)}$ whose frequency is determined by the difference of
eigenenergies.  A careful analysis of the local dynamics reveals that
the oscillation of $\expect{D(t)}$ can be traced back to an admixture
of two singlet states in the charge sector $Q=0$ (the quantum
number $Q$ measures the deviation from half filling
\cite{KrishWilWilson80a}) in the ground-state wave function.  The
difference of the eigenenergies of those states labeled by $r=1$ and
$r=2$ in Table I of Ref.\ \cite{KrishWilWilson80a} coincide
with the oscillation frequency extracted from $\expect{D(t)}$.

When we release the constraint of a vanishing band width, additional
DOFs of the Wilson chain need to be included in the analysis.  Adding a
single additional Wilson chain link, \ie $H_{\rm 0} \to H_{\rm 1}$,
immediately causes a more complex response due to the splitting of
these eigenfrequencies.  Since $V_{\rm 0}/D \approx 1$ for $\Gn/D =
1$, the energy splitting of the eigenenergies due to an increasing
Wilson chain length must be smaller than $D$ and, therefore, $V_{\rm
0}$ and $U$.  Hence the short-time dynamics on time scales much shorter
than $t \Uf$ is dominated by energy differences coming from a
slightly modified local dynamics.  This is the origin of the
pronounced minimum observed in $\expect{D(t)}$ in Fig.\
\ref{fig:iqSCtoSCD1}.

When the local excitation energies $E_{\rm d}$ and $E_{\rm d} + U$ are
lying inside the band continuum, \ie $|E_{\rm d}|, |E_{\rm d} + U| \le
D$, we observe thermalization.  Once these local energies exceed the
band continuum, bound states are formed outside of the band which
contribute to the expansion of the initial ground state but cannot
provide a relaxation channel.  Even though we remain in the SC phase,
$\expect{D(t)}$ cannot thermalize and remain oscillatory.  This is the
case for $\Uf/\Gn = 3, 4$.

\subsubsection{Quenches across the QCP}
\label{sec:interactionquenchesover}

\begin{figure}[tb]
  \centering
  \includegraphics[width=\linewidth]{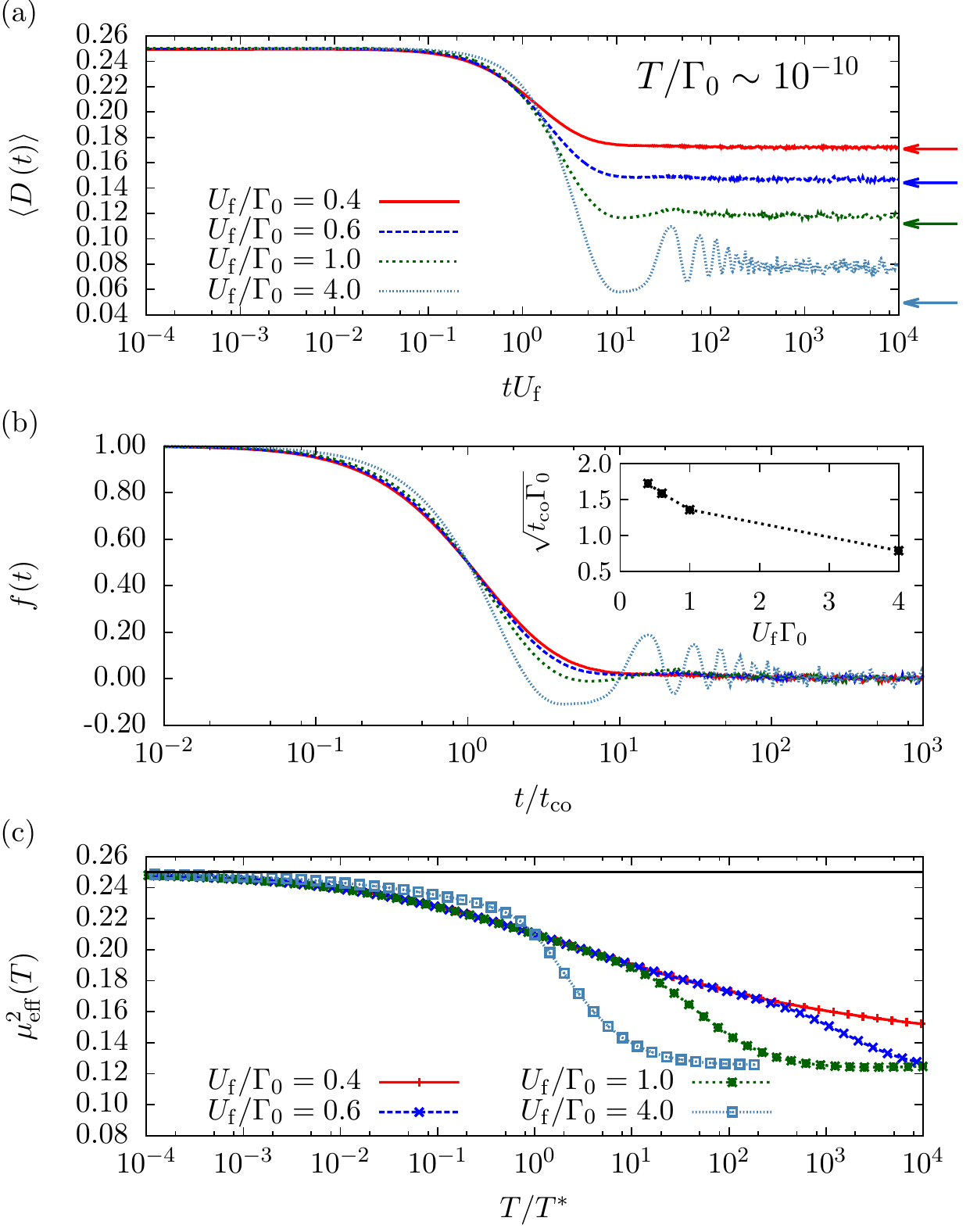}
  \caption{%
    (Color online)
     $\expect{D(t)}$ for quenches from the SC FP ($\Ui = 0$) into the LM
    phase for different Coulomb repulsions
    $\Uc/\Gn \approx 0.3392 < \Uf/\Gn = 0.4, 0.6, 1, 4$ for a
    wide band $D/\Gn = 100$.  (a) $\expect{D(t)}$ versus $t \Uf$ 
    at $T/\Gn \sim 10^{-10}$.  As a guide to the eye the
    equilibrium expectation values after the quench are marked by
    the arrows at the right side of the graph. 
    (b) Same data rescaled with Eq.\ \eqref{eq:f} to $f(t)$ versus
    $t/\tco$. In the inset the dependence of $\tco$ by $\Uf$.
    (c) The corresponding effective local moment of
    the impurity spin $\mueff(T)$ for the parameters $\Uf$ used in (a).
  }
  \label{fig:iqSCtoLM}
\end{figure}

In this section we investigate interaction quenches across the QCP
from the SC into the LM phase.  Initially the system is prepared in
the SC FP at $t=0$ by setting $\Ui = \epsilon_{\rm i} = 0$.

For Coulomb repulsions $\Uc < \Uf$ the equilibrium properties of
$H_{\rm f}$ governing the real-time dynamics belong to the LM phase.
The LM FP can be described by an effective local moment comprising a
linear combination of the impurity spin and the conduction electron
bath spins in addition to a decoupled remaining free effective
conduction electron band.  Therefore, a finite equilibrium double
occupancy $\expect{D}_{\rm eq}$ remains present even in the LM phase.

Similarly to the Kondo temperature $T_K$ that determines the low-energy
crossover to the SC fixed point, the characteristic temperature $T^*$
governs the crossover to the LM FP for $U > \Uc$.  We have defined
$T^*$ as $\mueff(T^*) = 0.21$ and plotted $\mueff(T)$ versus $T/T^*$
to illustrate the universality in Fig.\ \ref{fig:iqSCtoLM}(b).  We
note that $T^* \propto ( U - \Uc )^{\nu(r)}$ with $\nu(r)=4.3$ for
$r=0.4$, and $T^*$ vanishes at $U = \Uc$.

In Fig.\ \ref{fig:iqSCtoLM}(a) we present $\expect{D(t)}$ for quenches
over the QCP with $\Uc/\Gn \approx 0.3392 < \Uf/\Gn = 0.4, 0.6, 1, 4$
for a wide band $D/\Gn = 100$ and temperature $T/\Gn \sim 10^{-10}$.
The data obtained at a temperature $T/\Gn \sim 10^{-6}$ remains
indistinguishable from the results depicted in Fig.\
\ref{fig:iqSCtoLM}(a) and, therefore, $T\to 0$.  Even though $T > T^*$
for the lowest $\Uf$ value, the real-time dynamics remains
temperature-independent and is only governed by the overlap of the initial ground
state with the eigenstates of the final Hamiltonian.

We have plotted the data as function of the dimensionless time $t \Uf$
to reveal the time scale of the short-time dynamics. For quenches
within the SC phase, we have demonstrated that the characteristic time
scale is independent of $\Uf$. For quenches across the QCP into the
LM phase, we find that the time scale is proportional to $1/\Uf$. The
real-time dynamics is dominated by the local dynamics since part of the
impurity DOFs decouple from the rest of the conduction band to
participate in the local moment formation.

The characteristic energy scale $T^*$ extracted from the universality
of $\mueff(T)$ in the LM phase depicted in Fig.\ \ref{fig:iqSCtoLM}(b)
increases with increasing distance $U - \Uc$ to the QCP.  The length
scale $\xi^* = v_{\rm F} / T^*$ used as an estimate for the spatial
extension of the local moment decreases with increasing $U$.  The
decoupled spin DOFs become more localized and, consequently, the
thermalization is increasingly suppressed with increasing $\Uf$.

For $\Uf/\Gn = 4$, we observe a decaying oscillatory behavior driven
by a frequency proportional to $\Uf$ with a strong deviation between
the long-time steady-state and the thermal equilibrium.  For the
smallest two interactions, $\Uf/\Gn = 0.4 ,0.6$, however, we found a
steady-state value that is very close to the thermodynamic result and,
therefore, can be considered as evidence for thermalization.

Although the equilibrium expectation value $\expect{D}_{\rm eq}$ is
continuously reduced with increasing $U$ even across the QCP $\Uc$,
Fig.\ \ref{fig:iqSCtoLM}(a) demonstrates nicely the qualitatively
different response for quenches across the QCP versus the previously
investigated dynamics within the SC phase in Sec.\
\ref{sec:interactionquencheswithin}.  While the characteristic time
scale only depends on $\Gn$ for quenches within the SC phase and shows
universal behavior as long as all local excitations remain in the bath
continuum, the real-time dynamics across the QCP is governed by
$1/\Uf$.

\begin{figure}[tb]
  \centering
  \includegraphics[width=\linewidth]{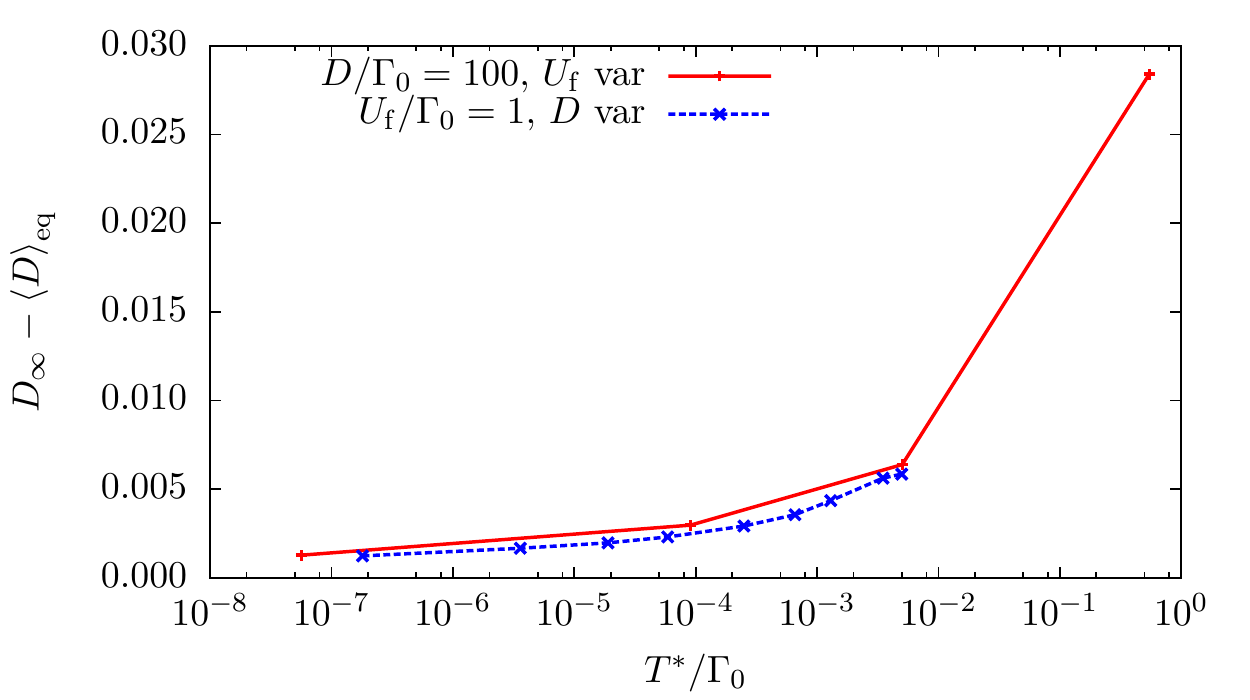}
  \caption{%
    (Color online)
    Difference between the long-time steady-state value $D_\infty$ and the thermodynamic equilibrium value
    $\expect{D}_{\rm eq}$ as function of $T^*$ either for fixed $D/\Gn$ and a variation of $\Uf$ (red curve)
    or a fixed $\Uf/\Gn$ and a variation of $D$ (blue curve).
  }
  \label{fig:iqSCtoLM_Tstar}
\end{figure}

In order to illustrate the connection between the spatial extension of the local moment and thermalization
we calculated the difference between the steady-state value $D_\infty$ defined as
\begin{align}
  D_\infty = \lim_{T \to \infty} \frac{1}{T} \int_0^T d t \expect{D(t)}
  \label{eq:def-D-oo}
\end{align}
and the thermal equilibrium value $\expect{D}_{\rm eq}$ obtained for $H_{\rm f}$.
The results are depicted versus $T^*/\Gn$ in Fig.\ \ref{fig:iqSCtoLM_Tstar} using the data of Fig.\ \ref{fig:iqSCtoLM}.
Close to the QCP, the deviation is less the 3\% given by the typical discretization error of the TD-NRG.
For increasing $\Uf$ the deviation increases, supporting the picture of an increasingly localized free moment,
leading to an increasing non decaying fraction of the local double occupancy.

We also have supplemented data for fixed $\Uf/\Gn = 1$ but a variation of $D$ as crosses connected with a blue line
(color online) in Fig.\ \ref{fig:iqSCtoLM_Tstar}.
Interestingly, the deviation $\Delta D = D_\infty - \expect{D}_{eq}$ follows the same  trend, and agrees within
the numerical error with the data for fixed $D/\Gn$ when approaching the QCP.

The deviation $\Delta D$ systematically increases with increasing $T^*$.
Once $T^*$ becomes large, the physics changes to an effectively decoupled free moment that is strongly localized
and therefore prevents thermalization of the local expectation values.

\begin{figure}[tb]
  \centering
  \includegraphics[width=\linewidth]{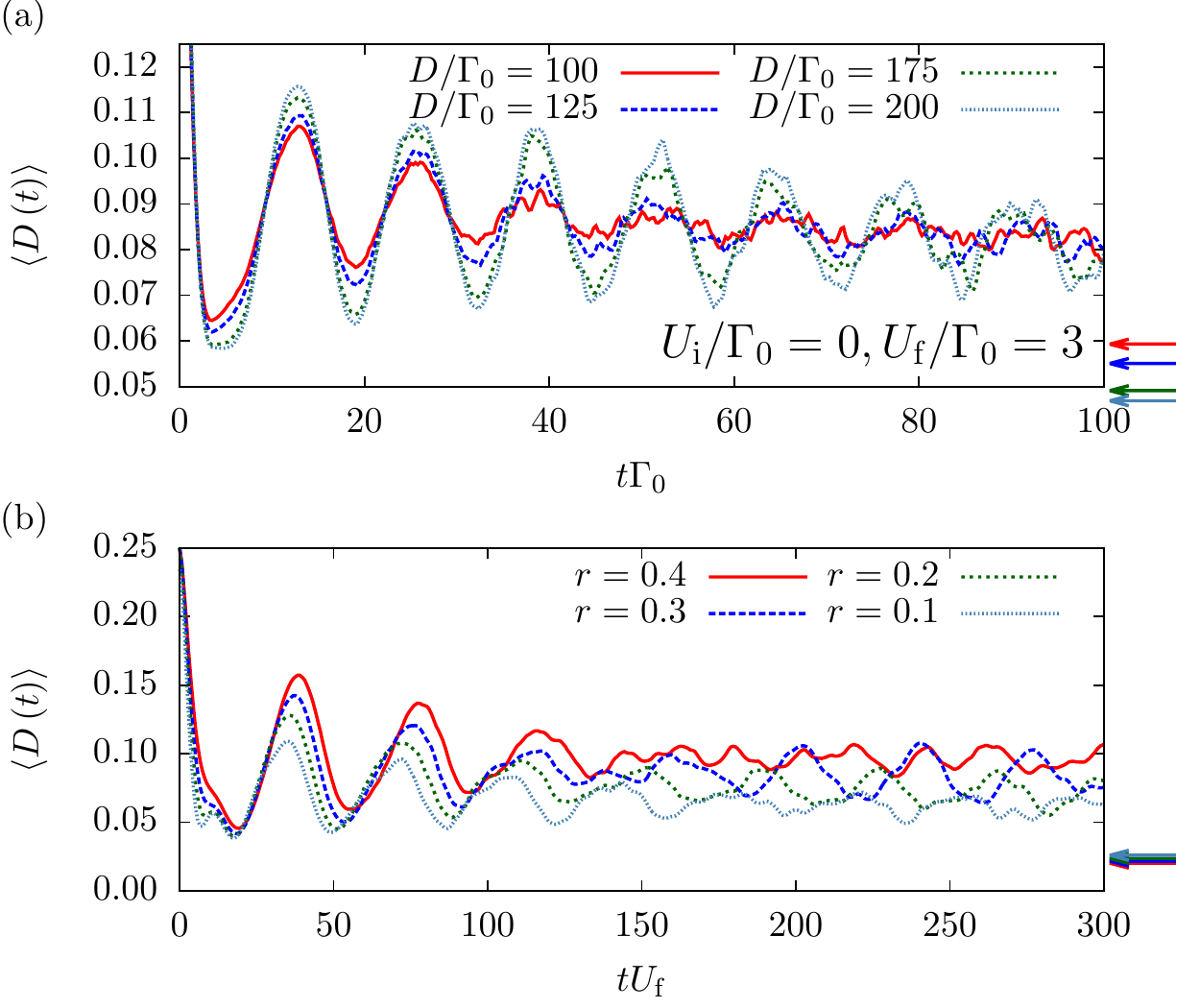}
  \caption{%
    (Color online)
    $\expect{D(t)}$ for interaction quenches from the SC FP into the LM phase.
    (a) $\expect{D(t)}$ versus $t \Gn$ for different band widths $D/\Gn = 100, 125, 175, 200$, $r=0.4$,
    and a fixed $\Uf/\Gn = 3 > \Uc$.
    (b) $\expect{D(t)}$ versus $t \Uf$ for different exponents $r$, fixed $D/\Gn = 100$, and
    adjusted Coulomb repulsion $\Uf(r=0.4)/\Gn = 14.72$, $\Uf(r=0.3)/\Gn = 17.73$, $\Uf(r=0.2)/\Gn = 21.06$,
    and $\Uf(r=0.1)/\Gn = 24.79$ such that $\expect{D(U,r)}_{\rm eq}\approx 0.02$.
    Equilibrium expectation values after the quench are marked by the arrows
    at the right side of the graphs.
  }
  \label{fig:iqSCtoLM_Dvar_rvar}
\end{figure}

Our results are in contrast to the results of Fig.\ 3 in Ref.\
\cite{Schiro2012} where these types of quenches were examined by a
time-dependent Gutzwiller approach.  Therein it was reported that the
double occupancy $\expect{D(t)}$ strongly oscillates and never reaches
a steady-state value at long times.

The reason for this disagreement is the nature of the Gutzwiller state
whose dynamics has been traced via a time-dependent Gutzwiller equation
\cite{Schiro2012}.  This ansatz wave function is not able to properly
represent the LM phase: $\expect{D}_{\rm eq}$ vanishes for $U >
\Uc$ and is taken as an indicator for the QPT while an equilibrium NRG
calculation \cite{GonzalezBuxtonIngersent1998} has proven the
continuity of $\expect{D}_{\rm eq}$ across $\Uc$; see also Fig.\
\ref{fig:equi-double-occupancy}.  While the NRG correctly contains the
spatially extended nature of the decoupled local moment, the
Gutzwiller state restricts the free moment to the local impurity site.
Therefore, the time-dependent Gutzwiller approach is restricted to the
strong-coupling regime.

In Fig.\ \ref{fig:iqSCtoLM_Dvar_rvar}(a) we examine the effect of the
bandwidth $D$ on the real-time dynamics using four different values
$D/\Gn = 100, 125, 175, 200$ for a constant ratio $\Uf/\Gn = 3$.  For
this series, $\Uc(D/\Gn=100)/\Gn = 0.3358$ has the largest value and
decreases further with increasing band width $D$.  So $\Uf$ exceeds
$\Uc$ almost by one decade even for the largest $\Uc$.

Therefore, we quench deeper into the LM phase, indicated by a
decreasing equilibrium double occupancy $\expect{D}_{\rm eq}$; see
arrows at right side of the graph.  With increasing band width $D$, the
amplitude of the damped oscillations are increased while the frequency
is only proportional to $\Uf$.

In Fig.\ \ref{fig:iqSCtoLM_Dvar_rvar}(b) we vary the bath exponent $r$
for a fixed ratio $D/\Gn = 100$ and choose $\Uf$ such that the
equilibrium expectation value $\expect{D(\Uf)}_{\rm eq} \approx 0.02$
remains nearly constant for all $r$ as indicated by the arrows on the
right side.  $\expect{D(t)}$ is plotted versus $t \Uf$ to remove the
leading order frequency dependency of the oscillations for different
Coulomb interactions $\Uf$.  The remaining small frequency shift with
increasing $r$ is related to the different ratios $\Uf/\Gf$: the
larger $r$, the larger the ratio $\Gf/\Uf$ and corrections of the
order $\sqrt{ 1 + 2 \Gf D / ( \pi U^2 ) }$ need to be taken into
account stemming from the energy difference of eigenstates of the
first Wilson shell as previously discussed in the context of Fig.\
\ref{fig:iqSCtoSCD1}.

\subsubsection{Quenches from the LM phase}

\begin{figure}[tb]
  \centering
  \includegraphics[width=\linewidth]{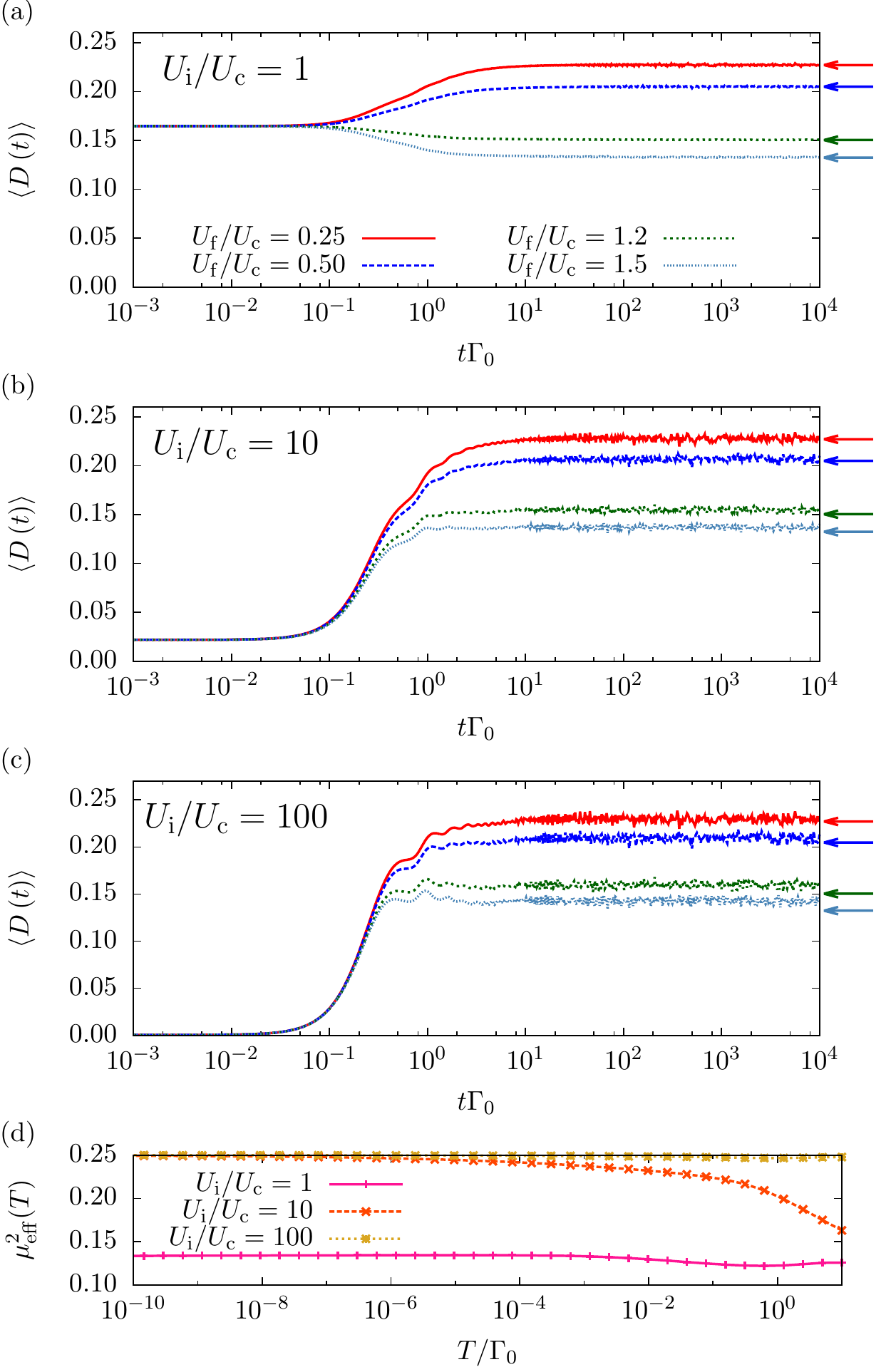}
  \caption{%
    (Color online)
    $\expect{D(t)}$ vs $t \Gn$ for four different values of $\Uf$.
    For quenches within the LM phase we set $\Uf/\Uc =1.2, 1.5$ and for quenches across the QCP we choose
    $\Uf/\Uc = 0.25, 0.5$.
    All data have been obtained for $D/\Gn = 10$.
    In the different panels (a)--(c) different initial Coulomb repulsions have been used:
    (a) $\Ui/\Gn \approx \Uc/\Gn = 1.5088$, (b) $\Ui/\Uc = 10$, and (c) $\Ui/\Uc =100$.
    (d) Effective moment $\mueff(T)$ for the different $\Ui$.
  }
  \label{fig:iqLMtoLMSC}
\end{figure}

In a reverse type of quench the initial $\Ui$, exceeding the critical
$\Uc$, is reduced to values $\Uf$ that are either larger than $\Uc$,
such that the systems remains in the LM phase, or $\Uf < \Uc$, for
quenching across the QCP into the symmetric SC phase.  The initial
value of $\expect{D(t=0)}$ depends on the distance $\Ui - \Uc$ and
reaches $\expect{D(0)} \to 0$ for $\Ui \to \infty$.  Since we always
maintain ph symmetry, $2\e_{\rm d}(t) + U(t) = 0$ holds at any time.

In order to investigate the dependency of $\Ui$, we calculate the
real-time dynamics for three different values $\Ui \approx \Uc$,
$\Ui/\Uc = 10$, and $\Ui/\Uc = 100$.  The results are depicted in
Figs.\ \ref{fig:iqLMtoLMSC}(a)--10(c).
Each graph shows $\expect{D(t)}$ for a fixed $\Ui$ and a series of
four values of $\Uf$: For quenches within the LM phase,
we use $\Uf/\Uc = 1.2, 1.5$ while for quenches across
the QCP into the SC phase we choose $\Uf/\Uc = 0.25, 0.5$.

The initial value $\expect{D(t=0)}$ decreases with increasing value
$\Ui$.  Starting with $\Ui/\Uc = 1$, we observe an increase of
$D_\infty$ for $\Uf < \Uc$ and a decrease for $\Uf > \Uc$ as expected.
Furthermore, $D_\infty \approx \expect{D}_{\rm eq}(\Uf)$ as indicated
by the arrows on the right side of Fig.\ \ref{fig:iqLMtoLMSC}(a) which
we interpret as indication for thermalization.

For Fig.\ \ref{fig:iqLMtoLMSC}(b), we have prepared the system with
$\Ui/\Uc = 10$, and quench to the same four final values of $\Uf$ as
in Fig.\ \ref{fig:iqLMtoLMSC}(a).  Again, the initial short-time
dynamics is $\Uf$-independent, starting from a much smaller initial
value.  The saturation is almost reached at $t \Gn \approx 1$.
However, we observe a slight deviation from the thermal equilibrium.
Those deviations remain negligibly small for quenches from the LM to
the SC phase but become visible for quenches within the LM phase: the
deviations are of the order of 7\%.  For the last value $\Ui/\Uc
=100$, starting deep in the LM phase with a completely suppressed
initial double occupancy, the short-time dynamics is again $\Uf$-independent
and the steady-state value is approached very fast,
governed by the time scale of $1 / \Gn$.  However, the deviation from
the thermal equilibrium is much more pronounced and reaches about 15\%
for $\Uf/\Uc = 1.5$.

\subsubsection{Quenches with particle-hole asymmetry}

\begin{figure}[tb]
  \centering
  \includegraphics[width=\linewidth]{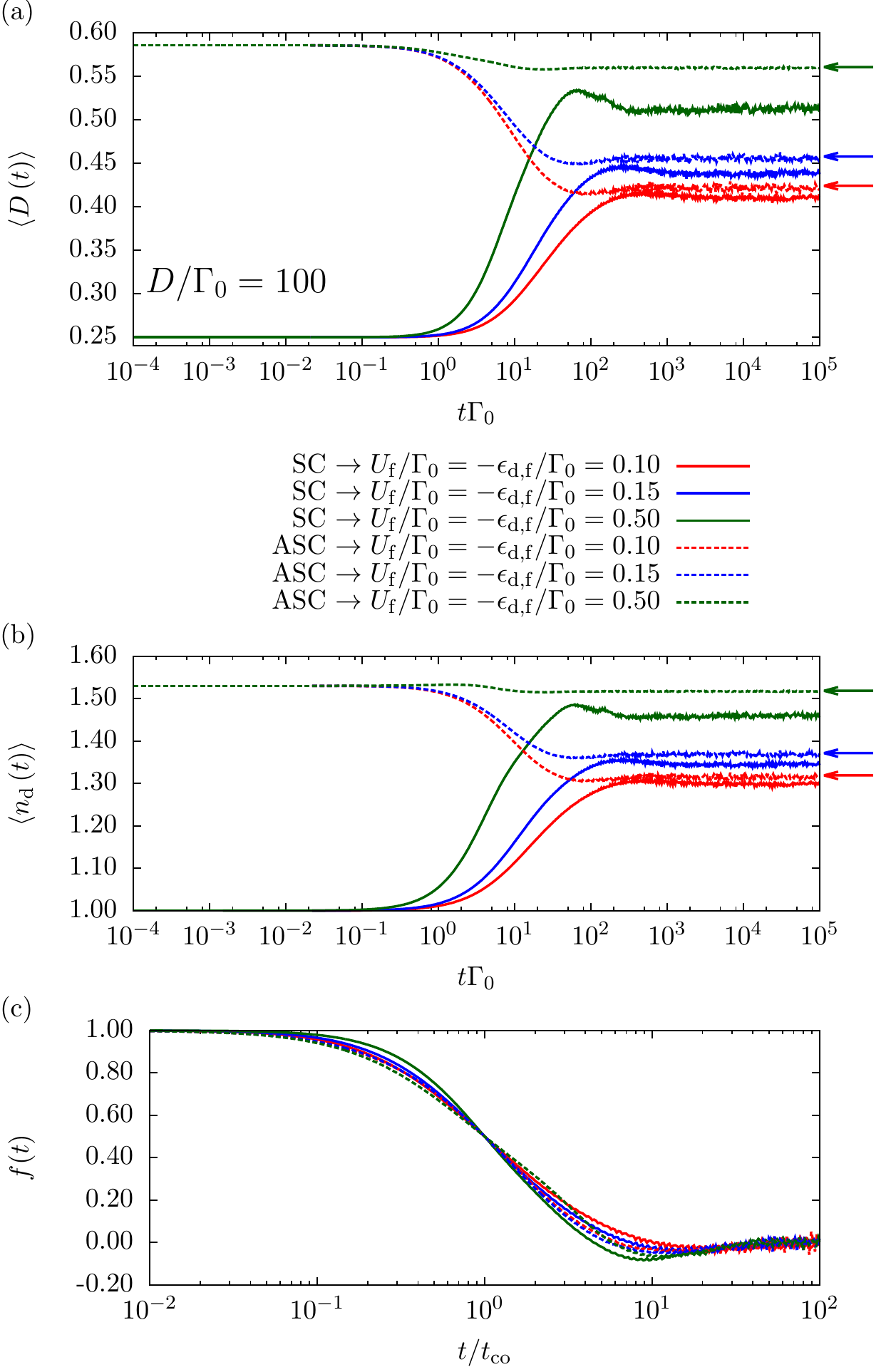}
  \caption{%
    (Color online)
    (a) $\expect{D(t)}$ vs $t \Gn$ for particle-hole asymmetry.
    We quench either from the SC FP ($U_{\rm i}=\epsilon_{\rm d,i}=0$) or the ASC FP
    ($U_{\rm i}=0,\epsilon_{\rm d,i}/\Gn=-0.1$) into or rather within the ASC phase.
    All data have been obtained for $D/\Gn = 100$ and $r=0.4$.
    (b) Change of the local occupancy $\expect{n_d(t)}$ vs time for the same parameter as
    in (a). In (c) the data of (a) rescaled via Eq.\ \eqref{eq:f} to $f(t)$ versus $t/\tco$
    with the crossover time scale $\tco$.
  }
  \label{fig:iqASC}
\end{figure}

So far we have restricted ourselves to particle-hole (ph) symmetry, namely
$U(t) = - 2 \epsilon_{\mathrm{d}}(t)$.
Now we investigate the influence of ph-symmetry breaking on the real-time dynamics.
We recall that the asymmetric strong-coupling (ASC) fixed point
\cite{GonzalezBuxtonIngersent1998,GlossopLogan2003,Vojta2006,FritzVojta2013,ChowdhuryIngerset2014}
differs from the SC FP by a complete screening of the impurity spin independently of $r$.
Furthermore, the equilibrium double occupancy $\expect{D}_{\rm eq}$ in the ASC FP depends on the
asymmetry $|U-2\epsilon_{\mathrm{d}}|$ and the band exponent $r$:
With increasing asymmetry or increasing $r$ the equilibrium double occupancy
($\expect{D}_{\rm eq}=1/4$ for ph symmetry) increases.

In Fig.\ \ref{fig:iqASC}(a) $\expect{D(t)}$ is shown for two different
initial conditions:
The dashed lines describes the time evolution starting from the
particle-hole asymmetric phase (ASC) using the parameters
$\Ui = 0, \epsilon_{\rm d,i}/\Gn = -0.1$
while the solid lines depict the time evolution starting from the
particle-hole symmetric point $\Ui = \epsilon_{\rm d,i} = 0$ as
initial condition.
The different colors distinguish the different values of the final Hamiltonian.
Since the filling is also changing with time, we augment the data by showing
the change of the impurity filling $\expect{n_d(t)}$ in  Fig.\ \ref{fig:iqASC}(b) for the same
parameter.

Although a steady state is always found, only quenches starting from a particle-hole symmetry
broken state thermalize as indicated by the arrows of equilibrium double occupancy value
on the right-hand side of Figs.\ \ref{fig:iqASC}(a) and \ref{fig:iqASC}(b).
The deviation of the steady-state value from the thermal value 
when changing the degree of particle-hole asymmetry is a shortcoming
of the TD-NRG \cite{AndersSchiller2006,EidelsteinGuettgeSchillerAnders2012,GuettgeAndersSchiller2013}
and has recently been investigated in great detail \cite{NghiemCosti2014}.

We have applied the same scaling procedure to the data of 
Fig.\ \ref{fig:iqASC}(a) as in  Fig.\ \ref{fig:iqSCtoSC_tstar} 
and have plotted the results in Fig.\ \ref{fig:iqASC}(c). Qualitatively 
a similar behavior is found. However, universality is lost, and the 
shape of the dimensionless function depends on the degree of
particle-hole asymmetry.

\subsection{Hybridization quenches}
\label{sec:hybridisationquenches}

Up to now, we have investigated interaction quenches.  In this
section, we focus on a second type of quench, the hybridization
quenches.  They are defined by switching the hybridization strength
$\Gn \to \Gamma(t) = \Theta(-t) \Gi + \Theta(t) \Gf$ between the
initial value $\Gi$ and the final value $\Gf$ at time $t=0$.  In
contrast to the previous sections, we keep here the Coulomb repulsion
$U_{\rm i}/D= U_{\rm f}/D= const$ and finite at all times and restrict
ourselves to particle-hole symmetry.  In order to have a well-defined,
unique reference point, we choose $\Gi = 0$ implying that the system
is initially prepared in the LM FP whose ground state is $U$-independent
with a double occupancy $\expect{D(t=0)}=0$ for all
parameters.

Since we change $\Gn$, we use the band width $D$ as unit of energy in
this section.  The terms short, intermediate, and long times will
correspond to $t D \ll 1$, $t D \sim 1$, and $t D \gg 1$.

As discussed in Sec.\ \ref{sec:overview-phases}, $H_{\rm f}$ has two
different low-energy FPs depending on $\Gf$.  For $\Gf < \Gc$, the LM
FP is reached at low temperature in equilibrium, while for $\Gf >
\Gc$, the system is described by the symmetric SC FP.  We have checked
that the equilibrium FP has been reached for all parameters such that
the real-time dynamics results correspond to $T \to 0$.

\subsubsection{Quenches within the LM phase}
\label{sec:hybridisationquencheswithin}

\begin{figure}[tb]
  \centering
  \includegraphics[width=\linewidth]{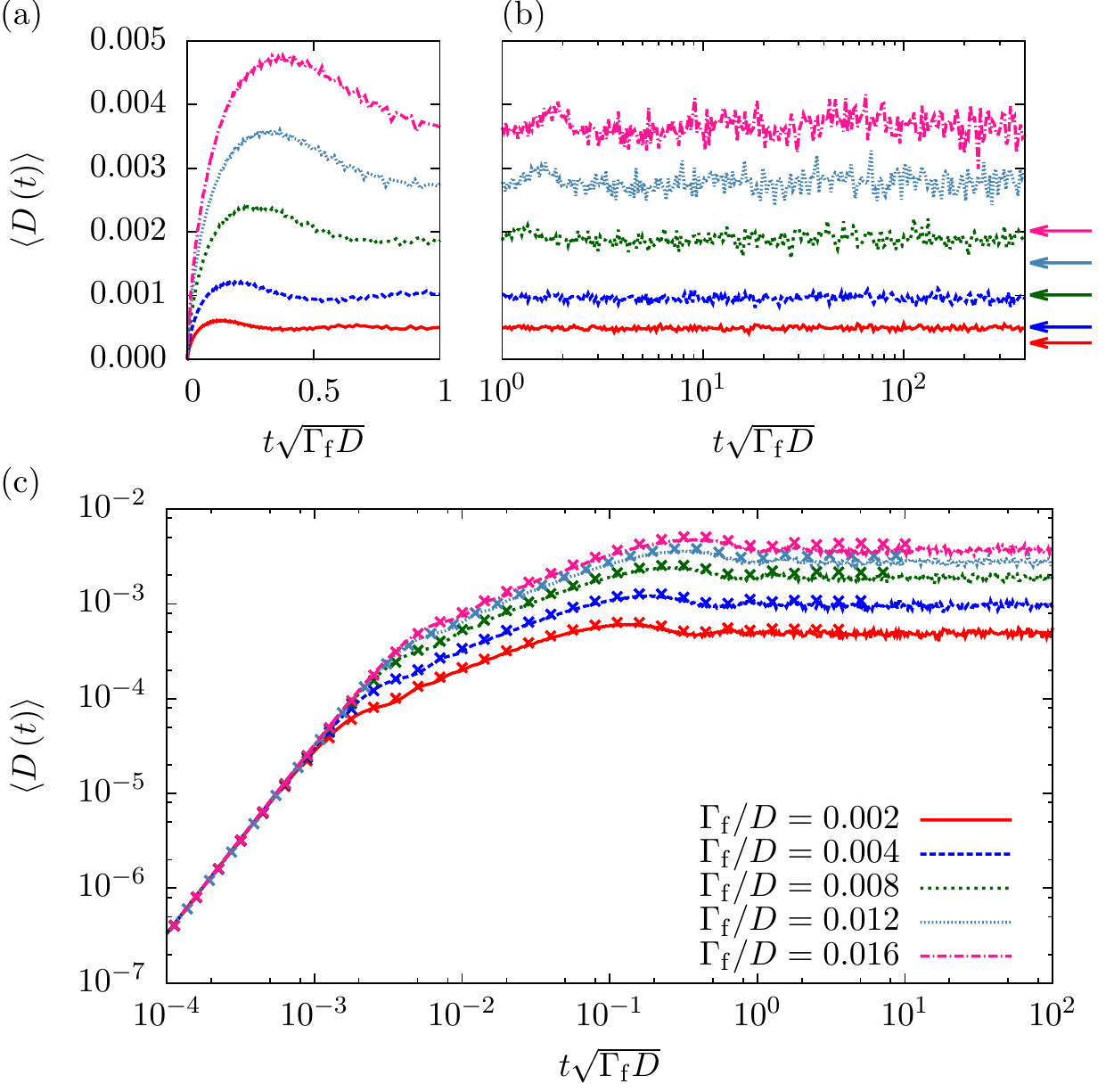}
  \caption{%
    (Color online)
    Quenches from the LM FP at time $t=0$ within the LM phase.
    The hybridization is $\Gf/D = 0.002, 0.004, 0.008, 0.012, 0.016 < \Gc/D \simeq 0.019$ with a constant Coulomb
    repulsion $U/D = \Ui/D = \Uf/D = 0.01$ and $r=0.4$.
     $\expect{D(t)}$ (a) for short times, (b) for long times on log time scale.
    As a guide to the eye the equilibrium expectation values after the quench are marked by the arrows at the right
    side of (b).
    For all hybridizations $\Gf$ the system does not thermalize.
    (c) Same data as (a) and (b) on a logarithmic $y$ scale.
    Additionally we added the result of the second-order perturbation theory [cf. Eq.\ \eqref{eq:ptfhq}]
    represented by the crosses.
  }
  \label{fig:hqLMtoLM}
\end{figure}

We start with quenches within the LM phase, \ie $\Gf < \Gc$, and
present data in the wide-band limit by setting $D/U = 100 = \text{constant}$.
The results for the time-dependent double occupancy $\expect{D(t)}$
are depicted for the short-time dynamics in Fig.\ \ref{fig:hqLMtoLM}(a)
and the long-time asymptotic in Fig.\ \ref{fig:hqLMtoLM}(b) for $\Gf/D
= 0.002, 0.004, 0.008, 0.012, 0.016 < \Gc/D \approx 0.019$.

The double occupancy $\expect{D(t)}$ rises quickly from its initial
value and exhibits a peak at intermediate times before it falls off
towards a steady-state value.  With increasing hybridization $\Gf$ the
peak height raises and is more pronounced.  Plotting all data versus
the dimensionless time $\tau = t \sqrt{ \Gf D }$ reveals that the
short-time dynamics is $U$-independent and only driven by the final
hybridization strength $\Gf$.

Although a steady-state is reached, the difference to the
thermodynamic equilibrium value -- indicated by the arrows in Fig.\
\ref{fig:hqLMtoLM}(b) -- is significant and increases with increasing
$\Gf$.  This indicates that the real-time dynamics is not governed by
the FP properties but from the overlap of the initial wave function
with the excited states of $H_{\rm f}$.

Deviations from the thermodynamic equilibrium could originate in
limitations of the method due to the discretization of the conduction
band continuum
\cite{AndersSchiller2006,EidelsteinGuettgeSchillerAnders2012,GuettgeAndersSchiller2013,NghiemCosti2014}.
The low-energy FP of $H_{\rm f}$, however, is still given by the LM FP
with a twofold-degenerate ground state so that the increase of the
deviation with increasing $\Gf$ is explained.  Varying $\Gf$ within
the LM phase has a profound impact on the physics. With increasing
$\Gf$ the system approaches $\Gc$ and, therefore, the
characteristic energy scale $T^*$ decreases. This implies that the
effective spin moment decoupling from the rest of the system becomes
more extended with increasing $\Gf$ and a larger fraction of the
impurity DOFs hybridize with the conduction band leading to an
continuous increase of $\expect{D}$. There are two states of finite
range that form the effective moment decoupling from the rest of the
chain.  An admixture of those states to the initial ground state
cannot relax further due to the decoupling of these states in the LM
regime. Consequently, we observe a state-state differing significantly
from the thermal state.

\subsubsection{Analytic result for the short-time dynamics}
\label{sec:analytic-result}

In order to shed some more light onto the dynamics for hybridization
quenches when $\Gf < \Gc$, we analytically calculate $\expect{D(t)}$
for the short-time dynamics.  Knowing all eigenstates and
eigenenergies for $\Gn = 0$, we expand the time-dependent density
operator in powers of the hybridization, and evaluate $\expect{D(t)}$
exactly up to second order -- see the Appendix for more details.

The short-time dynamics of the double occupancy is given by the analytic expression
\begin{align}
  \expect{D(t)} &= \frac{\Gf D}{\pi} t^2
    + \frac{2 \Gf (1+r)}{\pi} \sum_{n=2}^{\infty} \frac{(-1)^{2n}}{(2n)!}
    \label{eq:ptfhq}  \\
    &\qquad \times \int_{-D}^0 \left| \frac{\epsilon}{D}\right|^r
    \left( \left( \epsilon - \epsilon_{\rm d} - U \right) t \right)^{2n} \text{ d} \epsilon \nonumber
    \quad .
\end{align}
which is asymptotically exact for $t \to 0$.  In leading order in $t$,
$\expect{D(t)}$ increases quadratically in time with a $U$-independent
prefactor $\Gf D / \pi$ defining the squared
characteristic time scale for the short-time dynamics.  Our analytical
result confirms the dimensionless time scale $\tau = t \sqrt{ \Gf D }$
used previously in Fig.\ \ref{fig:hqLMtoLM} to reveal the universality
in the short-time TD-NRG response.  The $U$-dependent corrections
enter the higher terms in $t$ and account for a weak oscillation.

In order to illustrate the leading $t^2$ behavior in the short-time
dynamics of the full TD-NRG calculations and their excellent agreement
with the analytical result for the full continuum, we present the
numerical results of Figs.\ \ref{fig:hqLMtoLM}(a) and \ref{fig:hqLMtoLM}(b)
in a log-log plot as Fig.\ \ref{fig:hqLMtoLM}(c).
We added the analytical results of
Eq.\ \eqref{eq:ptfhq} for selected times as crosses in the same color.
Numerics and analytics coincide perfectly for short and intermediate
times even up to $t \sqrt{ \Gf D } \approx 1$.  The analytical result
does not only describe the leading order $O(t^2)$ term correctly but
also accounts for the deviation from the parabola starting at $t
\sqrt{ \Gf D } \approx 10^{-3}$ and the long-time steady state.  Of
course the deviation between the analytic steady-state value and the
TD-NRG results is stronger with increasing $\Gf$ as expected from the
perturbative nature of the analytical approach.

\subsubsection{Energy flow in the LM phase}
\label{sec:energy-flow}

\begin{figure}[tb]
  \centering
  \includegraphics[width=\linewidth]{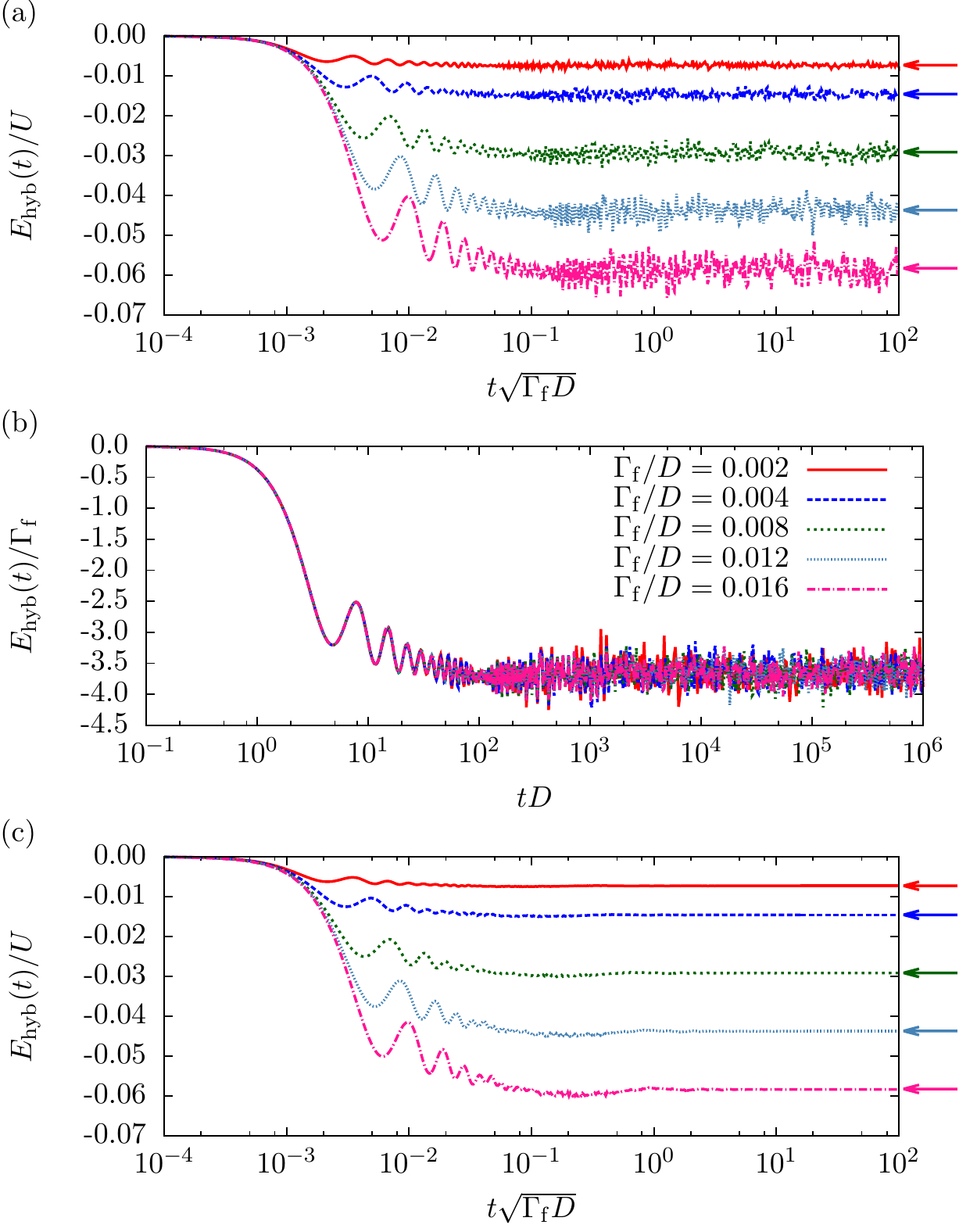}
  \caption{%
    (Color online) 
    The hybridization energy $E_{\rm hyb}(t)$ vs time for hybridization
    quenches within the LM phase using a constant Coulomb repulsion
    $\Ui/D = \Uf/D = U/D = 0.01$ and various hybridization strengths 
    $\Gf/U  = 0.2, 0.4, 0.8, 1.2, 1.6 < \Gc/U \simeq 1.9$.
    (a)  $E_{\rm hyb}(t)/U$ vs the dimensionless time
    $\tau = t \sqrt{ \Gf D }$ characterizing the dynamics of the 
    double occupancy, and (b) $E_{\rm hyb}(t)/\Gf$ vs $t D$.
    (c) Same quenches as (a) but calculated with an additional TD-NRG
    broadening $\alpha = 0.4$.
  }
  \label{fig:hqLMtoLM_hyb}
\end{figure}

Another interesting question arises as to whether the deviation
$\Delta O=O_\infty  -\expect{O}_{eq}$
of the long-time limit $O_\infty$ \cite{Reimann2008,RigolETH2012},
\begin{align}
  O_\infty &= \lim_{T\to \infty} \frac{1}{T}\int_0^T dt \expect{O(t)} \nonumber \\
           &= \sum_{m}^{N} \sum_{\substack{r,s\\ E_r=E_s}}^{\rm trun} \; O_{r,s}^m \rho^{\rm red}_{s,r}(m)
\end{align}
from its thermal expectation value $\expect{O}_{eq}$ with respect
to the final Hamiltonian is operator-dependent, and therefore,
different physical properties show different thermalization behavior
for the same quench.
Furthermore, it has been conjectured that the choice of a Wilson chain might be
not suitable for properly describing hybridization quenches \cite{Rosch2012}
because the NRG chain might not be able to serve as a heat reservoir for
larger changes in the hybridization energy. However, Wilson's NRG
as well as the TD-NRG target only the local dynamics and corresponding
bath expectation values do not have any physical meaning:
A discretized finite bath, as
used in any NRG calculation, has always only finite energy content
while the thermodynamic bath provides a reservoir with infinitely
large energy.
Since a quantum state by itself can never thermalize when subject to
the energy conserving dynamics defined by the Schr\"odinger equation,
one needs to restrict such an investigation to quantum impurity
expectation values.
The deviation from a corresponding equilibrium NRG calculation using
identical discretization parameters serves as a criterion of how well
the quantum impurity subsystem is able to thermalize.

Both questions can be addressed by investigating the local energy flow
for quenches within the LM phase for the quench parameters presented
in Fig.\ \ref{fig:hqLMtoLM}.
For those parameters we have established already above that thermalization
of the double occupancy $\expect{D(t)}$ is absent, and switching on
the hybridization results in an overestimation of the double occupancy
compared to the thermal equilibrium that increases with $\Gf$.
For particle-hole symmetry, $E_{\rm d} n_{\rm d}$ is discontinuous
at $t=0$, since work has been performed on the system, but afterwards
$E_{\rm d} n_{\rm d}$ remains constant for $t>0$.
In addition, the contribution of $U \expect{D(t)}$ to the impurity energy
is very small compared to $E_{\rm hyb}(t) = \expect{H_{\rm hyp}(t)}$.
Therefore, the main energy flow of the impurity is governed by the
hybridization energy $E_{\rm hyb}(t)$ which is initially zero for $t \le 0$.
Note that the hybridization energy $E_{\rm hyb}(t)$ involves in addition
to the impurity operator only the local host degree of freedom on the
first Wilson shell \cite{KrishWilWilson80a,*KrishWilWilson80b,BullaCostiPruschke2008}
and, therefore, is a local operator as required for the 
TD-NRG \cite{AndersSchiller2005,AndersSchiller2006}.

The results of $E_{\rm hyb}(t)$ are depicted in Fig.\ \ref{fig:hqLMtoLM_hyb}:
The arrows on the right side in Fig.\ \ref{fig:hqLMtoLM_hyb}(a) -- depicting
the thermal expectation value of the hybridization energy -- indicate clearly
that $E_{\rm hyb}(t)$ is approaching its thermal equilibrium in the long-time
limit, even though the system remains in the LM regime and $\expect{D(t)}$
does not equilibrate.
$E_{\rm hyb}(t)$ has been measured in units of the constant $U$ and plotted
versus $\tau = t \sqrt{ \Gf D }$.
Clearly the dynamics is not governed by the characteristic time scale of the
double occupancy.

Since the hybridization energy is proportional to $\Gf$,
\begin{align}
  \sum_{k\sigma } V_k \expect{d^\dagger_\sigma c_{k\sigma}} \propto \Gf
  \; ,
\end{align}
we have divided out the leading order prefactor of $E_{\rm hyb}$, $\Gf$,
and have plotted $E_{\rm hyb}/\Gf$ vs $tD$ in Fig.\ \ref{fig:hqLMtoLM_hyb}(b).
Surprisingly, the short-time dynamics of $E_{\rm hyb}/\Gf$ appears to be
universal and is governed by the band width $D$ while the overall magnitude
of $E_{\rm hyb}$ is determined by $\Gf$ for a finite $r=0.4$.
While for $r=0$ $E_{\rm hyb}/\Gf\propto \ln(D/\Gf)$ in equilibrium as it
can be shown by a simple analytic calculation, this is not the case for
$r>0$: The larger $r$ the less $E_{\rm hyb}/\Gf$ depends on the ratio
$D/\Gf$ for the wide-band limit.
Since the hybridization energy approaches its steady-state value very fast,
apparently this expectation value is not influenced by the buildup of
low-energy correlations characterizing the LM FP.

On the first side this surprising thermalization of $E_{\rm hyb}(t)$ seems
to contradict the increasing deviation $\Delta D = D_\infty - \expect{D}_{\rm eq}$
within the LM phase.
The physical content of the two operators, however, is different.
The impurity DOFs provide a major contribution to the effective moment formation
which decouples from the rest of the system in the LM phase, while $E_{\rm hyb}$
probes the coupling to the full bath continuum.
The number of bath DOFs contributing to the local moment formations are of
measure zero in the integration over all $k$ states so that the main error
in $E_{\rm hyb}(t)$ can be traced back to TD-NRG discretization errors.
Furthermore, the real-time dynamics of $E_{\rm hyb}$ is an ultrafast process
governed by the band width as shown in Fig.\ \ref{fig:hqLMtoLM_hyb} and has
approached its steady-state value on a time scale where long-time correlations
have not had a chance to develop.

Our findings indicate that the magnitude $\Delta O$ strongly depends on
the operator $\hat O$ and the details of matrix elements $O^{m}_{r,s}$
selecting the states contributing to the long-time limit as well as defining
the oscillation frequency distribution governing the real-time dynamics
via Eq.\ \eqref{eqn:time-evolution-intro}.

\subsubsection{Real-time dynamics of the impurity magnetization}

\begin{figure}[tb]
  \centering
  \includegraphics[width=\linewidth]{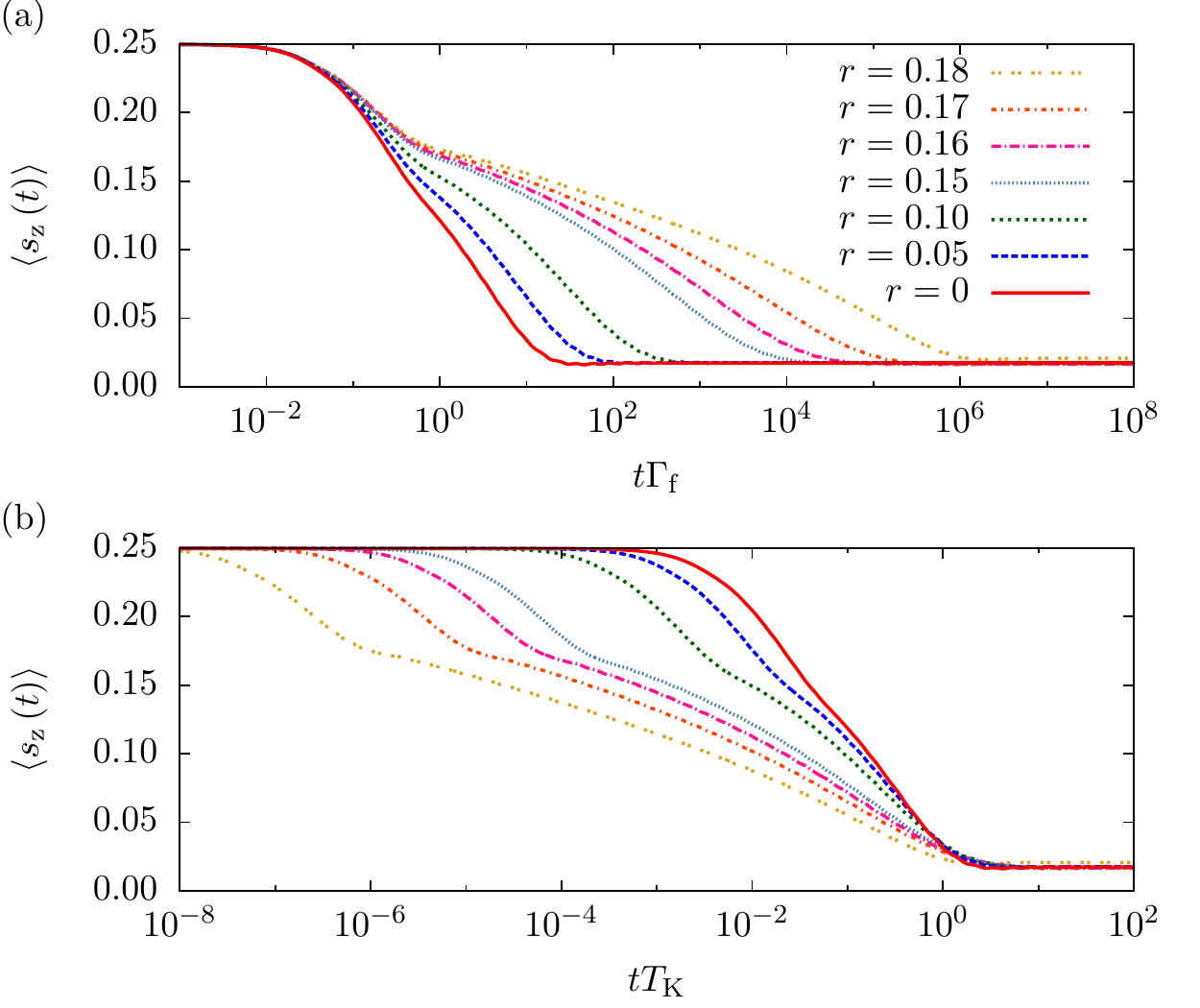}
  \caption{%
    (Color online)
    (a) Impurity magnetization $\expect{S_{\rm z}(t)}$ vs (a) $t\Gf$
    and (b) vs $t\TK$ for a fixed $U/\Gf=6$ and different bath exponents $r$.
    We quench $\epsilon_{\rm d,i}/2\Gf = H_{\rm i}/\Gf = 1 \to
    \epsilon_{\rm d,f}/\Uf = -1/2, H_{\rm f} = 0$ and switch on
    the hybridization $\Gf$ from $\Gi = 0$ at $T/\Gf = 7 \cdot 10^{-7}$.
    The NRG parameters: $\Lambda = 1.6$, $N_{\rm S} = 1500$, and $\alpha = 1$.
  }
  \label{fig:hqSz}
\end{figure}

By switching on an initial local magnetic field $H_{\rm i} = \Gf > 0$,
we break the spin symmetry and induce a spin polarization in the
initially decoupled impurity of $\expect{s_z} = 0.25$ independent of $r$,
where $\expect{s_z}= \expect{n_\uparrow - n_\downarrow}/2$.

$s_z(t) = \expect{s_z(t)}$ is shown in Fig.\ \ref{fig:hqSz} after switching
off the external magnetic field $H$ for times $t>0$.
Note that we have used a broadening factor $\alpha=1$ in Eq.\ \eqref{eq:gamma_m}
\cite{AndersSchiller2005,AndersSchiller2006} in this calculations
to illustrate the difference to the $z$ averaging.
Results are shown for a fixed $U/\Gf = 6$ but different $r$ for particle-hole
symmetry breaking initial conditions $\epsilon_{\rm d,i}/2\Gf = H_{\rm i}/\Gf = 1$
which imply $s_z(0)=0.25$.
We also have performed calculations for particle-hole symmetric initial conductions
where the level position remains unaltered; \ie $\epsilon_{\rm d,i} = \epsilon_{\rm d,f} =-U/2$
and consequently $s_z(0) = 0.5$ -- not shown.
Since $s_z(t)/2$ coincides perfectly with the data of Fig.\ \ref{fig:hqSz}(a),
the time-dependency of the spin decay is controlled  by the type of quench and
not by the initial amount of spin polarization.

Since $U > \Gf$, the short-time spin decay is governed by the Schrieffer-Wolff
exchange interaction \cite{SchriefferWol66}.
After that fast initial spin decay which is independent of $r$, 
the spin relaxation slows down as the Kondo correlations start to build up
\cite{NordlanderEtAl1999}.
The details of the spin decay strongly depend on the bath exponent $r$.
This is not surprising since (i) the nature of the SC FP is different for each $r$
and (ii) the critical coupling $\Uc$ decreases with increasing $r$.
In Fig.\ \ref{fig:hqSz}(b), we have plotted the data as a function of the dimensionless
time $t\TK$ to reveal the long-time behavior.
We have defined the Kondo temperature via the screening of the effective moment
\cite{Wilson75,AndersSchiller2005}, \ie $\mueff(\TK) = r/8 + 0.07$, where we have
factored in the $r$-dependent residual moment $\mueff(0)=r/8$.
Similar to the findings in Fig.\  3 of Ref.\ \cite{AndersSchiller2005}, $\TK$ plays
a role in determining different decay regimes in the real-time dynamics.
For $t\TK < 1$ no universal behavior is found, but the potential universal behavior
of $s_z(t)$ for very long times $t\TK \gg 1$ is obstructed by the finite resolution
of the TD-NRG.

For $U < \Gf$ but $U$ close to $\Uc$ at larger $r > 0.3$ the above analysis does
not hold. Here the spin decay is governed by the charge-fluctuation scale $\Gf$
and remains almost independent of $\TK$  -- not shown here.
At short time scales controlled by high-energy excitation the existence of the
pseudogap is less relevant and the dynamics is that of a weakly correlated system.
A more in depth analysis of the complex spin dynamics in the pg-SIAM will be
presented elsewhere.

\subsubsection{Quenches across the QCP}
\label{sec:hybridisationquenchesover}

\begin{figure}[tb]
  \centering
  \includegraphics[width=\linewidth]{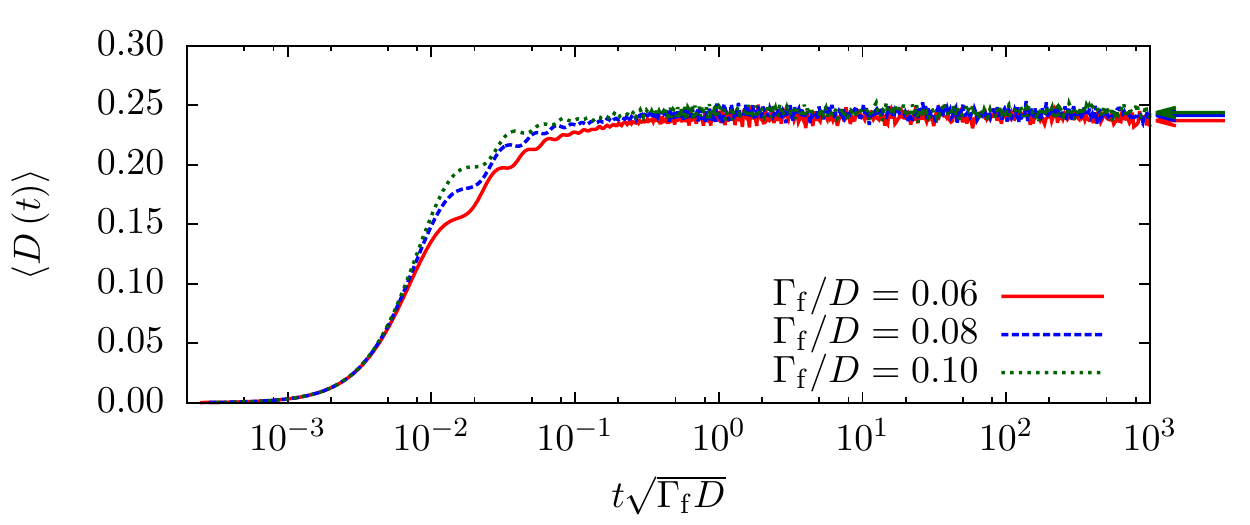}
  \caption{%
    (Color online) 
    $\expect{D(t)}$ for hybridization quenches into the SC phase for a
    constant Coulomb repulsion $U/D = \Ui/D = \Uf/D =0.01$.
    The hybridzation strength $\Gf$ is set well above the critical value,
    $ \Gf/D = 0.06, 0.08, 0.1>\Gc/D \simeq 0.019 $.
    Arrows at the right side mark the equilibrium expectation values
    after the quench.
  }
  \label{fig:hqLMtoSC_D100}
\end{figure}

Now we proceed with hybridization quenches across the QCP, from the
initial LM FP with $\Gi=0$ into the symmetric SC phase with $\Gc < \Gf$.

The time-dependent double occupancy $\expect{D(t)}$ is depicted in
Fig.\ \ref{fig:hqLMtoSC_D100} for $\Gc/D < \Gf/D = 0.06,0.08,0.1$.
As for $\Gf < \Gc$, $\expect{D(t)}$ shows a fast rise with an universal
time scale $1/\sqrt{ \Gf D }$ and saturates in the long-time limit at
a stationary value which is very close to the thermodynamic limit.  In
contrary to the LM FP within the LM phase quench, the thermodynamic
value $\expect{D}_{\rm eq}$ deviates only very little from the free
value of $\expect{D}_{\rm eq}=1/4$.

At intermediate times, $t \sqrt{\Gf D} \approx 10^{-2}$ damped
oscillations with an oscillation frequency $\propto U$ are
superimposed onto the continuous rise of $\expect{D(t)}$ from the
initial value $\expect{D(0)} = 0$ to the final steady-state value. For
the real-time dynamics of the double occupancy the Kondo scale is
irrelevant: The dynamics is governed by the short-time scale
$1/\sqrt{\Gf D}$, with superimposed oscillations depending on $U$.

\subsubsection{Comparision with Gutzwiller results}
\label{sec:hybridisation-quench-LM-gutzwiller}

\begin{figure}[tb]
  \centering
  \includegraphics[width=\linewidth]{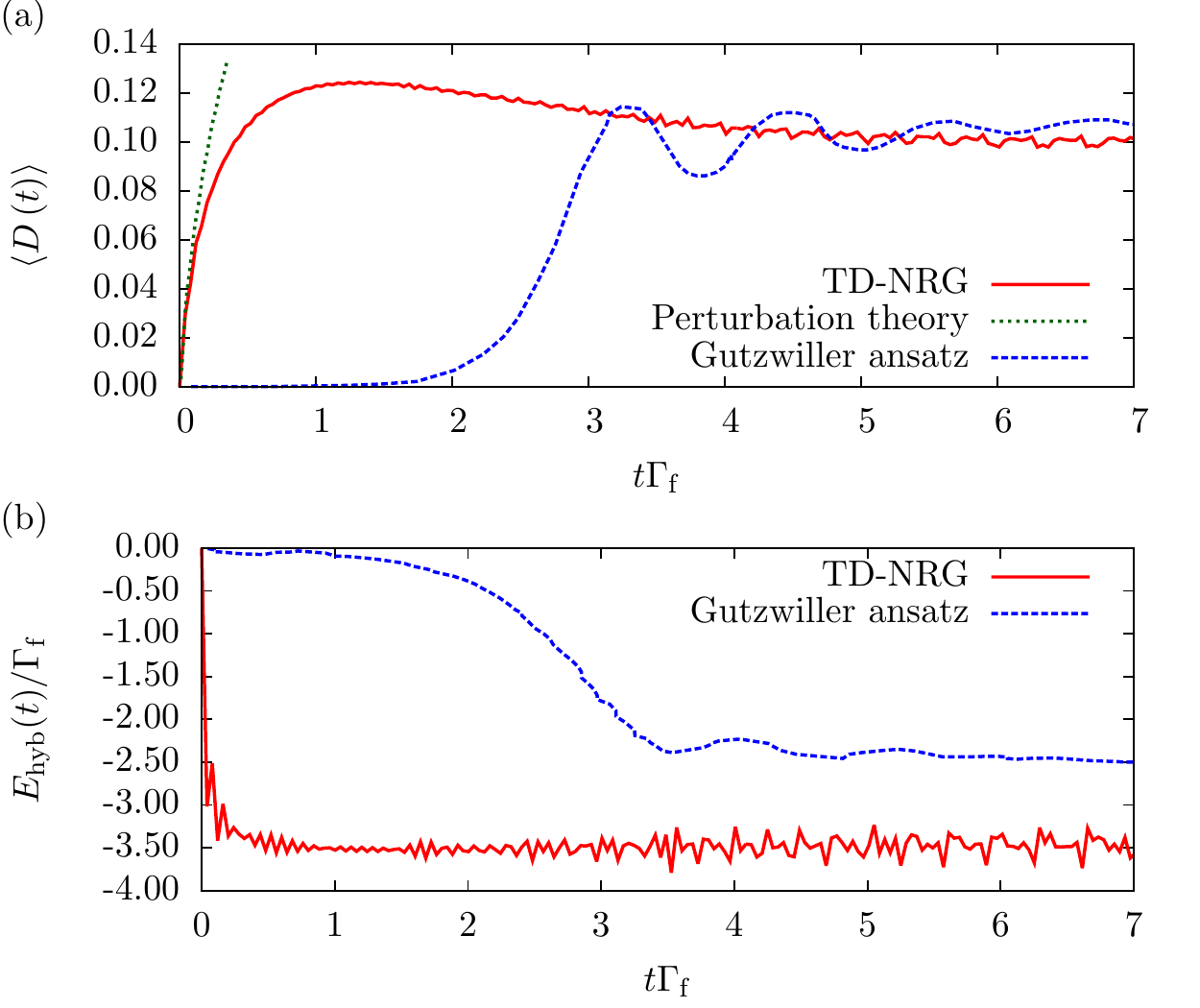}
  \caption{%
    (Color online)
    Comparison between the TD-NRG (solid line) and time-dependent Gutzwiller
    ansatz (dashed line) for a hybridization quench,
    \ie $\Gi = 0 \to \Gf/D = 0.01$, and $r=0.4$.
    The Gutzwiller data taken from Ref.\ \cite{Schiro2012} has been
    obtained for $U/\Gf = 9$.
    (a) $\expect{D(t)}$ and (b) the time-dependent hybridization energy
    $E_{\rm hyp}(t)$ for the NRG parameters $\Uc/\Gn < \Ui/\Gf = \Uf/\Gf = 1.6$
    and augmented with the Gutzwiller data.
    The NRG parameters have been chosen such that 
    $D_\infty^{\rm TD-NRG} \approx D_\infty^{\rm Gutz}$ for a
    quantitative comparison.
  }
  \label{fig:hqLMtoLM_schiro}
\end{figure}

In Sec.\ \ref{sec:equilibriumproperties} we have pointed out that the
local equilibrium properties obtained by the accurate NRG approach
deviate significantly from the approximate Gutzwiller approach.  This
discrepancy can be traced back to the physical content of the
Gutzwiller ansatz wave function that focuses only on the local spin
moment formation on the impurity site while the NRG ground state
correctly accounts for the extended nature of the decoupling moment.
Consequently, the local nature of Gutzwiller ansatz largely
overestimates the critical $\Uc$ as already shown in Fig.\
\ref{fig:equi-double-occupancy}.

A choice of identical model parameters always yields different local
properties in the two different methods. In order to make a useful
comparison between the approaches, we have chosen the NRG parameters
such that the equilibrium double occupancies approximately agree
between both approaches, \ie $D_\infty^{\rm TD-NRG} \approx
D_\infty^{\rm Gutz}$, as well as both approaches describe the same type
of quench.

The Gutzwiller data are taken from Fig.\ 2 in Ref.\
\cite{Schiro2012} calculated for a hybridization quench in the
wide-band limit with $U/\Gf = 9$, and $r=0.4$, where a hybridization
quench from the initial LM phase into the SC phase across the QCP has
been investigated.

While the equilibrium Gutzwiller approach predicts a screened local
moment and hence describes the SC phase, these parameters would be
located deeply in the LM phase in a NRG calculation.  Therefore, we
had to reduce $U$ to $U/\Gn = 1.6< \Uc/D$ for the TD-NRG calculation
to also maintain the NRG dynamics in the SC regime.

We compare the double occupancy $\expect{D(t)}$ from the TD-NRG and
the Gutzwiller approach in Fig.\ \ref{fig:hqLMtoLM_schiro}(a).  The
Gutzwiller approach predicts a very long silent phase for
$0 < t < 1/\Gf$ after the quench, before $\expect{D(t)}$ steeply
rises for $1 < t\Gf$ and then approaches oscillatory the equilibrium
value at long times.  In contrast to these approximate results, the
TD-NRG data perfectly agree with the analytical prediction of Eq.\
\eqref{eq:ptfhq} for the short-time dynamics, predicting a fast
quadratic raising of $\expect{D(t)} \approx t^2 \Gf D / \pi$ for short
times before higher order terms cause a convergence to a finite value.

The origin of the long silent phase in the Gutzwiller approach is
directly related to the slow increase of $|E_{\rm hyp}|$ caused by the
strong analytic restriction of the parameter space in the ansatz as
can be seen from the equation of motion stated in the Supplemental Material
of Ref.\ \cite{Schiro2012}. In the accurate TD-NRG, however, the
energy flow away from the impurity is very fast and occurs on the time
scale of the inverse band width.

\subsubsection{Quenches for exponents $r>1/2$}

\begin{figure}[tb]
  \centering
  \includegraphics[width=\linewidth]{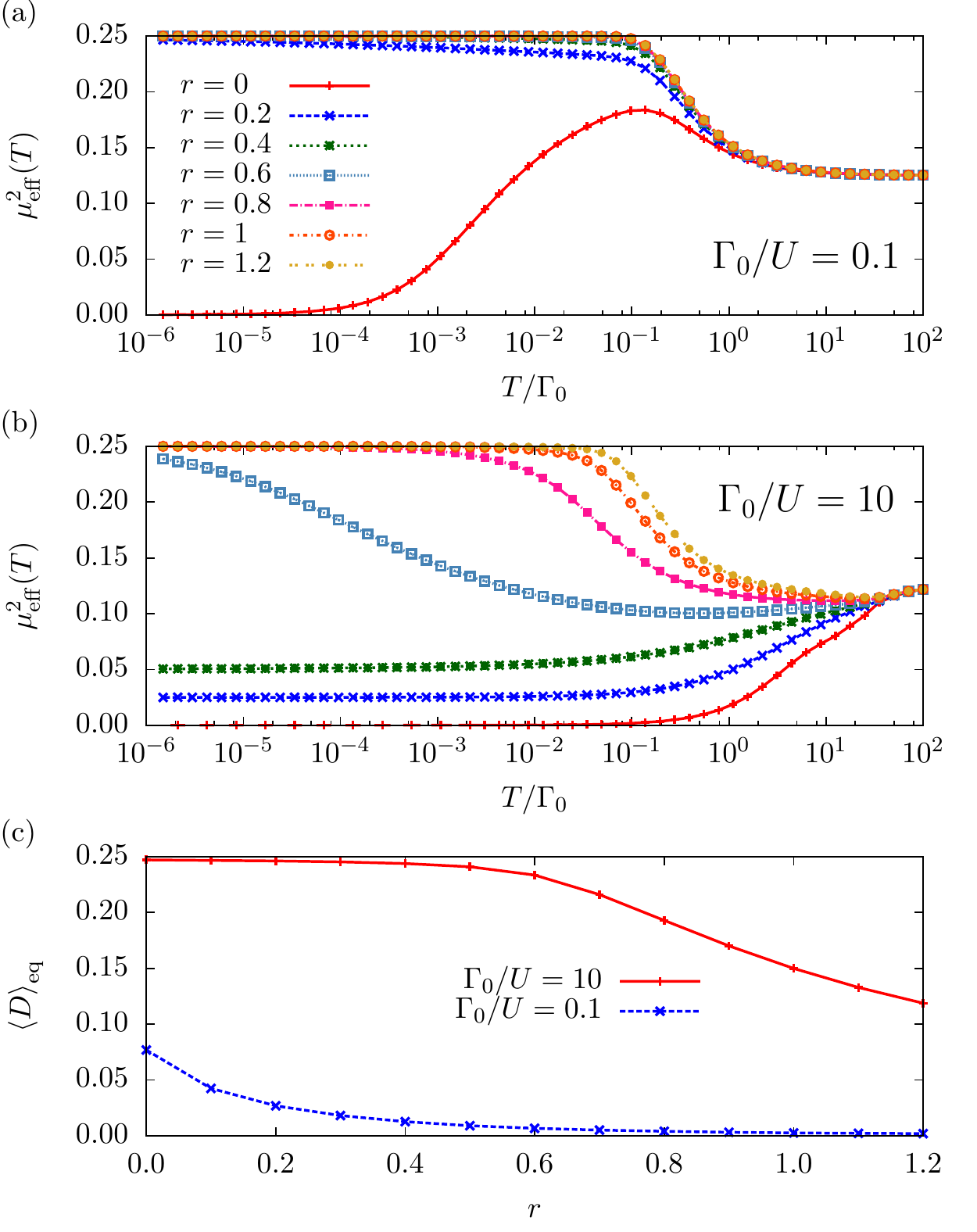}
  \caption{%
    (Color online)
    Equilibrium properties of the pg-SIAM in dependence of the bath exponent $r$ and fixed
    $D/\Gn = 100$. (a) $\mueff(T)$ in a strongly correlated regime $\Gn/U = 0.1$,
    and (b) in a weakly correlated regime $\Gn/U = 10$.
    (c) The equilibrium double occupancy $\expect{D}_{\rm eq}$ for both regimes for $T \to 0$.
  }
  \label{fig:hqRvar_eq}
\end{figure}

So far we restricted ourselves to exponents $0 < r < 1/2$ since a QCP
can be found only for those values if $U > 0$.  For $1/2 < r $ and $ U
> 0$, the system always approaches the LM FP for $T \to 0$ in
equilibrium \cite{GonzalezBuxtonIngersent1998}.

Let us briefly review the equilibrium properties of these regimes for
varying the bath exponent $r$ and a fixed $D/\Gn = 100$ depicted in
Fig.\ \ref{fig:hqRvar_eq}.  The effective local moment $\mueff(T)$ is
plotted versus $T$ in the strongly correlated regime $\Gn/U = 0.1$ in
Fig.\ \ref{fig:hqRvar_eq}(a) and in the weakly correlated regime
$\Gn/U = 10$ in Fig.\ \ref{fig:hqRvar_eq}(b) for
$r=0,0.2,\cdots,1.2$.  For $r=0$ the system always approaches the SC
FP with $\mueff(0) = \lim_{T \to 0} \mueff(T) = 0$ while for $0.2\le r
< 1/2$, the symmetric SC FP with $\mueff(0) = r/8$ for $\Gn/U = 10$,
and the LM FP $\mueff(0) = 1/4$ for $\Gn/U = 0.1$ is reached.  For $1/2 < r$, the
system always flows to the LM FP, and the crossover scale $T^*$ is
increasing with increasing $r$.

The corresponding equilibrium double occupancy $\expect{D}_{\rm eq}$
versus $r$ is shown in Fig.\ \ref{fig:hqRvar_eq}(c) for $T\to 0$.  In
the weakly correlated regime $\Gn/U = 10$ the double occupancy
$\expect{D}_{\rm eq}$ remains close to the uncorrelated value of
$0.25$ and only weakly dependent on $r$ for $0 < r < 1/2$.  The slow
decrease of $\expect{D}_{\rm eq}$ is related to a reduced screening of
the impurity with increasing $r$; cf. increasing residual local
moment in Fig.\ \ref{fig:hqRvar_eq}(b).  For $1/2 < r$ the double
occupancy $\expect{D}_{\rm eq}$ declines much faster with increasing
$r$ since the system approaches the LM FP for $T\to 0$.  For the
strongly correlated regime $\Gn/U = 0.1$, the double occupancy
$\expect{D}_{\rm eq}$ is already strongly suppressed for small $r$.

\begin{figure}[tb]
  \centering
  \includegraphics[width=\linewidth]{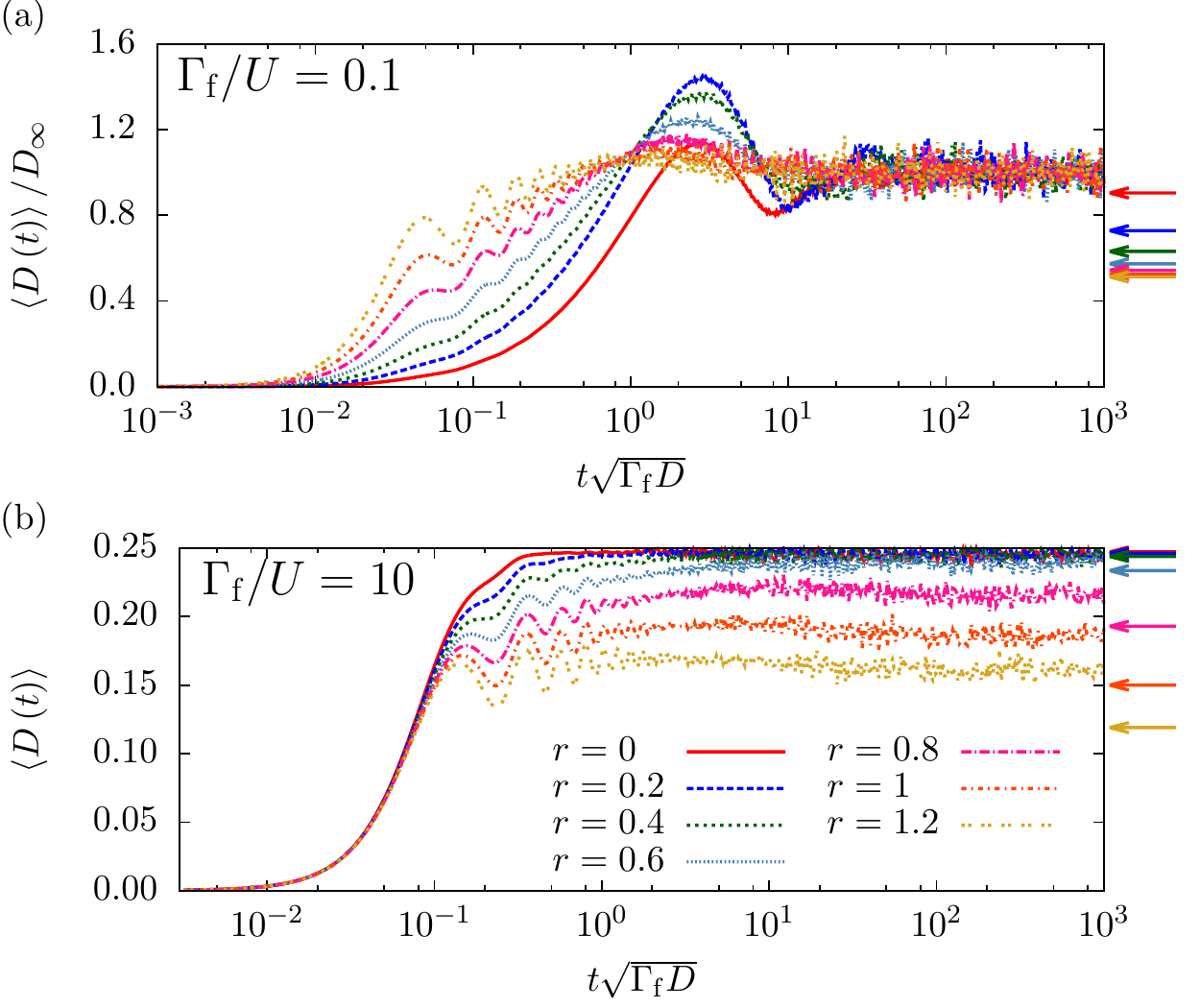}
  \caption{%
    (Color online)
    $\expect{D(t)}$ versus the dimensionless time $t \sqrt{ \Gf D }$   
    for different bath exponents $r$ in hybridization quenches
    from $\Gi=0$ to (a) $\Gf/U=0.1$ and (b) $\Gf/U=10$ 
    and fixed $D/U=100$, and $U = \Ui = \Uf = const$.
     In (a) we divided $\expect{D(t)}$ by $D_{\infty}$ (cf. Eq. \eqref{eq:def-D-oo}) to enlighten the oscillatory
    behavior independently of the absolute value of $\expect{D(t)}$.
    The equilibrium expectation values after the quench are marked by the arrows at the right side of the graph.
  }
  \label{fig:hqRvar_td}
\end{figure}

Now we present our results for hybridization quenches as a function of
$r$ for fixed $D/\Gi=D/\Gf=100$ and fixed $\epsilon_{\rm d} = -U/2$.  The TD-NRG
results for the double occupancy $\expect{D(t)}$ are depicted for two
values for $\Gf/\Uf = 0.1, 10$ and $r=0, 0.2, 0.4, 0.6, 0.8, 1.0, 1.2$
in Fig.\ \ref{fig:hqRvar_td}.

With the exception of $r=0$, Fig.\ \ref{fig:hqRvar_td}(a) shows
quenches within the LM phase.  The double occupancy $\expect{D(t)}$
rises from its initial value towards its steady-state value which
differs significantly from its thermal equilibrium.  Since the order
of magnitude of $\expect{D}_{\rm eq}$ is depicted in Fig.\
\ref{fig:hqRvar_eq}(c) we have divided out its steady-state value
$D_\infty$ and also indicated the ratio $\expect{D}_{\rm eq}/D_\infty$
as arrows on the right side with the same color to illuminate the
short-time oscillatory behavior independently of the absolute value of
$\expect{D(t)}$.  Due to the rescaling, the time scale $\tau =
1/\sqrt{\Gf D}$ governing the overall response is not directly
discernible as in Fig.\ \ref{fig:hqRvar_td} (b).

On an intermediate time scale, we observe a damped oscillatory
behavior for small $r$.  These oscillations are more strongly pronounced
with increasing $r$ due to a stronger concentrated local moment
formation around the impurity, and therefore the local dynamics between
the impurity an the first Wilson shell is more emphasized.  The
deviation of the equilibrated steady-state value and the thermodynamic
equilibrium, indicated by the arrows on the right of the graph, is
increasing with increasing $r$ which is also correlated with the
increasing localization of the effective moment.

For the weakly correlated regime, depicted in Fig.\
\ref{fig:hqRvar_td}(b), we find thermalization only for quenches into
the SC phase.  For $1/2 < r$, the steady-state value of
$\expect{D(t)}$ is reduced with increasing $r$, and the difference to
the thermal equilibrium is also continuously increasing due to quenching
into the LM phase.  Plotting the data versus $t\sqrt{\Gf D}$
clearly reveals the universal $r$-independent time scale governing the
short-time dynamics, as predicted by our perturbation theory.  For
quenches remaining in the LM phase, we observe short-time oscillations
whose damping decreases with increasing $r$.  This oscillations have
the same frequency dependency as in Fig.\ \ref{fig:hqRvar_td}(a)
proportional to the Coulomb repulsion $U$.

\section{Conclusion}

We have analyzed in detail the impurity dynamics of the pg-SIAM after 
local quenches.

In Sec.\ \ref{sec:equilibriumproperties} we have summarized the known
equilibrium properties of the pseudogap model.  There we have
discussed the limitations of a Gutzwiller ansatz
\cite{Gutzwiller1965,SeiboldLorenzana2001,LanataStrand2012,Schiro2012}
which only allows a decoupling spin moment forming on the impurity
site in the LM phase whereas the NRG is able to generate the correct
ground state describing a spatially extended spin-moment formation.

We have distinguished two quench types: the interaction quench in
Sec.\ \ref{sec:interactionquenches} and the hybridization quench in
Sec.\ \ref{sec:hybridisationquenches}.

For the former we have found a universal curve for interaction quenches
within the SC phase: All dependencies of the Coulomb repulsion $\Uf$,
the band width $D$, and the bath exponent $r$ 
are included in  the
scaling function $f(t)$ defined by Eq.\ \eqref{eq:f} and the crossover time scale
$\tco$ approximated by Eq.\ \eqref{eq:11}.

From the temperature-dependent effective moment $\mueff(T)$ in the LM
regime we have defined the crossover scale $T^*$ that depends on the
distance $U-\Uc$ and vanishes for $\Uc$.  Since $T^*$ is directly
related to the NRG iteration $N^*$ beyond which the LM FP is approached,
$1/T^*$ can be interpreted as a characteristic length scale of the
local-moment formation: the larger $T^*$ will be, the more localized
the effective spin moment decoupling from the rest of system will be.
Moreover, we have demonstrated that the difference between the
steady-state value $D_\infty$ and thermodynamic expectation value
$D_{\rm eq}$ only depends on $T^*$ independently of varying $\Uf$ at
fixed $D$ or varying $D$ for fixed $\Uf$.
This spatial dependency of the decoupled moment encoded in the NRG
ground state cannot be accounted for in the Gutzwiller approach
\cite{Schiro2012} since the wave-function ansatz restricts the moment
formation onto the impurity site.

In the presented hybridization quenches we always start from a
decoupled impurity at $t=0$ and switch on $H_{\rm hyp}$ with different
coupling strength. For a small hybridization, the system remains in
the LM phase while a large hybridization drives the system across
the QCP into the SC phase.
We have gauged the quality of our numerical results by our analytic
second-order perturbation theory, which becomes exact for times $t\to
0$, and found excellent agreement between the analytics and the
numerics in the applicability range of the perturbation theory. We
could show that the short-time dynamics is governed by the time scale
$1/\sqrt{\Gf D}$ which is independent of the Coulomb interaction.

By comparison of our data with the recent Gutzwiller results
\cite{Schiro2012} we could demonstrate the shortcomings of this
variational ansatz that already strongly deviates from the
asymptotically exact perturbative result for short-time scales.  After
illustrating the major difference of the physical content in the
ground-state wave function between the approximate Gutzwiller approach
and the exact NRG ground state, the differences between the
equilibrium as well as the nonequilibrium results of both methods
become transparent.

Thermalization was observed within the errors of the TD-NRG
\cite{EidelsteinGuettgeSchillerAnders2012,GuettgeAndersSchiller2013}
for quenches within or into the SC phase.  Due to the extended nature of
the decoupling effective spin in the LM phase, we still found a
steady-state value for $\expect{D(t)}$ that, however, increasingly
deviates from the thermodynamic expectation value.  Interestingly, we
found a thermalization of the hybridization energy $E_{\rm hyb}$ which
accounts for the major local energy change after the quench. We also
provide a short overview for hybridization quenches for
exponents $1/2 < r$ where only a LM FP is found for $U>0$.

\section{Acknowledgments}

We thank Florian Gebhard and J\"org B\"unnemann for fruitful discussions on the nature of the Gutzwiller
wave function for quantum impurity systems. 
We acknowledge financial support by the German-Israel Foundation through Grant No.~1035/2009 
and by the Deutsche Forschungsgemeinschaft through AN 275/7-1.
CPU time was partially granted by the NIC, FZ J\"ulich, under project No. HHB00.

\appendix
\section{Perturbation theory for hybridization quenches}
\label{app:pt}

In general, we assume that the system is in equilibrium for $t<0$ and that its dynamics is governed by the initial
Hamiltonian $H_{\rm 0}$ which is switched to $H_{\rm f} = H_{\rm 0} + H_{\rm p}$ by the
additional perturbation $H_{\rm p}$.

The real-time dynamics of any operator $\hat O$ can be calculated using a time-dependent density operator $\hat \rho(t)$,
\begin{align}
  \expect{ O(t) } = {\rm Tr} \left\{ \hat \rho(t) \hat O \right\}
                        = {\rm Tr} \left\{ \hat \rho^I(t) \hat O^I(t) \right\}
  \label{a:pT1}
\end{align}
where we have transformed all the operators into the interaction picture
$\hat O^I(t) =\exp(i H_{\rm 0} t) \hat O \exp(-i H_{\rm 0} t)$ in the last step.

The time evolution of the density operator is calculated by integrating the von Neumann equation
\begin{align}
  \frac{\partial \hat \rho^I}{\partial t} = i [\hat \rho^I(t), H_{\rm p}^I(t)] 
\end{align}
to
\begin{align}
  \hat \rho^I(t) = \hat \rho_{\rm 0} + i \int_0^t {\rm d} \tau_1 [\hat \rho^I (\tau_1), H_{\rm p}^I(\tau_1)]
  \label{eq:A3}
\end{align}
where we have used the boundary condition $\hat \rho^I(0) = \hat \rho_{\rm 0} = \exp(-\beta H_{\rm 0})/Z_{\rm 0}$
and $Z_{\rm 0} = {\rm Tr} \left\{ \exp(-\beta H_{\rm 0}) \right\}$.

Iterating Eq.\ \eqref{eq:A3} once more and substituting the expression into \eqref{a:pT1} the time evolution
of an expectation value up to second order of the perturbation is given by the general expression
\begin{align}
  &\expect{ O(t)} \approx {\rm Tr} \left\{ \hat \rho_{\rm 0} \hat O \right\}
    + i \int_0^t {\rm d} \tau_1 {\rm Tr} \left\{ \hat \rho_{\rm 0} [ H_{\rm p}^I(\tau_1),
        \hat O^I(t) ] \right\} \nonumber \\
    &\; - \int_0^t {\rm d} \tau_1 \int_0^{\tau_1} {\rm d} \tau_2 {\rm Tr} \left\{ \hat \rho_{\rm 0} 
        [ H_{\rm p}^I(\tau_2), [ H_{\rm p}^I(\tau_1), \hat O^I(t) ] ] \right\}
   \; .
  \label{eq:A4}
\end{align}

For a hybridization quench, we start with a decoupled impurity and switch on the hybridization
$H_{\rm p} = H_{\rm hyb}$ at $t=0$.
Although the Hamiltonian
\begin{align}
  H_{\rm 0} = \sum_{\sigma k} \epsilon_k c_{\sigma k}^\dagger c_{\sigma k}^{\phantom{\dagger}}
        + \sum_{\sigma} \epsilon_{\rm d} d_{\sigma}^\dagger d_{\sigma}^{\phantom{\dagger}}
        + U d_{\downarrow}^\dagger d_{\downarrow}^{\phantom{\dagger}}
            d_{\uparrow}^\dagger d_{\uparrow}^{\phantom{\dagger}}
\end{align}
is not bilinear, it can be diagonalized exactly and all expectation values with
respect to $\hat \rho_{\rm 0}$ are known.
The transformation of $H_{\rm hyb}$ into the interaction picture only requires
\begin{align}
  c_{\sigma k}^I(t) = c_{\sigma k} e^{-i\epsilon_k t}
  \quad , \quad
  c_{\sigma k}^{\dagger I}(t) = c_{\sigma k}^\dagger e^{i\epsilon_k t}
\end{align}
and
\begin{align}
  d_\sigma^I(t) &= |0\rangle\langle \sigma|e^{-i\epsilon_d t}-\sigma|-\sigma\rangle\langle 2|e^{-i(\epsilon_d+U) t} \; ,  \\
  d_\sigma^{\dagger I}(t) &= |\sigma\rangle\langle 0|e^{i\epsilon_d t}-\sigma|2\rangle\langle -\sigma|e^{i(\epsilon_d+U) t} \; .
\end{align}

Now we specialize to a particle-hole symmetric model, $\epsilon_{\rm d} = -U/2$ and $n_\sigma = 1/2$.
For $T \to 0$, the initial double occupancy ${\rm Tr} \left\{ \hat \rho_{\rm 0} \hat D \right\} = 0$;
likewise also the first-order contribution in Eq.\ \eqref{eq:A4} vanishes since
$\text{Tr} \left\{ \hat \rho_{\rm 0} [ H_{\rm hyb}^I(t), \hat D] \right\} = 0$.

After evaluating the commutators and the traces of the second-order contribution for $T\to 0$,
we are left with
\begin{align}
  \expect{D(t)} = \sum_k V^2_{k} f(\epsilon_k) \left( I^+(\epsilon_k) + I^-(\epsilon_k) \right)
\end{align}
where the two integrals
\begin{align}
  I^{\pm}(\epsilon_k) = \int_0^t {\rm d} t_1 \int_0^{t_1} {\rm d} t_2
        &~ {\bf e}^{\pm i ( \epsilon_k - \epsilon_{\rm d} - U ) t_2} \\
        &\quad \times {\bf e}^{\mp i ( \epsilon_k - \epsilon_{\rm d} - U ) t_1}
  \nonumber
\end{align}
can be evaluated analytically.

Using the pseudogap density of states we rewrite the sum over $k$ as an integral over $\epsilon$,
\begin{align}
  \expect{D(t)} = \int_{-D}^D \Gamma(\epsilon)
      \frac{ f(\epsilon) [ 1 - \cos( ( \epsilon - \epsilon_{\rm d} - U ) t ) ] }
        { ( \epsilon - \epsilon_{\rm d} - U )^2 } {\rm d} \epsilon
\end{align}
which requires numerical evaluation.

For $T \to 0$, we can perform the integration analytically in a series expansion
and obtain the final result
\begin{align}
  \expect{D(t)} &= \frac{\Gn D}{\pi} t^2
    + \frac{2 \Gn (1+r)}{\pi} \sum_{n=2}^{\infty} \frac{(-1)^{2n}}{(2n)!} \\
    &\qquad \times \int_{-D}^0 \left| \frac{\epsilon}{D}\right|^r
    \left( \left( \epsilon - \epsilon_{\rm d} - U \right) t \right)^{2n} \text{ d} \epsilon \nonumber
    \quad .
\end{align}

%


\begin{thebibliography}{50}%
\makeatletter
\providecommand \@ifxundefined [1]{%
 \@ifx{#1\undefined}
}%
\providecommand \@ifnum [1]{%
 \ifnum #1\expandafter \@firstoftwo
 \else \expandafter \@secondoftwo
 \fi
}%
\providecommand \@ifx [1]{%
 \ifx #1\expandafter \@firstoftwo
 \else \expandafter \@secondoftwo
 \fi
}%
\providecommand \natexlab [1]{#1}%
\providecommand \enquote  [1]{``#1''}%
\providecommand \bibnamefont  [1]{#1}%
\providecommand \bibfnamefont [1]{#1}%
\providecommand \citenamefont [1]{#1}%
\providecommand \href@noop [0]{\@secondoftwo}%
\providecommand \href [0]{\begingroup \@sanitize@url \@href}%
\providecommand \@href[1]{\@@startlink{#1}\@@href}%
\providecommand \@@href[1]{\endgroup#1\@@endlink}%
\providecommand \@sanitize@url [0]{\catcode `\\12\catcode `\$12\catcode
  `\&12\catcode `\#12\catcode `\^12\catcode `\_12\catcode `\%12\relax}%
\providecommand \@@startlink[1]{}%
\providecommand \@@endlink[0]{}%
\providecommand \url  [0]{\begingroup\@sanitize@url \@url }%
\providecommand \@url [1]{\endgroup\@href {#1}{\urlprefix }}%
\providecommand \urlprefix  [0]{URL }%
\providecommand \Eprint [0]{\href }%
\providecommand \doibase [0]{http://dx.doi.org/}%
\providecommand \selectlanguage [0]{\@gobble}%
\providecommand \bibinfo  [0]{\@secondoftwo}%
\providecommand \bibfield  [0]{\@secondoftwo}%
\providecommand \translation [1]{[#1]}%
\providecommand \BibitemOpen [0]{}%
\providecommand \bibitemStop [0]{}%
\providecommand \bibitemNoStop [0]{.\EOS\space}%
\providecommand \EOS [0]{\spacefactor3000\relax}%
\providecommand \BibitemShut  [1]{\csname bibitem#1\endcsname}%
\let\auto@bib@innerbib\@empty
\bibitem [{\citenamefont {Elzerman}\ \emph {et~al.}(2004)\citenamefont
  {Elzerman}, \citenamefont {Hanson}, \citenamefont {van Beveeren},
  \citenamefont {Witkamp}, \citenamefont {Vandersypen},\ and\ \citenamefont
  {Kouvenhoven}}]{Kouwenhoven2004}%
  \BibitemOpen
  \bibfield  {author} {\bibinfo {author} {\bibfnamefont {J.~M.}\ \bibnamefont
  {Elzerman}}, \bibinfo {author} {\bibfnamefont {R.}~\bibnamefont {Hanson}},
  \bibinfo {author} {\bibfnamefont {L.~H.~W.}\ \bibnamefont {van Beveeren}},
  \bibinfo {author} {\bibfnamefont {B.}~\bibnamefont {Witkamp}}, \bibinfo
  {author} {\bibfnamefont {L.~M.~K.}\ \bibnamefont {Vandersypen}}, \ and\
  \bibinfo {author} {\bibfnamefont {L.~P.}\ \bibnamefont {Kouvenhoven}},\
  }\href@noop {} {\bibfield  {journal} {\bibinfo  {journal} {Nature}\ }\textbf
  {\bibinfo {volume} {430}},\ \bibinfo {pages} {431} (\bibinfo {year}
  {2004})}\BibitemShut {NoStop}%
\bibitem [{\citenamefont {Anderson}(1961)}]{Anderson61}%
  \BibitemOpen
  \bibfield  {author} {\bibinfo {author} {\bibfnamefont {P.~W.}\ \bibnamefont
  {Anderson}},\ }\href@noop {} {\bibfield  {journal} {\bibinfo  {journal}
  {Phys. Rev.}\ }\textbf {\bibinfo {volume} {124}},\ \bibinfo {pages} {41}
  (\bibinfo {year} {1961})}\BibitemShut {NoStop}%
\bibitem [{\citenamefont {Krishna-murthy}\ \emph
  {et~al.}(1980{\natexlab{a}})\citenamefont {Krishna-murthy}, \citenamefont
  {Wilkins},\ and\ \citenamefont {Wilson}}]{KrishWilWilson80a}%
  \BibitemOpen
  \bibfield  {author} {\bibinfo {author} {\bibfnamefont {H.~R.}\ \bibnamefont
  {Krishna-murthy}}, \bibinfo {author} {\bibfnamefont {J.~W.}\ \bibnamefont
  {Wilkins}}, \ and\ \bibinfo {author} {\bibfnamefont {K.~G.}\ \bibnamefont
  {Wilson}},\ }\href@noop {} {\bibfield  {journal} {\bibinfo  {journal} {Phys.
  Rev. B}\ }\textbf {\bibinfo {volume} {21}},\ \bibinfo {pages} {1003}
  (\bibinfo {year} {1980}{\natexlab{a}})}\BibitemShut {NoStop}%
\bibitem [{\citenamefont {Krishna-murthy}\ \emph
  {et~al.}(1980{\natexlab{b}})\citenamefont {Krishna-murthy}, \citenamefont
  {Wilkins},\ and\ \citenamefont {Wilson}}]{KrishWilWilson80b}%
  \BibitemOpen
  \bibfield  {author} {\bibinfo {author} {\bibfnamefont {H.~R.}\ \bibnamefont
  {Krishna-murthy}}, \bibinfo {author} {\bibfnamefont {J.~W.}\ \bibnamefont
  {Wilkins}}, \ and\ \bibinfo {author} {\bibfnamefont {K.~G.}\ \bibnamefont
  {Wilson}},\ }\href@noop {} {\bibfield  {journal} {\bibinfo  {journal} {Phys.
  Rev. B}\ }\textbf {\bibinfo {volume} {21}},\ \bibinfo {pages} {1044}
  (\bibinfo {year} {1980}{\natexlab{b}})}\BibitemShut {NoStop}%
\bibitem [{\citenamefont {Bulla}\ \emph {et~al.}(2008)\citenamefont {Bulla},
  \citenamefont {Costi},\ and\ \citenamefont
  {Pruschke}}]{BullaCostiPruschke2008}%
  \BibitemOpen
  \bibfield  {author} {\bibinfo {author} {\bibfnamefont {R.}~\bibnamefont
  {Bulla}}, \bibinfo {author} {\bibfnamefont {T.~A.}\ \bibnamefont {Costi}}, \
  and\ \bibinfo {author} {\bibfnamefont {T.}~\bibnamefont {Pruschke}},\
  }\href@noop {} {\bibfield  {journal} {\bibinfo  {journal} {Rev.~Mod.~Phys.}\
  }\textbf {\bibinfo {volume} {80}},\ \bibinfo {pages} {395} (\bibinfo {year}
  {2008})}\BibitemShut {NoStop}%
\bibitem [{\citenamefont {Bulla}\ \emph {et~al.}(1997)\citenamefont {Bulla},
  \citenamefont {Pruschke},\ and\ \citenamefont
  {Hewson}}]{BullaPruschkeHewson1997}%
  \BibitemOpen
  \bibfield  {author} {\bibinfo {author} {\bibfnamefont {R.}~\bibnamefont
  {Bulla}}, \bibinfo {author} {\bibfnamefont {T.}~\bibnamefont {Pruschke}}, \
  and\ \bibinfo {author} {\bibfnamefont {A.~C.}\ \bibnamefont {Hewson}},\
  }\href {http://stacks.iop.org/0953-8984/9/i=47/a=014} {\bibfield  {journal}
  {\bibinfo  {journal} {Journal of Physics: Condensed Matter}\ }\textbf
  {\bibinfo {volume} {9}},\ \bibinfo {pages} {10463} (\bibinfo {year}
  {1997})}\BibitemShut {NoStop}%
\bibitem [{\citenamefont {Gonzalez-Buxton}\ and\ \citenamefont
  {Ingersent}(1998)}]{GonzalezBuxtonIngersent1998}%
  \BibitemOpen
  \bibfield  {author} {\bibinfo {author} {\bibfnamefont {C.}~\bibnamefont
  {Gonzalez-Buxton}}\ and\ \bibinfo {author} {\bibfnamefont {K.}~\bibnamefont
  {Ingersent}},\ }\href {\doibase 10.1103/PhysRevB.57.14254} {\bibfield
  {journal} {\bibinfo  {journal} {Phys. Rev. B}\ }\textbf {\bibinfo {volume}
  {57}},\ \bibinfo {pages} {14254} (\bibinfo {year} {1998})}\BibitemShut
  {NoStop}%
\bibitem [{\citenamefont {Glossop}\ and\ \citenamefont
  {Logan}(2003)}]{GlossopLogan2003}%
  \BibitemOpen
  \bibfield  {author} {\bibinfo {author} {\bibfnamefont {M.~T.}\ \bibnamefont
  {Glossop}}\ and\ \bibinfo {author} {\bibfnamefont {D.~E.}\ \bibnamefont
  {Logan}},\ }\href {http://stacks.iop.org/0953-8984/15/i=44/a=007} {\bibfield
  {journal} {\bibinfo  {journal} {Journal of Physics: Condensed Matter}\
  }\textbf {\bibinfo {volume} {15}},\ \bibinfo {pages} {7519} (\bibinfo {year}
  {2003})}\BibitemShut {NoStop}%
\bibitem [{\citenamefont {Vojta}(2006)}]{Vojta2006}%
  \BibitemOpen
  \bibfield  {author} {\bibinfo {author} {\bibfnamefont {M.}~\bibnamefont
  {Vojta}},\ }\href {\doibase 10.1080/14786430500070396} {\bibfield  {journal}
  {\bibinfo  {journal} {Philosophical Magazine}\ }\textbf {\bibinfo {volume}
  {86}},\ \bibinfo {pages} {1807} (\bibinfo {year} {2006})}\BibitemShut
  {NoStop}%
\bibitem [{\citenamefont {Withoff}\ and\ \citenamefont
  {Fradkin}(1990)}]{WithoffFradkin1990}%
  \BibitemOpen
  \bibfield  {author} {\bibinfo {author} {\bibfnamefont {D.}~\bibnamefont
  {Withoff}}\ and\ \bibinfo {author} {\bibfnamefont {E.}~\bibnamefont
  {Fradkin}},\ }\href {\doibase 10.1103/PhysRevLett.64.1835} {\bibfield
  {journal} {\bibinfo  {journal} {Phys. Rev. Lett.}\ }\textbf {\bibinfo
  {volume} {64}},\ \bibinfo {pages} {1835} (\bibinfo {year}
  {1990})}\BibitemShut {NoStop}%
\bibitem [{\citenamefont {Chen}\ and\ \citenamefont
  {Jayaprakash}(1995)}]{Chen1995}%
  \BibitemOpen
  \bibfield  {author} {\bibinfo {author} {\bibfnamefont {K.}~\bibnamefont
  {Chen}}\ and\ \bibinfo {author} {\bibfnamefont {C.}~\bibnamefont
  {Jayaprakash}},\ }\href {http://stacks.iop.org/0953-8984/7/i=37/a=003}
  {\bibfield  {journal} {\bibinfo  {journal} {Journal of Physics: Condensed
  Matter}\ }\textbf {\bibinfo {volume} {7}},\ \bibinfo {pages} {L491} (\bibinfo
  {year} {1995})}\BibitemShut {NoStop}%
\bibitem [{\citenamefont {Ingersent}(1996)}]{Ingersent1996}%
  \BibitemOpen
  \bibfield  {author} {\bibinfo {author} {\bibfnamefont {K.}~\bibnamefont
  {Ingersent}},\ }\href {\doibase 10.1103/PhysRevB.54.11936} {\bibfield
  {journal} {\bibinfo  {journal} {Phys. Rev. B}\ }\textbf {\bibinfo {volume}
  {54}},\ \bibinfo {pages} {11936} (\bibinfo {year} {1996})}\BibitemShut
  {NoStop}%
\bibitem [{\citenamefont {Wilson}(1975)}]{Wilson75}%
  \BibitemOpen
  \bibfield  {author} {\bibinfo {author} {\bibfnamefont {K.~G.}\ \bibnamefont
  {Wilson}},\ }\href@noop {} {\bibfield  {journal} {\bibinfo  {journal} {Rev.
  Mod. Phys.}\ }\textbf {\bibinfo {volume} {47}},\ \bibinfo {pages} {773}
  (\bibinfo {year} {1975})}\BibitemShut {NoStop}%
\bibitem [{\citenamefont {Anders}\ and\ \citenamefont
  {Schiller}(2005)}]{AndersSchiller2005}%
  \BibitemOpen
  \bibfield  {author} {\bibinfo {author} {\bibfnamefont {F.~B.}\ \bibnamefont
  {Anders}}\ and\ \bibinfo {author} {\bibfnamefont {A.}~\bibnamefont
  {Schiller}},\ }\href@noop {} {\bibfield  {journal} {\bibinfo  {journal}
  {Phys. Rev. Lett.}\ }\textbf {\bibinfo {volume} {95}},\ \bibinfo {pages}
  {196801} (\bibinfo {year} {2005})}\BibitemShut {NoStop}%
\bibitem [{\citenamefont {Anders}\ and\ \citenamefont
  {Schiller}(2006)}]{AndersSchiller2006}%
  \BibitemOpen
  \bibfield  {author} {\bibinfo {author} {\bibfnamefont {F.~B.}\ \bibnamefont
  {Anders}}\ and\ \bibinfo {author} {\bibfnamefont {A.}~\bibnamefont
  {Schiller}},\ }\href@noop {} {\bibfield  {journal} {\bibinfo  {journal}
  {Phys. Rev. B}\ }\textbf {\bibinfo {volume} {74}},\ \bibinfo {pages} {245113}
  (\bibinfo {year} {2006})}\BibitemShut {NoStop}%
\bibitem [{\citenamefont {Nghiem}\ and\ \citenamefont
  {Costi}(2014)}]{NghiemCosti2014}%
  \BibitemOpen
  \bibfield  {author} {\bibinfo {author} {\bibfnamefont {H.~T.~M.}\
  \bibnamefont {Nghiem}}\ and\ \bibinfo {author} {\bibfnamefont {T.~A.}\
  \bibnamefont {Costi}},\ }\href {\doibase 10.1103/PhysRevB.89.075118}
  {\bibfield  {journal} {\bibinfo  {journal} {Phys. Rev. B}\ }\textbf {\bibinfo
  {volume} {89}},\ \bibinfo {pages} {075118} (\bibinfo {year}
  {2014})}\BibitemShut {NoStop}%
\bibitem [{\citenamefont {Schir\'o}(2012)}]{Schiro2012}%
  \BibitemOpen
  \bibfield  {author} {\bibinfo {author} {\bibfnamefont {M.}~\bibnamefont
  {Schir\'o}},\ }\href {\doibase 10.1103/PhysRevB.86.161101} {\bibfield
  {journal} {\bibinfo  {journal} {Phys. Rev. B}\ }\textbf {\bibinfo {volume}
  {86}},\ \bibinfo {pages} {161101} (\bibinfo {year} {2012})}\BibitemShut
  {NoStop}%
\bibitem [{\citenamefont {Gutzwiller}(1965)}]{Gutzwiller1965}%
  \BibitemOpen
  \bibfield  {author} {\bibinfo {author} {\bibfnamefont {M.~C.}\ \bibnamefont
  {Gutzwiller}},\ }\href {\doibase 10.1103/PhysRev.137.A1726} {\bibfield
  {journal} {\bibinfo  {journal} {Phys. Rev.}\ }\textbf {\bibinfo {volume}
  {137}},\ \bibinfo {pages} {A1726} (\bibinfo {year} {1965})}\BibitemShut
  {NoStop}%
\bibitem [{\citenamefont {Schickling}\ \emph {et~al.}(2012)\citenamefont
  {Schickling}, \citenamefont {Gebhard}, \citenamefont {B\"unemann},
  \citenamefont {Boeri}, \citenamefont {Andersen},\ and\ \citenamefont
  {Weber}}]{Schickling2012}%
  \BibitemOpen
  \bibfield  {author} {\bibinfo {author} {\bibfnamefont {T.}~\bibnamefont
  {Schickling}}, \bibinfo {author} {\bibfnamefont {F.}~\bibnamefont {Gebhard}},
  \bibinfo {author} {\bibfnamefont {J.}~\bibnamefont {B\"unemann}}, \bibinfo
  {author} {\bibfnamefont {L.}~\bibnamefont {Boeri}}, \bibinfo {author}
  {\bibfnamefont {O.~K.}\ \bibnamefont {Andersen}}, \ and\ \bibinfo {author}
  {\bibfnamefont {W.}~\bibnamefont {Weber}},\ }\href {\doibase
  10.1103/PhysRevLett.108.036406} {\bibfield  {journal} {\bibinfo  {journal}
  {Phys. Rev. Lett.}\ }\textbf {\bibinfo {volume} {108}},\ \bibinfo {pages}
  {036406} (\bibinfo {year} {2012})}\BibitemShut {NoStop}%
\bibitem [{\citenamefont {Seibold}\ and\ \citenamefont
  {Lorenzana}(2001)}]{SeiboldLorenzana2001}%
  \BibitemOpen
  \bibfield  {author} {\bibinfo {author} {\bibfnamefont {G.}~\bibnamefont
  {Seibold}}\ and\ \bibinfo {author} {\bibfnamefont {J.}~\bibnamefont
  {Lorenzana}},\ }\href {\doibase 10.1103/PhysRevLett.86.2605} {\bibfield
  {journal} {\bibinfo  {journal} {Phys. Rev. Lett.}\ }\textbf {\bibinfo
  {volume} {86}},\ \bibinfo {pages} {2605} (\bibinfo {year}
  {2001})}\BibitemShut {NoStop}%
\bibitem [{\citenamefont {Lanat\`a}\ and\ \citenamefont
  {Strand}(2012)}]{LanataStrand2012}%
  \BibitemOpen
  \bibfield  {author} {\bibinfo {author} {\bibfnamefont {N.}~\bibnamefont
  {Lanat\`a}}\ and\ \bibinfo {author} {\bibfnamefont {H.~U.~R.}\ \bibnamefont
  {Strand}},\ }\href {\doibase 10.1103/PhysRevB.86.115310} {\bibfield
  {journal} {\bibinfo  {journal} {Phys. Rev. B}\ }\textbf {\bibinfo {volume}
  {86}},\ \bibinfo {pages} {115310} (\bibinfo {year} {2012})}\BibitemShut
  {NoStop}%
\bibitem [{\citenamefont {Fritz}\ and\ \citenamefont
  {Vojta}(2013)}]{FritzVojta2013}%
  \BibitemOpen
  \bibfield  {author} {\bibinfo {author} {\bibfnamefont {L.}~\bibnamefont
  {Fritz}}\ and\ \bibinfo {author} {\bibfnamefont {M.}~\bibnamefont {Vojta}},\
  }\href {http://stacks.iop.org/0034-4885/76/i=3/a=032501} {\bibfield
  {journal} {\bibinfo  {journal} {Reports on Progress in Physics}\ }\textbf
  {\bibinfo {volume} {76}},\ \bibinfo {pages} {032501} (\bibinfo {year}
  {2013})}\BibitemShut {NoStop}%
\bibitem [{\citenamefont {Lo}\ \emph {et~al.}(2014)\citenamefont {Lo},
  \citenamefont {Guo},\ and\ \citenamefont {Anders}}]{Lo-graphene-2014}%
  \BibitemOpen
  \bibfield  {author} {\bibinfo {author} {\bibfnamefont {P.-W.}\ \bibnamefont
  {Lo}}, \bibinfo {author} {\bibfnamefont {G.-Y.}\ \bibnamefont {Guo}}, \ and\
  \bibinfo {author} {\bibfnamefont {F.~B.}\ \bibnamefont {Anders}},\ }\href
  {\doibase 10.1103/PhysRevB.89.195424} {\bibfield  {journal} {\bibinfo
  {journal} {Phys. Rev. B}\ }\textbf {\bibinfo {volume} {89}},\ \bibinfo
  {pages} {195424} (\bibinfo {year} {2014})}\BibitemShut {NoStop}%
\bibitem [{\citenamefont {Fritz}\ and\ \citenamefont
  {Vojta}(2004)}]{FritzVojta2004}%
  \BibitemOpen
  \bibfield  {author} {\bibinfo {author} {\bibfnamefont {L.}~\bibnamefont
  {Fritz}}\ and\ \bibinfo {author} {\bibfnamefont {M.}~\bibnamefont {Vojta}},\
  }\href {\doibase 10.1103/PhysRevB.70.214427} {\bibfield  {journal} {\bibinfo
  {journal} {Phys. Rev. B}\ }\textbf {\bibinfo {volume} {70}},\ \bibinfo
  {pages} {214427} (\bibinfo {year} {2004})}\BibitemShut {NoStop}%
\bibitem [{\citenamefont {Vojta}\ and\ \citenamefont
  {Fritz}(2004)}]{VojtaFritz2004}%
  \BibitemOpen
  \bibfield  {author} {\bibinfo {author} {\bibfnamefont {M.}~\bibnamefont
  {Vojta}}\ and\ \bibinfo {author} {\bibfnamefont {L.}~\bibnamefont {Fritz}},\
  }\href {\doibase 10.1103/PhysRevB.70.094502} {\bibfield  {journal} {\bibinfo
  {journal} {Phys. Rev. B}\ }\textbf {\bibinfo {volume} {70}},\ \bibinfo
  {pages} {094502} (\bibinfo {year} {2004})}\BibitemShut {NoStop}%
\bibitem [{\citenamefont {Schneider}\ \emph {et~al.}(2011)\citenamefont
  {Schneider}, \citenamefont {Fritz}, \citenamefont {Anders}, \citenamefont
  {Benlagra},\ and\ \citenamefont {Vojta}}]{Schneider2011}%
  \BibitemOpen
  \bibfield  {author} {\bibinfo {author} {\bibfnamefont {I.}~\bibnamefont
  {Schneider}}, \bibinfo {author} {\bibfnamefont {L.}~\bibnamefont {Fritz}},
  \bibinfo {author} {\bibfnamefont {F.~B.}\ \bibnamefont {Anders}}, \bibinfo
  {author} {\bibfnamefont {A.}~\bibnamefont {Benlagra}}, \ and\ \bibinfo
  {author} {\bibfnamefont {M.}~\bibnamefont {Vojta}},\ }\href {\doibase
  10.1103/PhysRevB.84.125139} {\bibfield  {journal} {\bibinfo  {journal} {Phys.
  Rev. B}\ }\textbf {\bibinfo {volume} {84}},\ \bibinfo {pages} {125139}
  (\bibinfo {year} {2011})}\BibitemShut {NoStop}%
\bibitem [{\citenamefont {Mazur}(1969)}]{Mazur1969}%
  \BibitemOpen
  \bibfield  {author} {\bibinfo {author} {\bibfnamefont {P.}~\bibnamefont
  {Mazur}},\ }\href {\doibase http://dx.doi.org/10.1016/0031-8914(69)90185-2}
  {\bibfield  {journal} {\bibinfo  {journal} {Physica}\ }\textbf {\bibinfo
  {volume} {43}},\ \bibinfo {pages} {533 } (\bibinfo {year}
  {1969})}\BibitemShut {NoStop}%
\bibitem [{\citenamefont {Suzuki}(1971)}]{Suzuki1971}%
  \BibitemOpen
  \bibfield  {author} {\bibinfo {author} {\bibfnamefont {M.}~\bibnamefont
  {Suzuki}},\ }\href {\doibase http://dx.doi.org/10.1016/0031-8914(71)90226-6}
  {\bibfield  {journal} {\bibinfo  {journal} {Physica}\ }\textbf {\bibinfo
  {volume} {51}},\ \bibinfo {pages} {277 } (\bibinfo {year}
  {1971})}\BibitemShut {NoStop}%
\bibitem [{\citenamefont {Uhrig}\ \emph {et~al.}(2014)\citenamefont {Uhrig},
  \citenamefont {Hackmann}, \citenamefont {Stanek}, \citenamefont {Stolze},\
  and\ \citenamefont {Anders}}]{UhrigHackmann2014}%
  \BibitemOpen
  \bibfield  {author} {\bibinfo {author} {\bibfnamefont {G.~S.}\ \bibnamefont
  {Uhrig}}, \bibinfo {author} {\bibfnamefont {J.}~\bibnamefont {Hackmann}},
  \bibinfo {author} {\bibfnamefont {D.}~\bibnamefont {Stanek}}, \bibinfo
  {author} {\bibfnamefont {J.}~\bibnamefont {Stolze}}, \ and\ \bibinfo {author}
  {\bibfnamefont {F.~B.}\ \bibnamefont {Anders}},\ }\href {\doibase
  10.1103/PhysRevB.90.060301} {\bibfield  {journal} {\bibinfo  {journal} {Phys.
  Rev. B}\ }\textbf {\bibinfo {volume} {90}},\ \bibinfo {pages} {060301}
  (\bibinfo {year} {2014})}\BibitemShut {NoStop}%
\bibitem [{\citenamefont {Rosch}(2012)}]{Rosch2012}%
  \BibitemOpen
  \bibfield  {author} {\bibinfo {author} {\bibfnamefont {A.}~\bibnamefont
  {Rosch}},\ }\href {\doibase 10.1140/epjb/e2011-20880-7} {\bibfield  {journal}
  {\bibinfo  {journal} {Eur. Phys. J. B}\ }\textbf {\bibinfo {volume} {85}},\
  \bibinfo {pages} {6} (\bibinfo {year} {2012})}\BibitemShut {NoStop}%
\bibitem [{\citenamefont {Kanao}\ \emph {et~al.}(2012)\citenamefont {Kanao},
  \citenamefont {Matsuura},\ and\ \citenamefont {Ogata}}]{Kanao2012}%
  \BibitemOpen
  \bibfield  {author} {\bibinfo {author} {\bibfnamefont {T.}~\bibnamefont
  {Kanao}}, \bibinfo {author} {\bibfnamefont {H.}~\bibnamefont {Matsuura}}, \
  and\ \bibinfo {author} {\bibfnamefont {M.}~\bibnamefont {Ogata}},\ }\href
  {\doibase 10.1143/JPSJ.81.063709} {\bibfield  {journal} {\bibinfo  {journal}
  {Journal of the Physical Society of Japan}\ }\textbf {\bibinfo {volume}
  {81}},\ \bibinfo {pages} {063709} (\bibinfo {year} {2012})}\BibitemShut
  {NoStop}%
\bibitem [{\citenamefont {Vojta}\ \emph {et~al.}(2010)\citenamefont {Vojta},
  \citenamefont {Fritz},\ and\ \citenamefont {Bulla}}]{VojtaFritzBulla2010}%
  \BibitemOpen
  \bibfield  {author} {\bibinfo {author} {\bibfnamefont {M.}~\bibnamefont
  {Vojta}}, \bibinfo {author} {\bibfnamefont {L.}~\bibnamefont {Fritz}}, \ and\
  \bibinfo {author} {\bibfnamefont {R.}~\bibnamefont {Bulla}},\ }\href
  {http://stacks.iop.org/0295-5075/90/i=2/a=27006} {\bibfield  {journal}
  {\bibinfo  {journal} {Europhys. Lett.}\ }\textbf {\bibinfo {volume} {90}},\
  \bibinfo {pages} {27006} (\bibinfo {year} {2010})}\BibitemShut {NoStop}%
\bibitem [{\citenamefont {Eidelstein}\ \emph {et~al.}(2012)\citenamefont
  {Eidelstein}, \citenamefont {Schiller}, \citenamefont {G\"uttge},\ and\
  \citenamefont {Anders}}]{EidelsteinGuettgeSchillerAnders2012}%
  \BibitemOpen
  \bibfield  {author} {\bibinfo {author} {\bibfnamefont {E.}~\bibnamefont
  {Eidelstein}}, \bibinfo {author} {\bibfnamefont {A.}~\bibnamefont
  {Schiller}}, \bibinfo {author} {\bibfnamefont {F.}~\bibnamefont {G\"uttge}},
  \ and\ \bibinfo {author} {\bibfnamefont {F.~B.}\ \bibnamefont {Anders}},\
  }\href {\doibase 10.1103/PhysRevB.85.075118} {\bibfield  {journal} {\bibinfo
  {journal} {Phys. Rev. B}\ }\textbf {\bibinfo {volume} {85}},\ \bibinfo
  {pages} {075118} (\bibinfo {year} {2012})}\BibitemShut {NoStop}%
\bibitem [{\citenamefont {Costi}\ \emph {et~al.}(1994)\citenamefont {Costi},
  \citenamefont {Hewson},\ and\ \citenamefont {Zlatic}}]{CostiHewsonZlatic94}%
  \BibitemOpen
  \bibfield  {author} {\bibinfo {author} {\bibfnamefont {T.~A.}\ \bibnamefont
  {Costi}}, \bibinfo {author} {\bibfnamefont {A.~C.}\ \bibnamefont {Hewson}}, \
  and\ \bibinfo {author} {\bibfnamefont {V.}~\bibnamefont {Zlatic}},\
  }\href@noop {} {\bibfield  {journal} {\bibinfo  {journal} {J. Phys.: Condens.
  Matter}\ }\textbf {\bibinfo {volume} {6}},\ \bibinfo {pages} {2519} (\bibinfo
  {year} {1994})}\BibitemShut {NoStop}%
\bibitem [{\citenamefont {Peters}\ \emph {et~al.}(2006)\citenamefont {Peters},
  \citenamefont {Pruschke},\ and\ \citenamefont
  {Anders}}]{PetersPruschkeAnders2006}%
  \BibitemOpen
  \bibfield  {author} {\bibinfo {author} {\bibfnamefont {R.}~\bibnamefont
  {Peters}}, \bibinfo {author} {\bibfnamefont {T.}~\bibnamefont {Pruschke}}, \
  and\ \bibinfo {author} {\bibfnamefont {F.~B.}\ \bibnamefont {Anders}},\
  }\href@noop {} {\bibfield  {journal} {\bibinfo  {journal} {Phys. Rev. B}\
  }\textbf {\bibinfo {volume} {74}},\ \bibinfo {pages} {245114} (\bibinfo
  {year} {2006})}\BibitemShut {NoStop}%
\bibitem [{\citenamefont {Weichselbaum}\ and\ \citenamefont {von
  Delft}(2007)}]{WeichselbaumDelft2007}%
  \BibitemOpen
  \bibfield  {author} {\bibinfo {author} {\bibfnamefont {A.}~\bibnamefont
  {Weichselbaum}}\ and\ \bibinfo {author} {\bibfnamefont {J.}~\bibnamefont {von
  Delft}},\ }\href@noop {} {\bibfield  {journal} {\bibinfo  {journal} {Phys.
  Rev. Lett.}\ }\textbf {\bibinfo {volume} {99}},\ \bibinfo {pages} {076402}
  (\bibinfo {year} {2007})}\BibitemShut {NoStop}%
\bibitem [{\citenamefont {Lanat\`a}(2010)}]{Lanata2010}%
  \BibitemOpen
  \bibfield  {author} {\bibinfo {author} {\bibfnamefont {N.}~\bibnamefont
  {Lanat\`a}},\ }\href {\doibase 10.1103/PhysRevB.82.195326} {\bibfield
  {journal} {\bibinfo  {journal} {Phys. Rev. B}\ }\textbf {\bibinfo {volume}
  {82}},\ \bibinfo {pages} {195326} (\bibinfo {year} {2010})}\BibitemShut
  {NoStop}%
\bibitem [{Note1()}]{Note1}%
  \BibitemOpen
  \bibinfo {note} {The precise mathematical definition of $T^*$ will be given
  in Sec.\ \ref {sec:interactionquenchesover}.}\BibitemShut {Stop}%
\bibitem [{\citenamefont {Barzykin}\ and\ \citenamefont
  {Affleck}(1996)}]{Barzykin1996}%
  \BibitemOpen
  \bibfield  {author} {\bibinfo {author} {\bibfnamefont {V.}~\bibnamefont
  {Barzykin}}\ and\ \bibinfo {author} {\bibfnamefont {I.}~\bibnamefont
  {Affleck}},\ }\href {\doibase 10.1103/PhysRevLett.76.4959} {\bibfield
  {journal} {\bibinfo  {journal} {Phys. Rev. Lett.}\ }\textbf {\bibinfo
  {volume} {76}},\ \bibinfo {pages} {4959} (\bibinfo {year}
  {1996})}\BibitemShut {NoStop}%
\bibitem [{\citenamefont {Barzykin}\ and\ \citenamefont
  {Affleck}(1998)}]{Barzykin1998}%
  \BibitemOpen
  \bibfield  {author} {\bibinfo {author} {\bibfnamefont {V.}~\bibnamefont
  {Barzykin}}\ and\ \bibinfo {author} {\bibfnamefont {I.}~\bibnamefont
  {Affleck}},\ }\href {\doibase 10.1103/PhysRevB.57.432} {\bibfield  {journal}
  {\bibinfo  {journal} {Phys. Rev. B}\ }\textbf {\bibinfo {volume} {57}},\
  \bibinfo {pages} {432} (\bibinfo {year} {1998})}\BibitemShut {NoStop}%
\bibitem [{\citenamefont {Affleck}\ and\ \citenamefont
  {Simon}(2001)}]{Affleck2001}%
  \BibitemOpen
  \bibfield  {author} {\bibinfo {author} {\bibfnamefont {I.}~\bibnamefont
  {Affleck}}\ and\ \bibinfo {author} {\bibfnamefont {P.}~\bibnamefont
  {Simon}},\ }\href {\doibase 10.1103/PhysRevLett.86.2854} {\bibfield
  {journal} {\bibinfo  {journal} {Phys. Rev. Lett.}\ }\textbf {\bibinfo
  {volume} {86}},\ \bibinfo {pages} {2854} (\bibinfo {year}
  {2001})}\BibitemShut {NoStop}%
\bibitem [{\citenamefont {S\o{}rensen}\ and\ \citenamefont
  {Affleck}(2005)}]{Affleck2005}%
  \BibitemOpen
  \bibfield  {author} {\bibinfo {author} {\bibfnamefont {E.~S.}\ \bibnamefont
  {S\o{}rensen}}\ and\ \bibinfo {author} {\bibfnamefont {I.}~\bibnamefont
  {Affleck}},\ }\href {\doibase 10.1103/PhysRevLett.94.086601} {\bibfield
  {journal} {\bibinfo  {journal} {Phys. Rev. Lett.}\ }\textbf {\bibinfo
  {volume} {94}},\ \bibinfo {pages} {086601} (\bibinfo {year}
  {2005})}\BibitemShut {NoStop}%
\bibitem [{\citenamefont {Affleck}\ \emph {et~al.}(2008)\citenamefont
  {Affleck}, \citenamefont {Borda},\ and\ \citenamefont
  {Saleur}}]{Affleck2008}%
  \BibitemOpen
  \bibfield  {author} {\bibinfo {author} {\bibfnamefont {I.}~\bibnamefont
  {Affleck}}, \bibinfo {author} {\bibfnamefont {L.}~\bibnamefont {Borda}}, \
  and\ \bibinfo {author} {\bibfnamefont {H.}~\bibnamefont {Saleur}},\ }\href
  {\doibase 10.1103/PhysRevB.77.180404} {\bibfield  {journal} {\bibinfo
  {journal} {Phys. Rev. B}\ }\textbf {\bibinfo {volume} {77}},\ \bibinfo
  {pages} {180404} (\bibinfo {year} {2008})}\BibitemShut {NoStop}%
\bibitem [{\citenamefont {Lechtenberg}\ and\ \citenamefont
  {Anders}(2014)}]{LechtenbergAnders2014}%
  \BibitemOpen
  \bibfield  {author} {\bibinfo {author} {\bibfnamefont {B.}~\bibnamefont
  {Lechtenberg}}\ and\ \bibinfo {author} {\bibfnamefont {F.~B.}\ \bibnamefont
  {Anders}},\ }\href {\doibase 10.1103/PhysRevB.90.045117} {\bibfield
  {journal} {\bibinfo  {journal} {Phys. Rev. B}\ }\textbf {\bibinfo {volume}
  {90}},\ \bibinfo {pages} {045117} (\bibinfo {year} {2014})}\BibitemShut
  {NoStop}%
\bibitem [{\citenamefont {Chowdhury}\ and\ \citenamefont
  {Ingersent}(2014)}]{ChowdhuryIngerset2014}%
  \BibitemOpen
  \bibfield  {author} {\bibinfo {author} {\bibfnamefont {T.}~\bibnamefont
  {Chowdhury}}\ and\ \bibinfo {author} {\bibfnamefont {K.}~\bibnamefont
  {Ingersent}},\ }\href@noop {} {\bibfield  {journal} {\bibinfo  {journal}
  {arXiv:1410.5546}\ } (\bibinfo {year} {2014})}\BibitemShut {NoStop}%
\bibitem [{\citenamefont {Guettge}\ \emph {et~al.}(2013)\citenamefont
  {Guettge}, \citenamefont {Anders}, \citenamefont {Schollwoeck}, \citenamefont
  {Eidelstein},\ and\ \citenamefont {Schiller}}]{GuettgeAndersSchiller2013}%
  \BibitemOpen
  \bibfield  {author} {\bibinfo {author} {\bibfnamefont {F.}~\bibnamefont
  {Guettge}}, \bibinfo {author} {\bibfnamefont {F.~B.}\ \bibnamefont {Anders}},
  \bibinfo {author} {\bibfnamefont {U.}~\bibnamefont {Schollwoeck}}, \bibinfo
  {author} {\bibfnamefont {E.}~\bibnamefont {Eidelstein}}, \ and\ \bibinfo
  {author} {\bibfnamefont {A.}~\bibnamefont {Schiller}},\ }\href@noop {}
  {\bibfield  {journal} {\bibinfo  {journal} {Phys. Rev. B}\ }\textbf {\bibinfo
  {volume} {87}},\ \bibinfo {pages} {115115} (\bibinfo {year}
  {2013})}\BibitemShut {NoStop}%
\bibitem [{\citenamefont {Reimann}(2008)}]{Reimann2008}%
  \BibitemOpen
  \bibfield  {author} {\bibinfo {author} {\bibfnamefont {P.}~\bibnamefont
  {Reimann}},\ }\href {\doibase 10.1103/PhysRevLett.101.190403} {\bibfield
  {journal} {\bibinfo  {journal} {Phys. Rev. Lett.}\ }\textbf {\bibinfo
  {volume} {101}},\ \bibinfo {pages} {190403} (\bibinfo {year}
  {2008})}\BibitemShut {NoStop}%
\bibitem [{\citenamefont {Rigol}\ and\ \citenamefont
  {Srednicki}(2012)}]{RigolETH2012}%
  \BibitemOpen
  \bibfield  {author} {\bibinfo {author} {\bibfnamefont {M.}~\bibnamefont
  {Rigol}}\ and\ \bibinfo {author} {\bibfnamefont {M.}~\bibnamefont
  {Srednicki}},\ }\href {\doibase 10.1103/PhysRevLett.108.110601} {\bibfield
  {journal} {\bibinfo  {journal} {Phys. Rev. Lett.}\ }\textbf {\bibinfo
  {volume} {108}},\ \bibinfo {pages} {110601} (\bibinfo {year}
  {2012})}\BibitemShut {NoStop}%
\bibitem [{\citenamefont {Schrieffer}\ and\ \citenamefont
  {Wolff}(1966)}]{SchriefferWol66}%
  \BibitemOpen
  \bibfield  {author} {\bibinfo {author} {\bibfnamefont {J.~R.}\ \bibnamefont
  {Schrieffer}}\ and\ \bibinfo {author} {\bibfnamefont {P.~A.}\ \bibnamefont
  {Wolff}},\ }\href@noop {} {\bibfield  {journal} {\bibinfo  {journal} {Phys.
  Rev.}\ }\textbf {\bibinfo {volume} {149}},\ \bibinfo {pages} {491} (\bibinfo
  {year} {1966})}\BibitemShut {NoStop}%
\bibitem [{\citenamefont {Nordlander}\ \emph {et~al.}(1999)\citenamefont
  {Nordlander}, \citenamefont {Pustilnik}, \citenamefont {Meir}, \citenamefont
  {Wingreen},\ and\ \citenamefont {Langreth}}]{NordlanderEtAl1999}%
  \BibitemOpen
  \bibfield  {author} {\bibinfo {author} {\bibfnamefont {P.}~\bibnamefont
  {Nordlander}}, \bibinfo {author} {\bibfnamefont {M.}~\bibnamefont
  {Pustilnik}}, \bibinfo {author} {\bibfnamefont {Y.}~\bibnamefont {Meir}},
  \bibinfo {author} {\bibfnamefont {N.~S.}\ \bibnamefont {Wingreen}}, \ and\
  \bibinfo {author} {\bibfnamefont {D.~C.}\ \bibnamefont {Langreth}},\
  }\href@noop {} {\bibfield  {journal} {\bibinfo  {journal} {Phys. Rev. Lett.}\
  }\textbf {\bibinfo {volume} {83}},\ \bibinfo {pages} {808} (\bibinfo {year}
  {1999})}\BibitemShut {NoStop}%
\end{thebibliography}
\end{document}